\documentclass[twocolumn]{openjournal}

\usepackage{lipsum}

\usepackage{xcolor}
\usepackage{textgreek}
\usepackage[utf8]{inputenc}
\usepackage[english]{babel}
\usepackage{amsmath}

\usepackage{hyperref}
\hypersetup{
    unicode, 
    colorlinks=true,
    linkcolor=linkcolor,
    citecolor=linkcolor,
    filecolor=linkcolor,
    urlcolor=linkcolor,
}
\usepackage{color,colortbl}
\definecolor{linkcolor}{rgb}{0.0,0.3,0.5}
\usepackage{tensind}
\tensordelimiter{?}
\DeclareGraphicsExtensions{.bmp,.png,.jpg,.pdf}
\usepackage{verbatim}
\usepackage[normalem]{ulem}
\usepackage{orcidlink}
\usepackage{soul}

\graphicspath{ {./figs/} }

\begin{document}
\title{White dwarf + M dwarf detached binaries in long period radio transients:\\observed binary parameters, evolution, and population constraints}

\author{\vspace{-30pt}Antonio C. Rodriguez$^{1,2}$ \orcidlink{0000-0003-4189-9668}}
\author{Kareem El-Badry$^{2}$ \orcidlink{0000-0002-6871-1752}}
\author{Iris de Ruiter$^{3,4}$ \orcidlink{0000-0002-4752-5467}}
\author{Kaustubh Rajwade$^{5,6}$ \orcidlink{0000-0002-8043-6909}}
\author{Edo Berger$^{1}$ \orcidlink{0000-0002-9392-9681}}
\author{Liam Connor$^{1}$ \orcidlink{0000-0002-7587-6352}}
\author{Natasha Hurley-Walker$^{7}$ \orcidlink{0000-0002-5119-4808}}

\email{Corresponding Author: antonio.c.rodriguez.1998@gmail.com}
\affiliation{$^1$Center for Astrophysics $|$ Harvard \& Smithsonian, 60 Garden Street, Cambridge, MA 02138, USA}
\affiliation{$^2$Department of Astronomy, California Institute of Technology, 1200 E. California Blvd., Pasadena, CA 91125, USA}
\affiliation{$^3$Sydney Institute for Astronomy, School of Physics, The University of Sydney, NSW 2006, Australia}
\affiliation{$^4$ARC Centre of Excellence for Gravitational Wave Discovery (OzGrav), Australia}
\affiliation{$^5$Astrophysics, Denys Wilkinson Building, University of Oxford, Keble Road, Oxford OX1 3RH, UK}
\affiliation{$^6$ Doctoral Training Centre, University of Oxford, Keble Road, Oxford OX1 3NP, UK}
\affiliation{$^7$International Centre for Radio Astronomy Research, Curtin University, Kent Street, Bentley WA, 6102, Australia}

\begin{abstract}
Long period radio transients (LPTs) are the slowest radio-pulsing sources ever found, with the current population spanning periods of seven minutes to over six hours. Two of the fourteen published LPTs, ILT J1101+5521 and GLEAM-X J0704--37, have been associated with an M dwarf closely orbiting a white dwarf (WD) through optical spectroscopy. Here, we present new Keck I/LRIS optical spectroscopy of ILT J1101+5521, which reveals H$\alpha$ emission from the M dwarf and confirms an orbital period nearly matching the radio period (2.092 hr). Radio pulses in both systems arrive just after maximum M dwarf redshift, assuming the radio period matches the orbital period. Based on \textit{Gaia} proper motions and systemic velocities, we find that these systems are \textrm{likely members of the Galactic thick disk}. Both systems harbor unusually massive and cool WDs, with $M_\mathrm{WD} \approx 0.84-1.0 M_\odot$ and $T_\mathrm{eff} \approx 5200-7300$ K, implying that their carbon-oxygen cores are nearly entirely crystallized. Both systems are unusually close to being face-on binaries ($i=13^\circ-28^\circ$), signaling that the production of coherent radio pulses may be a strongly inclination-dependent phenomenon. We present MESA models that show that the M dwarf in each system will fill its Roche lobe within $\sim1$ Gyr, becoming a cataclysmic variable.  Finally, we place lower limits on the space density of WD + M dwarf LPTs ($\rho \gtrsim 10^{-8}\;\mathrm{pc}^{-3}$); based on the broader population of WD + M dwarf binaries, we estimate that there are 100 (1000) WD + M dwarf LPTs within 2 kpc if current radio findings are 100\% (10\%) complete. Current and upcoming radio surveys will be sensitive to many such systems, and M dwarf optical counterparts out to $\sim$2 kpc will be detectable with the Rubin Observatory Legacy Survey of Space and Time (LSST).
\end{abstract}


\maketitle

\section{Introduction}
\label{sec:intro}
Long-period radio transients (LPTs) have rapidly grown as an exciting new phenomenon in radio astronomy. At the time of writing, fourteen such sources have been published (Table \ref{tab:list}; see \cite{2026rea} for a recent review). As a whole, they appear to be concentrated in the Galactic plane and have moderate radio dispersion measures, pointing to a Galactic origin. While both a neutron star (NS) and white dwarf (WD) origin has been argued for LPTs, there is evidence pointing to two separate classes of LPTs: one class in which the radio position is associated with a clear optical counterpart amenable to spectroscopy \citep[e.g.][]{2025deruiter, 2024discovery, 2025rodriguez_gleamx}; and another class in which no optical or near-infrared counterpart (at best, a marginal detection) has been seen even with 8--10m class telescopes \citep[e.g.][ and references in Table \ref{tab:list}]{2025pelisoli_limits}. \textrm{However, the mechanism(s) behind the periodic emission in LPTs is yet to be fully determined, which will ultimately dictate how many subclasses of LPTs indeed exist.}

LPTs show promise in shining a new light on the origin and evolution of magnetism in compact objects. The discovery of pulsars led to a rich study of electromagnetism in relativistic environments \citep[e.g.][]{1968gold, 1968pacini}, and the association of millisecond pulsars with binary companions and globular clusters demonstrated that neutron stars can sustain magnetic fields for billions of years \citep[e.g.][]{1986kulkarni, 1994pulsars}. Explanations for the LPTs unassociated with optical counterparts has ranged from white dwarf pulsars \citep{2005zhang, 2022katz} to hot subdwarf pulsars \citep{2022loeb} to precessing black hole jets \citep{2025nathanail} to magnetars \citep[e.g.][]{2023beniamini, 2024cooper, 2025cary_lu, 2025mao_magnetar}. It has also been proposed that if all LPTs are compact object binaries, they could be strong sources of millihertz gravitational waves observable by future space-based interferometry missions like \textit{LISA} \citep{2025suvorov}. However, theoretical work has been most successful for LPTs associated with optical counterparts --- it has been proposed that the LPT phenomenon in WD + M dwarf binaries is due to electron-cyclotron maser emission \citep[ECME;][]{2025qu_zhang, 2025zhong}. Indeed, evolutionary models have been put forth explaining how WD + M dwarf LPTs fit into the picture of WD binary evolution and magnetic field generation \citep[e.g.][]{2025yang, 2025horvath}.

Today, the origin of WD magnetism remains uncertain, particularly in close binary systems. Over 35\% of accreting WDs (also known as cataclysmic variables, or CVs) in multiwavelength, volume-limited surveys, have been found to host $B= 1-230$ MG magnetic fields \citep{2020pala, 2024rodriguez_survey}. However, the incidence of WD magnetism in progenitor systems (detached WD + main-sequence binaries) has been estimated to be only 2\% \citep[e.g.][]{2011zorotovic, 2021schreiber}.  Because accreting WDs are further evolved and host cooler WDs, this has led to the idea that magnetism arises as a result of a crystallization-driven dynamo aided by spin-up once mass transfer onto the WD begins, as the WD cools. Crystallization in WDs occurs as the carbon-oxygen core cools, leading to an oxygen-enhanced solid inner core and carbon-enhanced outer liquid envelope. The initial energetics to justify this model, the incorporation into the full picture of WD binary evolution, and detailed calculations surrounding the convective mechanism are presented by \cite{2017isern, 2021schreiber, 2022ginzburg}, respectively. The association of some LPTs with close WD binaries provides an exciting new opportunity to study the origin and evolution of WD magnetism.

In this paper, we focus on the two LPTs that have been associated with WDs, ILT J1101+5521 (ILT J1101) and GLEAM-X J0704--37 (GLEAM-X J0704)\footnote{\textrm{While this paper was under review, ASKAP J174508.9-505149 was published \citep{2026rose}, an LPT with a 1.34 radio period associated with a white dwarf accreting from an M dwarf companion, known as a cataclysmic variable. Since the focus of this paper is on \textit{detached} WD + M dwarfs, no major conclusions have been affected}}. These two sources have been associated with a clear optical counterpart, for which optical spectroscopy has revealed an M dwarf orbit on a period nearly equal to the radio pulse period \citep{2025deruiter, 2024discovery, 2025rodriguez_gleamx}. The main objective of our work is to jointly analyze the binary parameters of ILT J1101 and GLEAM-X J0704 and place them into the context of WDs in close binaries.

In Section \ref{sec:data}, we present new phase-resolved spectroscopy of ILT J1101 acquired with the Keck I telescope, confirming the RV variability detected by \cite{2025deruiter}, with higher precision. In Section \ref{sec:parameters}, we present the binary parameters of ILT J1101 and GLEAM-X J0704, revealing both to be massive, cool WDs whose cores are largely crystallized. In Section \ref{sec:evolution}, we present MESA evolutionary models which show that ILT J1101 and GLEAM-X J0704 will commence accretion in less than a Hubble time, becoming cataclysmic variables. Finally, in Section \ref{sec:population}, we assess the completeness of current radio surveys to finding WD + M dwarf LPTs, estimating a lower limit for the local space density to be $\rho \gtrsim 10^{-8}\;\mathrm{pc}^{-3}$. We also provide open questions and potential new directions for the discovery and characterization of WD + M dwarf LPTs.

\begin{table*}[]
    \centering
    \begin{tabular}{l|c|c|c}
         LPT Name & Radio Period (min) & Counterpart? & Reference \\\hline
GCRT J1745-3009 & 76.2 & None & \cite{2005hyman} \\
GLEAM-X J162759.5-523504.3 & 18.18 & None & \cite{2022hurley-walker} \\
GPM J1839-10 & 21.97 & None & \cite{2023hurley-walker} \\
ASKAP J193505.1+214841.0 & 53.76 & None & \cite{2024caleb} \\
CHIME J0630+25 & 7.017 & None & \cite{2024dong} \\
ILT J1101+5521 & 125.5& M dwarf & \cite{2025deruiter} \\
GLEAM-X J0704--37 & 174.9 & M dwarf & \cite{2024discovery} \\
ASKAP/DART J1832-0911 & 44.27 & X-ray source & \cite{2024wang, 2024li} \\
ASKAP J183950.5-075635.0 & 387 & None & \cite{2025lee}\\
CHIME J1634+44/ILT J163430+445010 & 14.02 & Faint optical/UV source? & \cite{2025dong_1634, 2025bloot_1634}\\
ASKAP J175534.9-252749.1 & 69.6 & None & \cite{2025mcsweeney}\\
ASKAP J144834-685644 & 93.85 & Fading optical source/X-ray & \cite{2025Anumarlapudi}\\
ASKAP J142431.2-612611 & 35.79 & None & \cite{2026Pritchard}\\
ASKAP J174508.9-505149 & 80.70 & Cataclysmic Variable & \cite{2026rose}\\

    \end{tabular}
    \caption{List of the fourteen known LPTs at the time of writing (reported in published form or as a preprint).}
    \label{tab:list}
\end{table*}

\section{Data}
\label{sec:data}
ILT J1101 and GLEAM-X J0704 were both observed for an entire radio pulse period with the Low Resolution Imaging Spectrometer \citep[LRIS;][]{lris} on the 10 m Keck I telescope on Mauna Kea in Hawai'i. The same spectroscopic setup and exposure time per spectrum was used for both objects, leading to a nearly identical resolution. Details surrounding the observations of GLEAM-X J0704 are in \cite{2025rodriguez_gleamx}. \textrm{In brief, Keck I/LRIS was also used to observe that source over an entire orbital period on one occasion, and over half an orbital period on a second night. A nearly identical instrument set up and exposure times were used to observe that source.}

\subsection{Keck I/LRIS Spectroscopy of ILT J1101+5521}
ILT J1101 was observed on 26 January 2025 (UT). Eight 900 second exposures were taken consecutively, with red and blue sides starting at the same time. The total observing time (2h 08m, including read-out time) covered an entire radio period of the object. The 600/4000 grism binned at 2x2 (spatial, spectral) and 400/8500 grating binned at 2x1 were used on the blue and red side, respectively. Seeing was variable during the observation, varying from 0.9" to 1.5"; the 1.0" long slit was used, leading to some slit losses. All LRIS data were wavelength calibrated with internal lamps, flat-fielded, and cleaned for cosmic rays using \texttt{lpipe}, a pipeline for LRIS optimized for long-slit spectroscopy \citep{2019perley_lpipe}. 

Spectra of ILT J1101 at all orbital phases are dominated by M dwarf features, including TiO headbands, as first noted by \cite{2025deruiter}. Additionally, we report a clear detection of H$\alpha$ that was marginally detected in the discovery paper, but also seen clearly in GLEAM-X J0704 \citep{2025rodriguez_gleamx}. All spectra are presented in Appendix Figure \ref{fig:all_spec}.

Radial velocities were obtained by cross-correlating a 50\;\AA\; window around the Na I doublet with an M5 template from the BT-DUSTY library of theoretical spectral atmospheres \citep{2011allard}. The mid-exposure time of all observations was corrected to the barycentric Julian date (BJD$_\mathrm{TDB}$) \textrm{in order to ensure consistency between the optical and radio observations taken from different terrestrial locations}. The RVs were fit to Equation \ref{eq:rv} assuming a circular orbit ($e=0$) \textrm{due to the compact nature of the orbit leading to circularization}:
\begin{equation}
    \mathrm{RV}_\mathrm{MD} = K_\mathrm{MD}\sin(2\pi(\phi-\phi_0)) + \gamma.
    \label{eq:rv}
\end{equation}
Here MD stands for M dwarf, RV is an individual RV, $K$ is the RV curve semi-amplitude, $\phi$ is the orbital phase, $\phi_0$ sets the zero-point of the orbit (inferior conjunction), and $\gamma$ is the systemic velocity. To constrain the parameter values,  a Markov chain Monte Carlo (MCMC) parameter exploration \citep{1970mcmc} was performed using the \texttt{emcee} package \citep{2019emcee}; the ensemble sampler was run with twelve walkers for 10,000 steps, taking the first half as the burn-in period. Table \ref{tab:params} shows the credible intervals of all parameters, with the most relevant being $K_\mathrm{MD} = 96.4\pm 6.3 \mathrm{km s}^{-1}$. 

\subsection{Orbital Period and Radio Pulse Arrival Phase}

To calculate orbital phases, the period was first fixed to be equal to the barycentric radio period of 7531.1699 s (I. de Ruiter and K. Rajwade, priv. comm.)\footnote{This period is slightly different than the topocentric period reported in \cite{2025deruiter} of 7531.7868 s}. To estimate the uncertainty in the orbital period, we recalculated the phases using a range of trial periods between 1.9 and 2.2 hours and evaluated their consistency with the radial velocity (RV) solution derived above. This Monte Carlo analysis constrains the orbital period to $P=7530 \pm 390 $ s.

\begin{figure*}
    \centering
\includegraphics[width=0.5\textwidth]{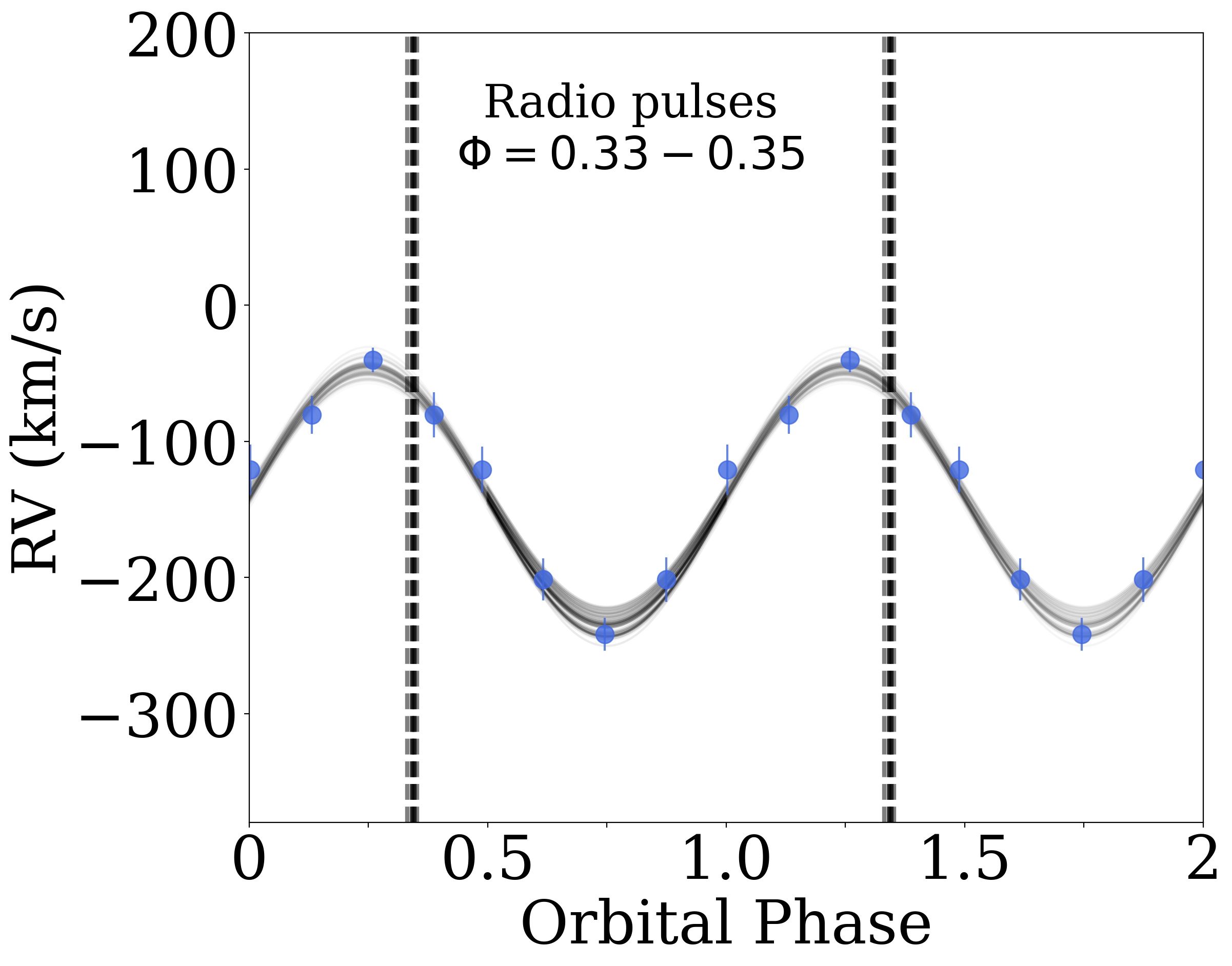}\includegraphics[width=0.5\textwidth]{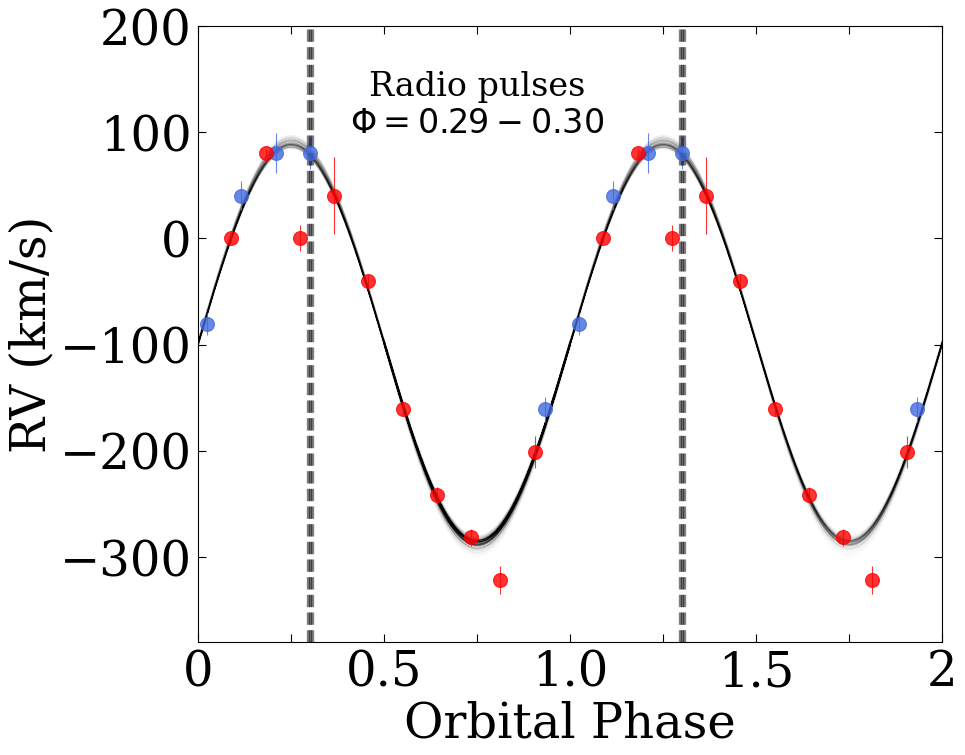}
    \caption{RV curves of the M dwarf in ILT J1101 (left) and GLEAM-X J0704 (right; different colors represent different nights). In both cases, the spectroscopic orbital period nearly matches the radio pulse period. Assuming that the two periods are \textit{exactly} equal, the radio pulse arrival phases are shown. Radio pulses arrive just after an extremum of the M dwarf RV (i.e. when a line between the WD and M dwarf is perpendicular to our line of sight). }
    \label{fig:rv}
\end{figure*}

In Figure \ref{fig:rv}, we show the RV curves of the M dwarfs in ILT J1101 and GLEAM-X J0704, along with the radio pulse arrival phases assuming the spectroscopic period and radio periods are exactly equal. Radio pulse arrival times were taken to be barycentric corrected pulse arrival times reported in the discovery papers by \cite{2025deruiter} and \cite{2024discovery}. In ILT J1101, radio pulses arrive around orbital phase 0.35, which is 0.1 of an orbit after RV maximum (i.e. when the M dwarf is moving away from our line of sight). In GLEAM-X J0704, very similar behavior is seen: radio pulses arrive at orbital phase 0.30, which is 0.05 of an orbit after RV maximum. Additionally, we note that in the WD pulsar 191213.72-441045.1 (henceforth J1912, \cite{2023j1912}), where the radio pulse period and orbital period are precisely known, radio pulse activity is highest at orbital phase 0.5 in Figure \ref{fig:rv}, when the \textrm{irradiated} face of the M dwarf faces Earth.

We note that the radio pulse arrival phases differ from what is presented in the discovery paper by \cite{2025deruiter}. In that work, RV phases are fixed to a slightly less accurate period (125.5 min), which introduce a significant error since radio pulses were observed years before optical observations. Nevertheless, our optical observations cannot constrain the orbital period to better than 5 percent. If the optical period does not \textit{exactly} match the radio period, then radio pulses arrive at different phases, and may be tied to an asynchronously spinning WD. Ultimately, additional observations, ideally carried across consecutive nights, are needed to better constrain the orbital period, which currently agrees with the radio period within its uncertainties.

\subsection{Kinematic Analysis}
With estimates of systemic velocity, $\gamma$ and precise proper motion measurements from \textit{Gaia} Data Release 3 for both systems, we conduct a kinematic analysis of ILT J1101 and GLEAM-X J0704 \citep{2023gaiadr3}. Distances to these objects are obtained from the analysis described in Section \ref{sec:parameters}. To summarize, SED analysis of ILT J1101 and GLEAM-X J0704 place them $300\pm30$pc and $380\pm10$ pc away, respectively.

We convert the 6-dimensional coordinates of each system into Galactic $U, V, W$ space velocities and compare them to two samples on a standard Toomre diagram \citep[e.g.][]{1964toomre, 2003bensby}: random samples from the \textit{Gaia} catalog out to 500 pc and detached WD + M dwarf binaries discovered by the Sloan Digital Sky Survey (SDSS) with a similar orbital period range \citep{2016zorotovic}.

\begin{figure}
    \centering
    \includegraphics[width=0.5\textwidth]{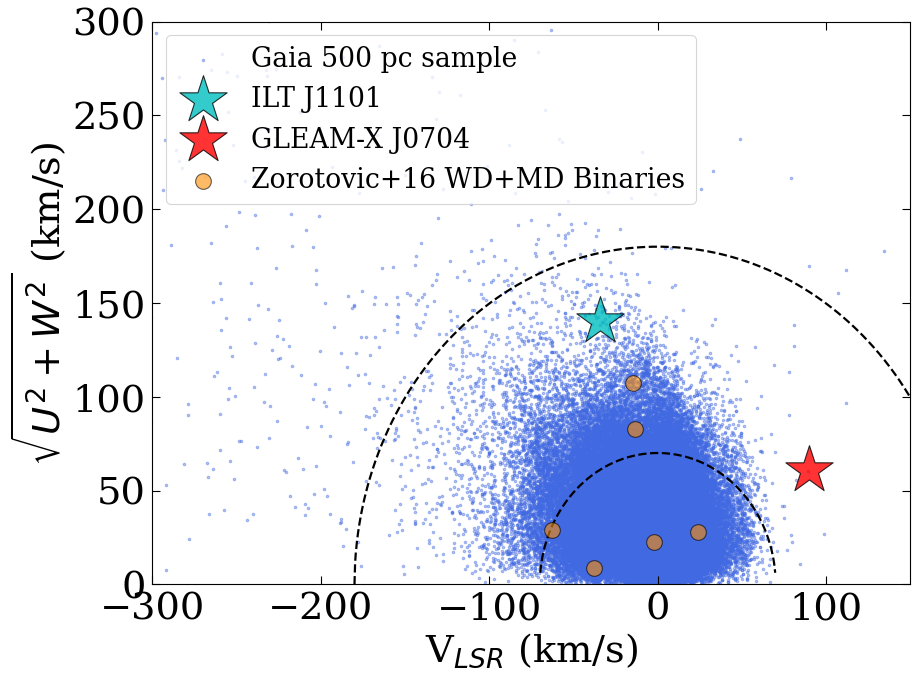}
    \caption{Unlike the majority of LPTs which reside in the Galactic Plane, a Toomre diagram of ILT J1101 and GLEAM-X J0704 shows they are located in the Galactic thick disk (dotted lines represent the thin and thick disk). Detached WD + M dwarf binaries in a similar orbital period range (2--3 hr) from the \cite{2016zorotovic} sample are kinematically colder than the WD + M dwarf LPTs, likely owing to their lower mass (and therefore younger) WDs. }
    \label{fig:toomre}
\end{figure}

The results of this analysis are shown in Figure \ref{fig:toomre}, with the thin disk defined as the semicircle within 70 km/s and the thick disk within 180 km/s. Both systems are well within the thick disk, which is made up of systems that are $\approx8-10$ Gyr old \citep[e.g.][]{2017kilic}, and noticeably older than those the SDSS sample likely due to the more massive WDs in LPTs (see Section \ref{sec:parameters}). Crucially, this \textrm{may establish} WD + M dwarf LPTs as a kinematically distinct type of LPT. 

Whereas the majority of other LPTs are located in the Galactic Plane ($b\lesssim5^\circ$), with the first system, GCRT J1745-3009 located near the Galactic Center\footnote{\textrm{All previous works have assumed that GCRT J1745-3009 is located near the Galactic Center (GC) due to its apparent proximity on the plane of the sky, though it may be a more nearby or distant source that happens to co-align with the GC. }}, ILT J1101 ($b\approx55^\circ$) and GLEAM-X J0704 ($b\approx-13^\circ$) are members of the thick disk and likely several Gyr older than the rest of the known LPTs, \textrm{if the more distant LPTs concentrated in the Plane are indeed Galactic thin disk members}. \textrm{Future observations, however, are needed to obtain space velocities for the more distant LPTs and definitively establish them as members of the thin disk.}

\section{Binary Parameters}
\label{sec:parameters}

We model the average spectrum of ILT J1101 using an identical process to GLEAM-X J0704 in \cite{2025rodriguez_gleamx}, but summarize the main points here. We use the following free parameters: the WD effective temperature and radius, $T_\mathrm{WD}$ and $R_\mathrm{WD}$; the M dwarf effective temperature and radius, $T_\mathrm{MD}$ and $R_\mathrm{MD}$; the distance to the system, $d$. We fix the reddening to the system, $E(B-V) = 0.01$ based on the \cite{2024edenhofer} dust map. $T_\mathrm{MD}$ and $R_\mathrm{MD}$ are related through the M dwarf isochrones (assuming an age of 10 Gyr) of \cite{2015baraffe}. 

 To create an M dwarf model,  $T_\mathrm{MD}$ is specified, interpolating through the BT-DUSTY library of theoretical stellar atmospheres \citep{2011allard}, assuming solar metallicity and $\log g = 5.0$. Small changes in the adopted metallicity have a negligible effect on inferred binary parameters. An inflation factor of 10\% is adopted for the radius of the M dwarf, as it has been shown that for binaries this close, fast rotation leads to increased magnetic activity and thus inflation \citep{2018parsons}. The flux at the stellar photosphere is multiplied by $(R_\mathrm{MD}/d)^2$ to obtain a flux as viewed from Earth. The same process is used for the WD, taking the theoretical (DA; H-rich) WD atmospheres of \citep{2010koester} and assuming $\log g = 8.0$.

A Markov chain Monte Carlo (MCMC) parameter exploration was performed using the \texttt{emcee} package \citep{2019emcee} to explore all four simultaneously, adopting uniform priors on all parameters: $5000 < T_\mathrm{WD}\;(\mathrm{K}) < 12000 $, $2500 < T_\mathrm{MD}\;(\mathrm{K})< 4000$, $0.008 < R_\mathrm{WD}\;(R_\odot)< 0.02$, $200 < d\;(\mathrm{pc}) < 1000$. The MCMC sampler was run for 2000 steps, taking half as the burn-in period. The parameter exploration converged, with a Gelman-Rubin statistic \citep[$\hat{R}$;][]{1992gelman} of 1.09 averaged over all chains. \cite{1992gelman} argue that values close to 1, with $\hat{R}\lesssim1.1$ being a typical threshold, indicate convergence for typical multivariate distributions. The corner plot and marginalized posterior distributions are shown in Appendix Figure \ref{fig:corner}. 

In Table \ref{tab:params}, the inferred binary parameters (median values and credible intervals: 16$^\mathrm{th}$ and 84$^\mathrm{th}$ percentiles of the posterior distributions) resulting from the MCMC analysis are shown. $M_\mathrm{WD}$ is obtained from the mass-radius relation of \cite{2020bedard}; $M_\mathrm{MD}$, from the M dwarf isochrones (assuming an age of 10 Gyr) of \cite{2015baraffe}, assuming solar metallicity. The binary separation, $a$, and inclination, $i$, are solved for using Kepler's laws, and the  M dwarf Roche lobe filling factor, $R_\mathrm{MD}/R_L$, (i.e., how much of its Roche lobe is filled), is solved for using the \cite{1983eggleton} relation
\begin{equation}
        \frac{R_L}{a} = \frac{0.49 q^{2/3}}{0.6 q^{2/3} + \ln\left(1 + q^{1/3}\right)},
        \label{eq:roche}
\end{equation}
where $q=M_\mathrm{MD}/M_\mathrm{WD}$ is the mass ratio of the two objects.

\begin{table*}[]
    \centering
\begin{tabular}{lcc}
\hline
Parameter & ILT J1101& GLEAM-X J0704 \\
\hline
$\Phi_\mathrm{radio}$ & $0.33-0.35$ & $0.29-0.30$\\
$P_\mathrm{radio}$ (s) & $7531.1699 \pm 0.0012$&  $10496.5522 \pm 0.0008$ \\
$P_\mathrm{orb}$ (s) & $7530 \pm 390$&  $10496 \pm 5$\\
$\phi_0$ & $0.931\pm0.001$ & $1.307 \pm 0.002$\\
$K_\mathrm{MD}$ (km s$^{-1}$) & $96.4\pm 6.3$ & $189.4 \pm 2.7$\\
$M_\mathrm{WD, min}$ ($M_\odot$) & $0.11$ & $0.22$\\
$\gamma$ (km s$^{-1}$) & $-138.4^{+4.9}_{-4.8}$ & $-98.9^{+1.7}_{-1.8}$\\
Gaia DR3 ID & 855784912471554176 & 5566254014771398912\\
Gaia G (mag) & 20.0 & 20.8 \\
\hline
$E(B-V)$ & $0.01\pm0.001$ & $0.119\pm0.001$\\
$T_\mathrm{WD}$ (K) & $5160^{+290}_{-130}$ & $7320^{+800}_{-590}$\\
$R_\mathrm{WD}$ ($R_\odot$) & $0.010^{+0.0010}_{-0.0016}$ & $0.0079^{+0.0014}_{-0.0015}$\\
$T_\mathrm{MD}$ (K) & $2960^{+40}_{-30}$ & $3010 \pm 20$\\
$R_\mathrm{MD}$ ($R_\odot$) & $0.156^{+0.007}_{-0.006}$ & $0.165^{+0.003}_{-0.004}$\\
$d$ (pc) & $300\pm30$ & $380\pm10$\\
\hline
$M_\mathrm{WD}$ ($M_\odot$) & $0.84^{+0.14}_{-0.13}$ & $1.02^{+0.12}_{-0.13}$\\
$M_\mathrm{MD}$ ($M_\odot$) & $0.128 \pm 0.005$ & $0.136 \pm 0.003$\\
$a$ ($R_\odot$) & $0.82^{+0.03}_{-0.05}$ & $1.07\pm0.04$\\
$q = M_\mathrm{MD}/M_\mathrm{WD}$ & $0.15^{+0.03}_{-0.02}$ & $0.133^{+0.019}_{-0.014}$\\
$R_\mathrm{MD}/R_L$ & $0.817 \pm 0.001$ & $0.680 \pm 0.001$\\
$i$ ($^\circ$) & $13\pm1$ & $28^{+2}_{-1}$\\
\hline
\end{tabular}

    \caption{All parameters for ILT J1101 and GLEAM-X J0704. The first level of parameters are just based on the RV shifts of the MD, while the next level is based on an MCMC parameter exploration of the average spectrum. The final level represents parameters derived from those used in the MCMC analysis.}
    \label{tab:params}
\end{table*}

\begin{figure*}
    \centering  \includegraphics[width=0.48\textwidth]{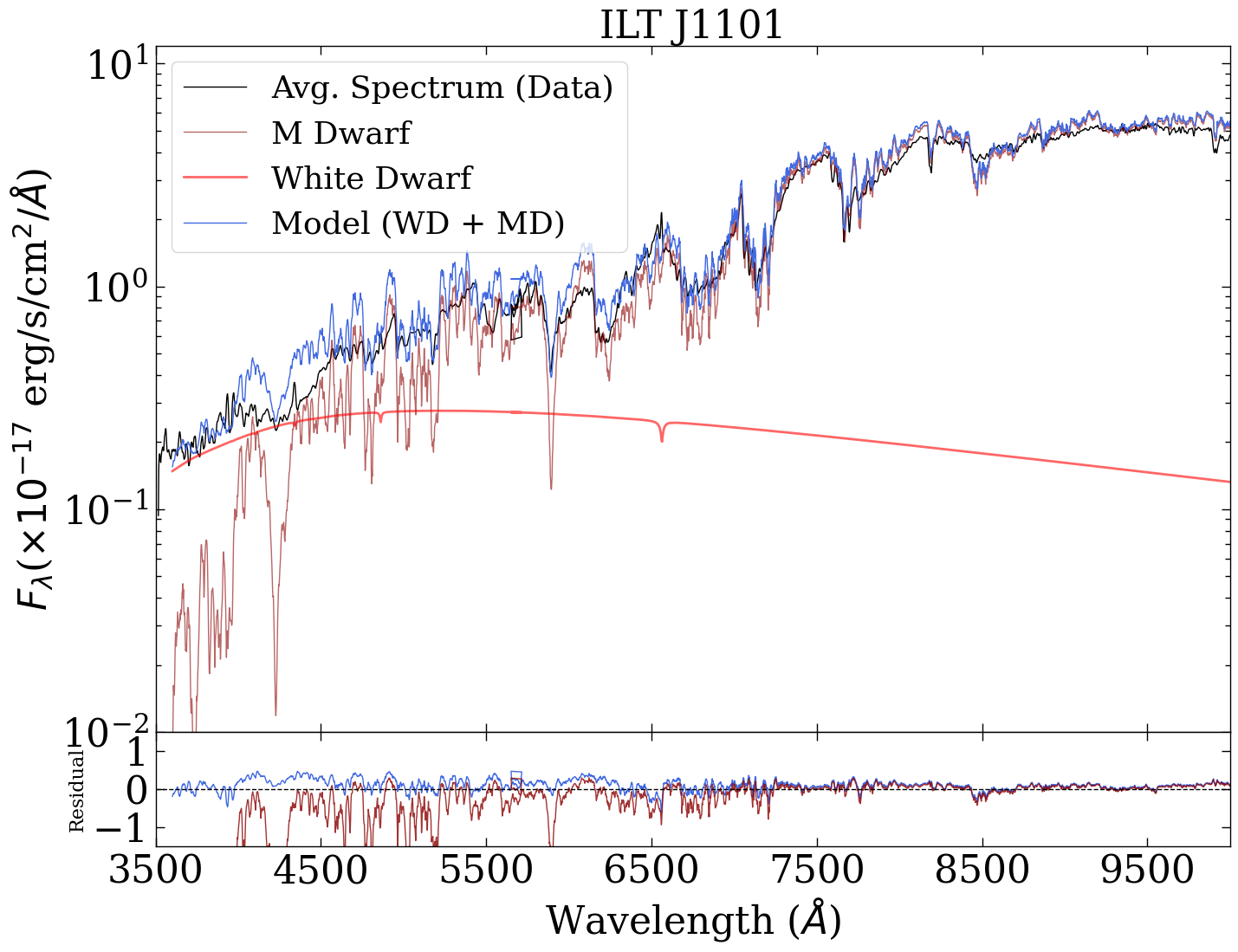}
\includegraphics[width=0.48\textwidth]{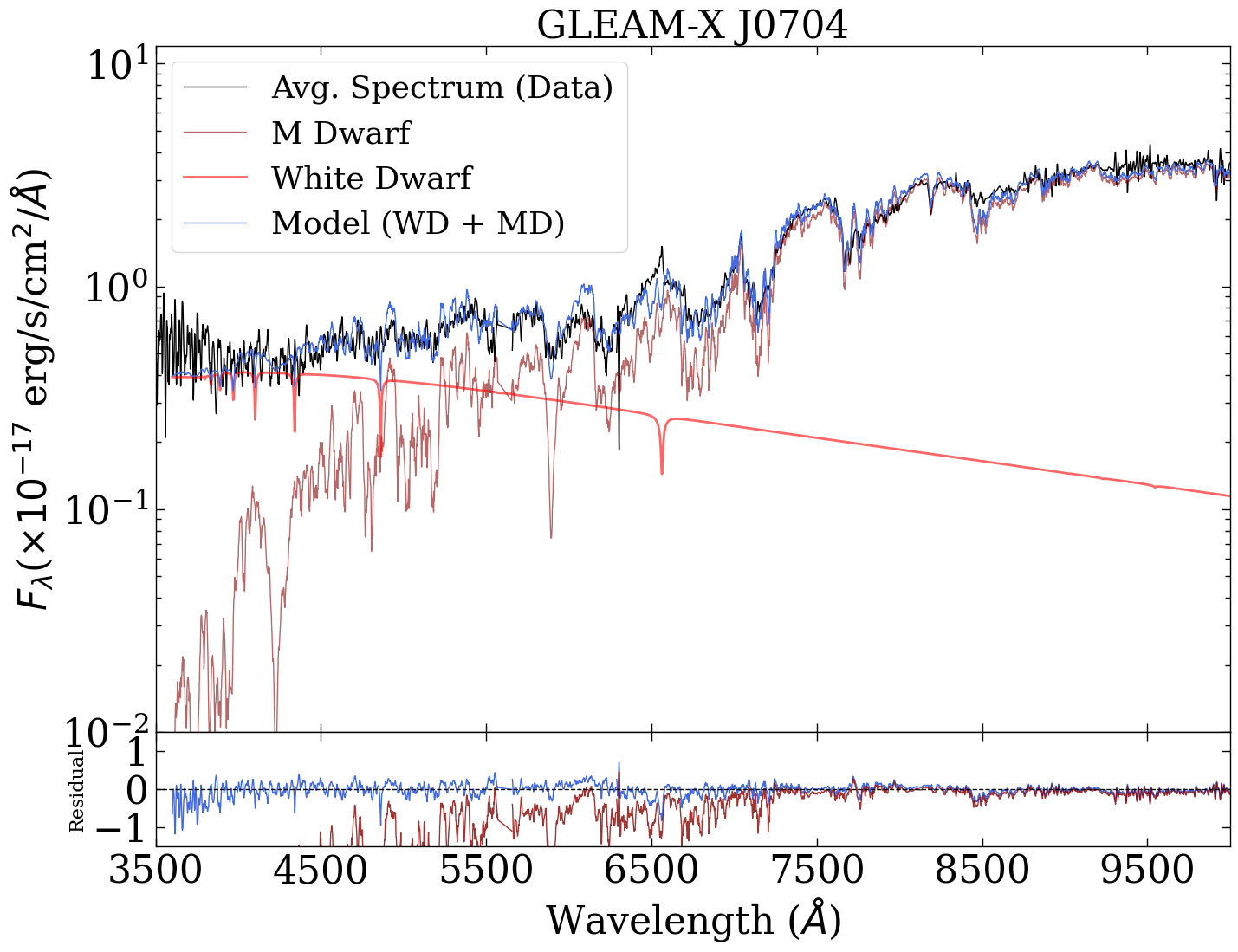}  
    \caption{Average Keck I/LRIS optical spectrum of ILT J1101 (left) and GLEAM-X J0704 (right) are shown (black) alongside the best model inferred from the MCMC parameter exploration. The M dwarf component is shown in dark red, the WD component in red, and the sum of the two in blue. The residual plots in the lower panel show that at the bluest end of the wavelength range, a single M dwarf model cannot fit the data, invoking the need for a WD companion.}
    \label{fig:sed}
\end{figure*}

Figure \ref{fig:sed} shows the average spectrum of each object, along with the best model from the MCMC parameter exploration (median values in Table \ref{tab:params}). The logarithmic scaling of the vertical axis along with the residual plot at the bottom of each panel clearly show that a single M dwarf model cannot reproduce the bluest ends of the spectrum due to the steep dropoff at those wavelengths. Furthermore, as argued in \cite{2025rodriguez_gleamx}, the presence of this blue continuum and good fit by a WD model rules out a NS nature for ILT J1101 and GLEAM-X J0704 --- a NS would be much bluer and much fainter at the inferred distance of these systems. Since there is no evidence for accretion in these systems, no disk model or accretion continuum could explain the blue continuum, only a WD.

\subsection{H$\alpha$ Emission}
\begin{figure}
    \centering
    \includegraphics[width=0.21\textwidth]{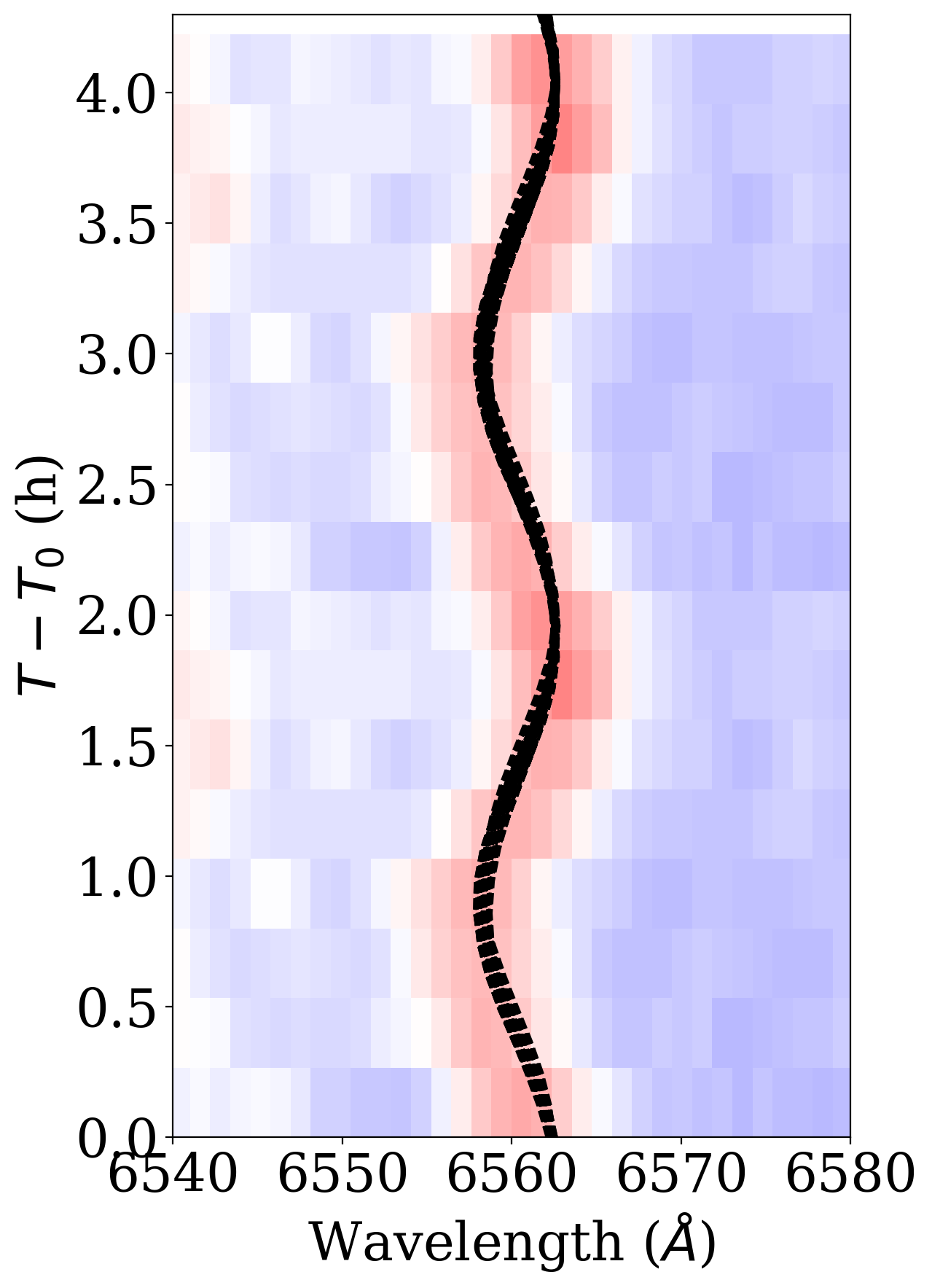}\includegraphics[width=0.29\textwidth]{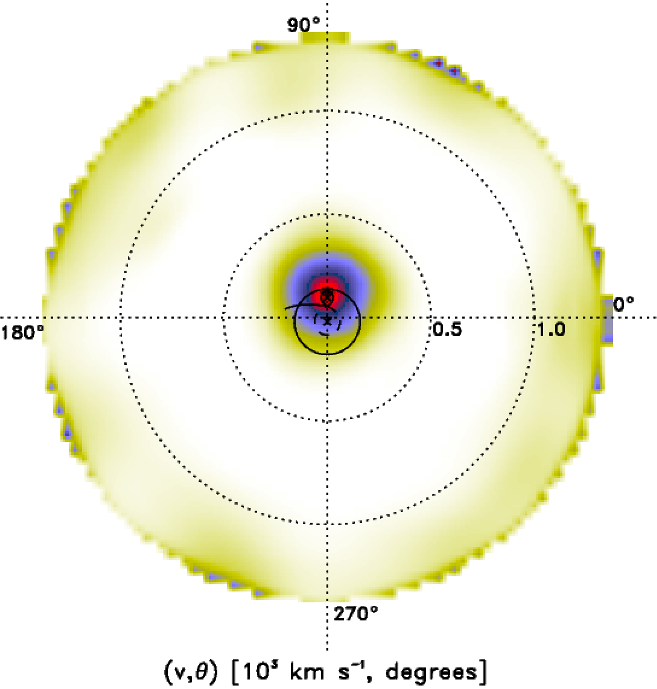}
    \caption{\textit{Left}: Trailed spectrum near the H$\alpha$ line reveals RV variability that matches that of the M dwarf Na I doublet to within 1$\sigma$ (two periods shown for clarity). \textit{Right}: The Doppler tomogram of H$\alpha$ emission reveals that it comes from the M dwarf, consistent with chromospheric activity.}
    \label{fig:halpha}
\end{figure}

 In Figure \ref{fig:halpha}, we show the trailed spectrum of the H$\alpha$ emission line along with the best-fit model, assumed to follow Equation \ref{eq:rv}. We find that all three parameters are consistent with the RV parameters of the Na I absorption doublet to within the uncertainties: $\phi_\mathrm{0, H\alpha} = 0.93\pm0.01$, $\gamma_\mathrm{H\alpha} = -140\pm10\;\textrm{km s}^{-1}$, and $K_\mathrm{H\alpha} = 99\pm 8\;\textrm{km s}^{-1}$ after undertaking a similar MCMC analysis to what was done for the Na I doublet while assuming a single Gaussian emission line model. The uncertainties are all larger than parameters derived from the Na I doublet due to the weak nature of the line.

 Figure \ref{fig:halpha} also shows a Doppler tomogram created using \texttt{doptomog} \citep{2015doptomog}, which maps flux as a function of phase onto polar coordinates, where the radius represents a projected velocity and the azimuthal coordinate is the orbital phase. A complete description of Doppler tomography is presented in \cite{2001marsh}. We show the projected Roche lobes of the WD and M dwarf derived from the parameters in Table \ref{tab:params}. Since the RV parameters of the H$\alpha$ line are consistent with the Na I doublet, it is not surprising that the Doppler map shows the H$\alpha$ emission from within the M dwarf Roche lobe. We infer that the H$\alpha$ emission stems from the M dwarf, likely due to chromospheric activity from being in such a compact orbit. This phenomenon is common for M dwarfs in the field as well as in close binaries \citep[e.g.][]{1998delfosse, 2017skinner, 2023elbadry_bd}. 
 
This result is different from what was observed in GLEAM-X J0704, where some of the H$\alpha$ emission traces the M dwarf, but the bulk of it traces a higher velocity component outside the M dwarf \citep{2025rodriguez_gleamx}.  Furthermore, the H$\alpha$ emission does not appear to originate from a particular part of the M dwarf (e.g. the face), as was found in the WD pulsars AR Sco and J1912 \citep{2016marsh, 2023j1912}. Further spectroscopic monitoring of both ILT J1101 and GLEAM-X J0704 will reveal if the origin of H$\alpha$ emission varies as a function of time and/or radio activity of the systems.

\subsection{\textrm{Nearly Face-On: A Potential Clue to the Mechanism Powering LPTs?}}

\begin{figure}
    \centering
    \includegraphics[width=0.45\textwidth]{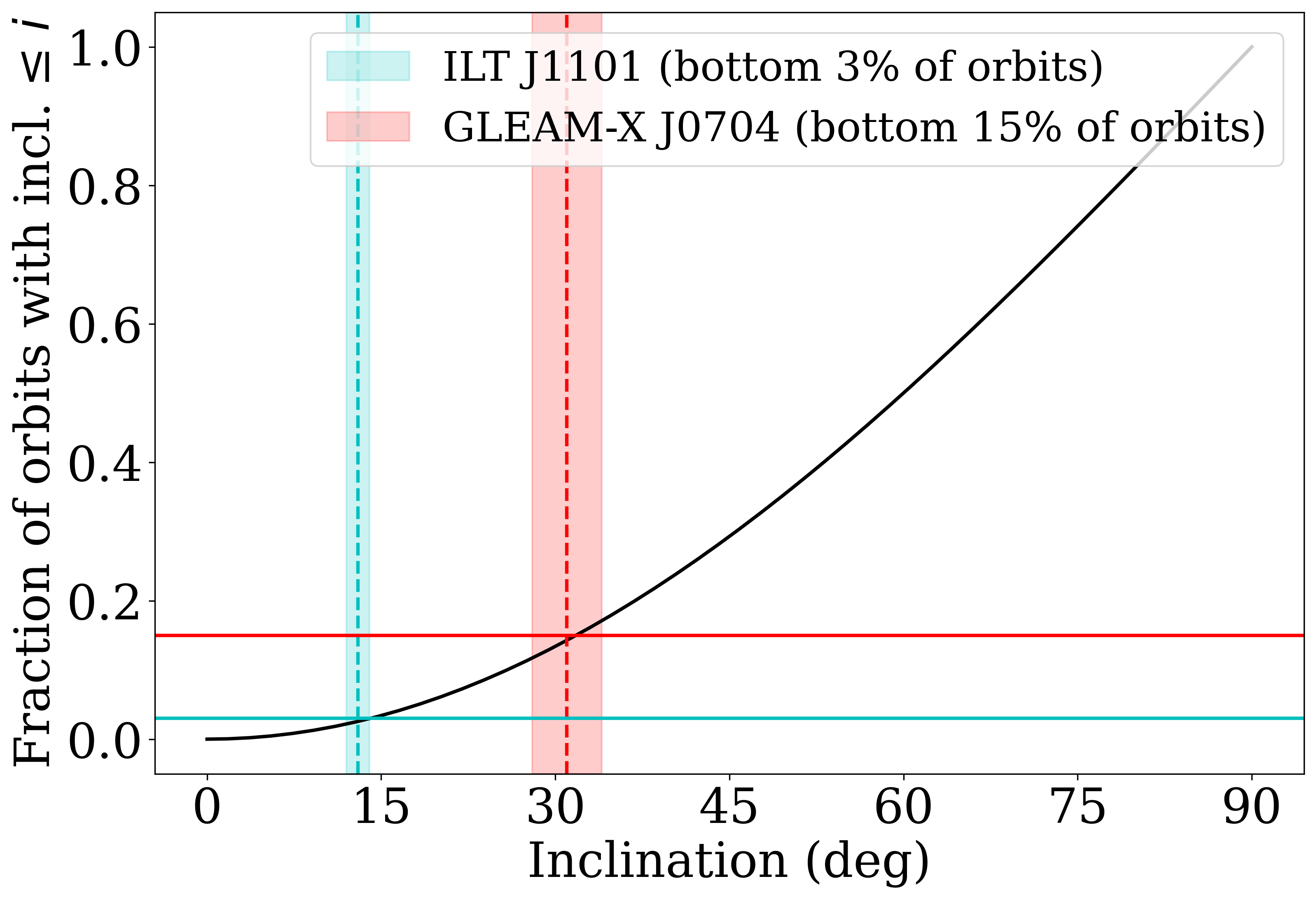}
    \caption{The LPT phenomenon in WD + M dwarf binaries appears to favor face-on inclinations. This is because the cumulative distribution of binary orbits on the sky means that the low inclinations of ILT J1101 and GLEAM-X J0704 are extremely unlikely.}
    \label{fig:inclination}
\end{figure}

Additionally, we note that the orbital inclinations of ILT J1101 and GLEAM-X J0704 are exceptionally small (close to face-on), hinting that the LPT phenomenon may be inclination-dependent. From an observer's vantage point on Earth, inclinations of binary orbits should be uniformly distributed in the cosine of the inclination, $\cos i$, between $\cos i = 0$ and $\cos i = 1$. A transformation of variables  reveals that the probability distribution function of $i$ should be $\sin i$. The ensuing cumulative distribution function (CDF), then, is $1-\cos i$. Figure \ref{fig:inclination} shows this CDF along with the inferred inclinations of ILT J1101 and GLEAM-X J0704. The former is in the lower 3 percent of orbital inclinations, while the latter is in the lower 15 percent. Curiously, in the case of the WD pulsar J1912, \cite{2023j1912} found a maximum inclination of $37^\circ$, placing all possible values for that system in the lower 20 percent of the CDF in Figure \ref{fig:inclination}.

Current models of the electron cyclotron maser emission (ECME) from these systems note a strong dependence on inclination. In the initial model proposed by \cite{2025qu_zhang}, radio pulses become more difficult to detect at face-on inclinations (closer to $i=0^\circ$). However, in the detailed models presented by \cite{2025zhong}, radio pulses are reproduced at luminosities matching observations, while adopting a modest inclination ($i=20^\circ$). Future theoretical work should aim to reproduce such pulses from low-inclination binaries and investigate if the LPT phenomenon in WD + M dwarf binaries is indeed most favorably seen at low inclinations. 

\subsection{Cool and Massive Crystallized WDs}
\begin{figure}
    \centering
    \includegraphics[width=0.5\textwidth]{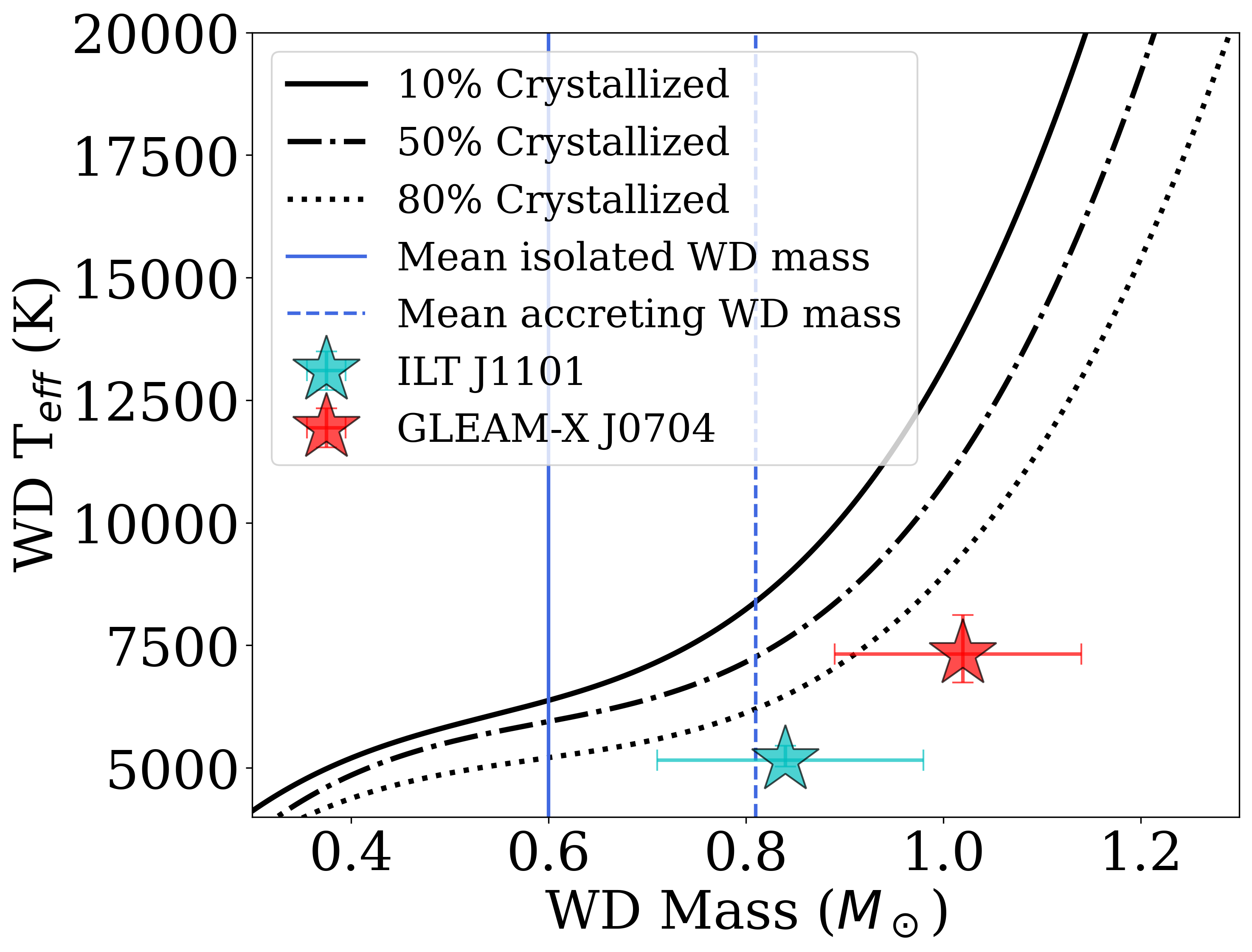}
    \caption{The cool and massive WDs in ILT J1101 and GLEAM-X J0704 have exceptionally crystallized cores. Recent work has suggested that WD crystallization in close binaries is an important driver of magnetism, which likely plays an important role in the production of the coherent radio pulses.}
    \label{fig:crystal}
\end{figure}

It is of special interest to note the relatively cool temperatures (5160K and 7320 K) and high masses ($0.84M_\odot$ and $1.02M_\odot$) of the WDs in ILT J1101 and GLEAM-X J0704. A series of papers in the last decade has suggested that crystallization-driven dynamos in cooling WDs are responsible for $\sim1-100$ MG magnetic fields in WDs in close binaries \citep{2017isern, 2021schreiber, 2022ginzburg}. In Figure \ref{fig:crystal}, we show the temperatures and masses of the WDs in both systems along with results from the WD cooling models of \cite{2020bedard}. Curves corresponding to 10, 50, and 80 percent of a crystallized core are shown, and both LPTs appear to be well over 80 percent crystallized. 

One key unknown factor in the crystallization-driven dynamo picture, however, is the spin of the WD. \cite{2022ginzburg} showed that in order to generate $\sim100$ MG magnetic fields, WDs must be spun up to min-long periods, which naturally occurs from accretion in CVs. However, that work also showed that $\sim1-10$ MG magnetic fields can arise from modest WD spins ($\sim$1--few hours). Since it appears that ILT J1101 and GLEAM-X J0704 have not been spun up through accretion in the past, determining their WD spin periods (and magnetic field strengths) in future work is of special importance for constraining the mechanism responsible for generating a strong magnetic field.

Not only that, but the WD mass of both systems is well above the mean mass of single WDs \citep[$\sim0.6M_\odot$; ][]{2007wd} and also above the mean mass of WDs in accreting WDs \citep[$\sim0.8M_\odot$;][]{2022pala}. We speculate that the radio production mechanism in WD + M dwarf LPTs favors close binary separations, $a$, and therefore lower values of mass ratio, $q$ (Equation \ref{eq:roche} is monotonically increasing as a function of $q$). If radio pulsations in the ECME model are indeed caused by the interaction between WD and M dwarf magnetic fields \citep{2025qu_zhang, 2025zhong}, then closer binary configurations (without Roche lobe overflow) will favor the LPT phenomenon.

\section{Evolutionary Scenarios of WD LPTs}
\label{sec:evolution}
\subsection{Origins: Products of a Common Envelope}
Due to their short orbital periods, it is most likely that ILT J1101 and GLEAM-X J0704 both originated from initial binary systems in which the progenitor to the WD was more massive than the MD. The WD progenitor then exhausted hydrogen burning, and while evolving through a red giant phase, its outer envelope filled its Roche lobe and surrounded the companion MD. Drag forces between the stars and the common envelope (CE) lead to rapid angular momentum loss (AML), which brings the binary closer together, leading to either the creation of a compact binary or a stellar merger. This rapid inspiral has been shown to last $\sim$100 days in modern simulations \citep[e.g.][]{1976paczynski, 2013ivanova}. 

However, it is difficult to constrain the initial properties of the WD progenitor from the current state of the system. Isolated WDs with similar masses to ILT J1101 and GLEAM-X J0704 originate from 4--6 $M_\odot$ stars, but the initial separation of the binary cannot be constrained \citep[e.g.][]{2000weidemann, 2018elbadry}. This is due to the uncertain physics of CE evolution, where the efficiency of energy dissipation is summarized by the so-called ``$\alpha$-parameter'' \citep[e.g.][]{2013ivanova}. For the purposes of this work, all that matters is that a WD and M dwarf emerged from the CE without merging, and AML led to the present state of the systems. Curious is the fact that both M dwarfs in ILT J1101 and GLEAM-X J0704 are on the low-mass end, even for M dwarfs in PCEBs \citep[e.g.][]{2026shariat}.

\subsection{Current State: No Evidence for Roche lobe Overflow }

ILT J1101 and GLEAM-X J0704 do not show any evidence for high accretion along three different lines of evidence: 1) optical spectroscopy, 2) X-ray non-detections, 3) inferred binary parameters. Optical spectroscopy, in both cases, reveals the absence of an accretion disk or column of any sort. The small equivalent width of the H$\alpha$ line is consistent with chromospheric activity of the M dwarf, induced by fast rotation due to its tidal lock in such a tight orbit. Furthermore, upper limits from X-ray measurements of $5\times10^{-14} \mathrm{erg s}^{-1}\mathrm{cm}^{-2}$ and $5\times10^{-15} \mathrm{erg s}^{-1}\mathrm{cm}^{-2}$ for ILT J1101 and GLEAM-X J0704, respectively, lead to upper limits on the X-ray luminosity of  $5\times10^{29} \mathrm{erg s}^{-1}$ and $8\times10^{28} \mathrm{erg s}^{-1}$, respectively \citep{2025deruiter, 2024discovery}. Only a small fraction of CVs have X-ray luminosities below the former value and effectively none in the case of the latter \citep[e.g.][]{2024rodriguez_survey}. Finally, based on the inferred binary parameters of each system, we show in Table \ref{tab:params} that ILT J1101 and GLEAM-X J0704 fill approximately 82 and 68 percent of their Roche lobes, respectively. 

Nonetheless, it is likely that wind accretion is occurring in both systems, much like in ``pre-polars" or ``low accretion rate polars'' \citep[LARPs;][]{2002schwope}. In pre-polars, a strongly magnetized WD spins in sync with the orbit, with cyclotron humps dominating the optical spectrum due to the $B\sim 10-100$ MG magnetic field strength of the WD. Because such features are seen, along with weak emission lines and in some cases, low-luminosity X-ray emission ($L_X\sim 10^{28}-10^{29}$ erg/s), it is believed that some of the M dwarf wind is accreted onto the WD \citep[e.g.][]{2002schwope}. While it is mysterious that cyclotron humps are not seen in ILT J1101 or GLEAM-X J0704, \cite{2025vanroestel} found that magnetic WDs with companions that have similar masses Roche lobe filling factors to the ones in ILT J1101 and GLEAM-X J0704 are most likely to show strong cyclotron emission patterns only at specific magnetic pole and orbital inclination configurations, with low inclinations generally disfavoring the cyclotron beaming towards our line of sight. 

\subsection{Future Evolution: Cataclysmic Variables in the Making}

We simulate the evolution of ILT J1101 and GLEAM-X J0704 using Modules for Experiments in Stellar Astrophysics \citep[MESA;][]{2011mesa, 2013mesa, 2015mesa, 2018mesa}. We set up a model following the steps outlined by \cite{2021el-badry}, with modifications made to the initial masses of each component to match those of ILT J1101 and GLEAM-X J0704. 

We initialize each binary at an orbital period of 0.5 days, and model the WD as a point mass. The initial masses and radii of the WD and M dwarf for ILT J1101 and GLEAM-X J0704 are set as the median values in Table \ref{tab:params}. The evolution of the system is dominated by two physical processes driving angular momentum loss (AML): magnetic braking (MB) and gravitational wave radiation (GWR), with latter dominating at smaller orbital separations. We assume the MB prescription of \cite{1983rappaport}, though at the current orbital separation of both systems, and at shorter periods, GWR is by far the dominant source of AML. 

\begin{figure*}
    \centering
    \includegraphics[width=0.9\textwidth]{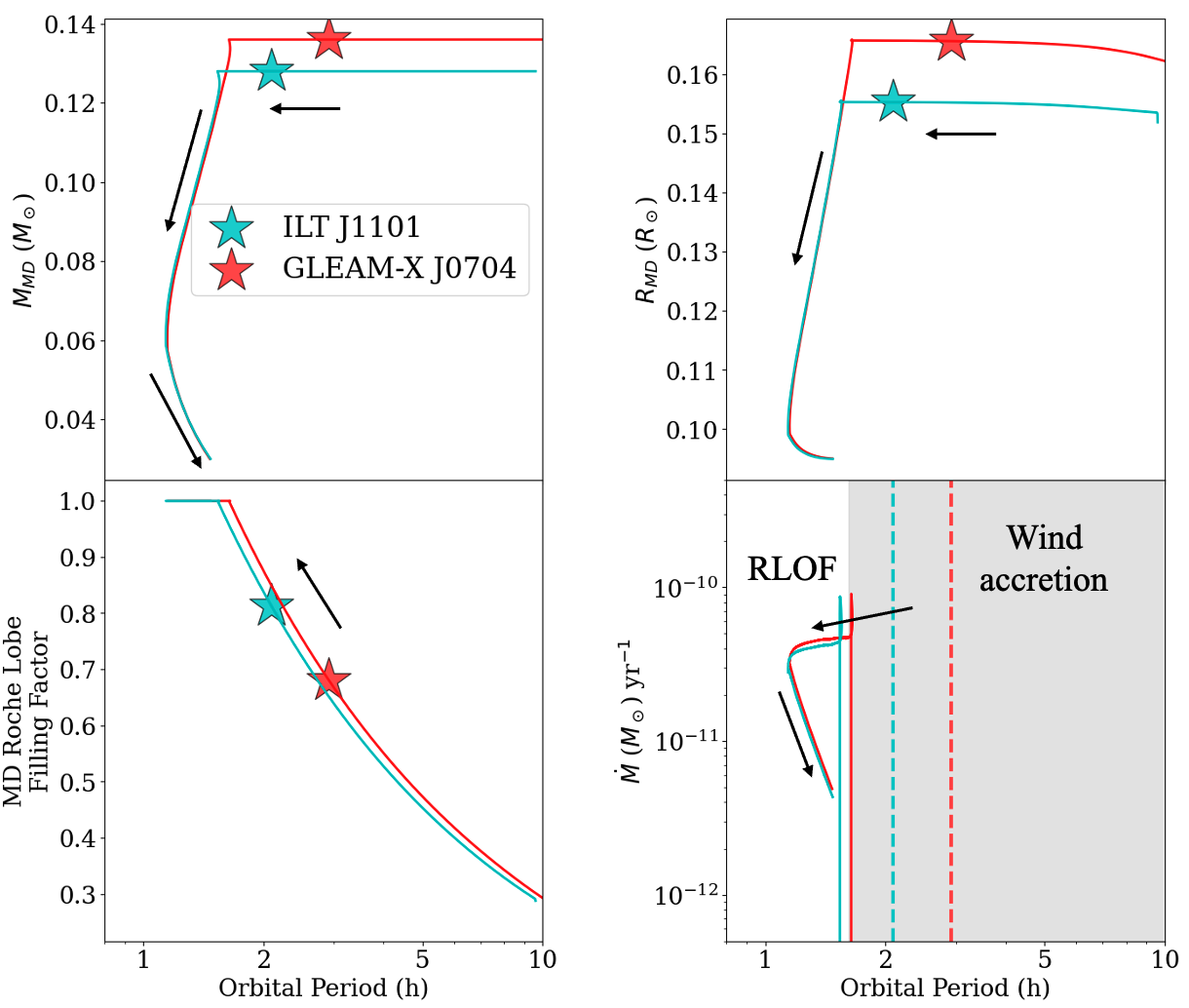}
    \caption{MESA models of the binary systems ILT J1101 and GLEAM-X J0704 show that their orbits will decay due to gravitational wave radiation, ultimately bringing the system into Roche lobe overflow and forming a CV. The upper panels show that once accretion begins, the mass and radius of the M dwarf companion will plummet, leading to the creation of a degenerate, brown dwarf-like object. The lower left panel shows that the M dwarf in ILT J1101 and GLEAM-X J0704 has filled 68 and 82 percent of its Roche lobe, respectively. The lower right panel shows that at their current orbital periods (dotted lines), each system is transferring a negligible amount of mass due to wind accretion, but that RLOF will begin when each system has an orbital period of $\sim$1.5 h.}
    \label{fig:mesa}
\end{figure*}

\begin{figure}
    \centering
    \includegraphics[width=0.5\textwidth]{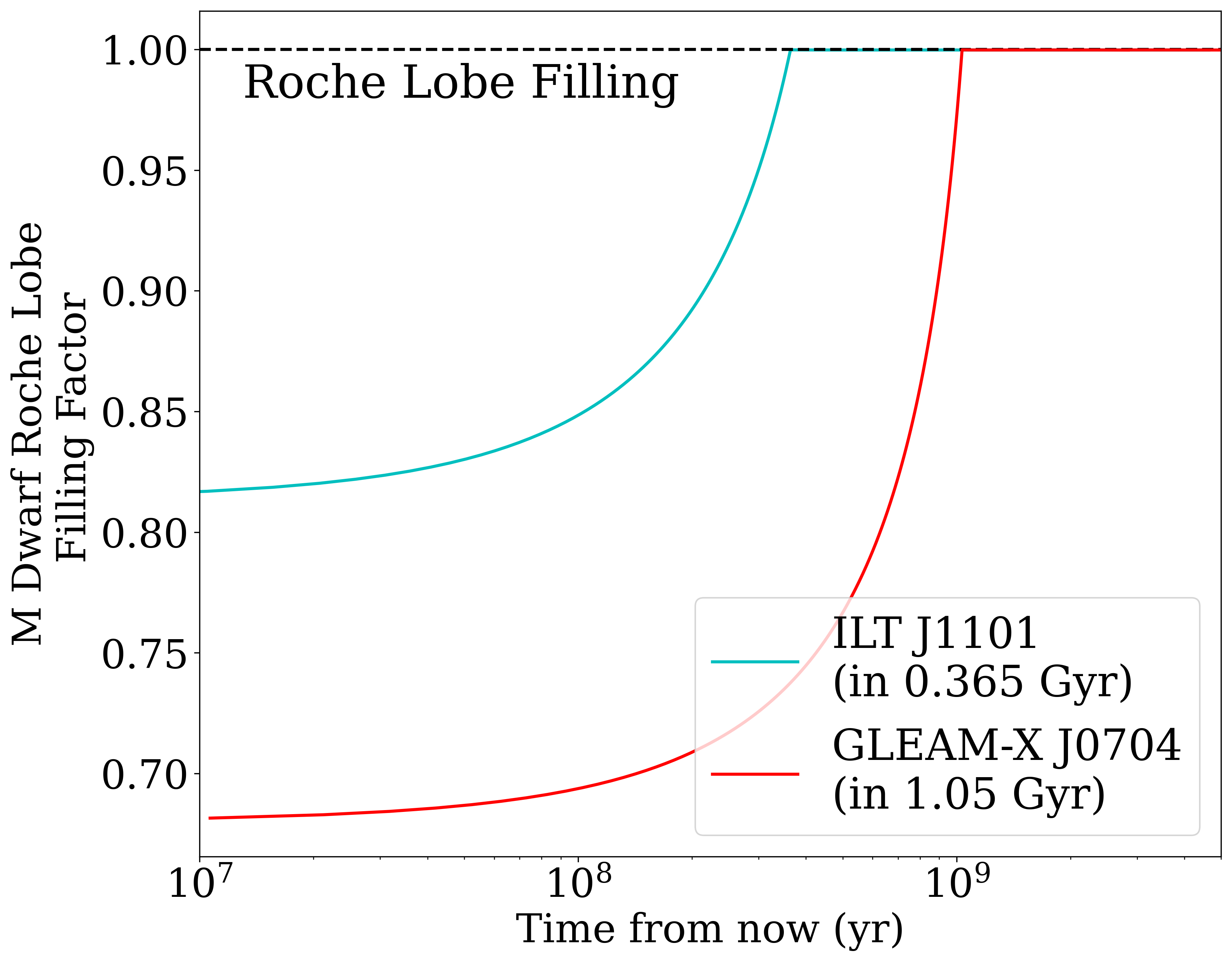}
    \caption{Based on the MESA models in Figure \ref{fig:mesa}, the orbits of ILT J1101 and GLEAM-X J0704 will shrink due to gravitational wave radiation. Both systems will become cataclysmic variables within a Hubble time, with accretion commencing in 0.365 and 1.05 Gyr in ILT J1101 and GLEAM-X J0704, respectively.}
    \label{fig:rl}
\end{figure}

Figure \ref{fig:mesa} shows the output of the MESA simulations, with arrows showing the evolutionary path that the binary will take. Stars represent the current state of the binary system. Figure \ref{fig:rl} shows that ILT J1101 and GLEAM-X J0704 will remain as detached binaries for the next 0.365 and 1.05 Gyr, after which the M dwarf will fill its Roche lobe and the system will become a CV. Depending on the magnetic field strength of the WD and evolution on the path to Roche lobe overflow (RLOF), the system will then become either an intermediate polar ($B_\textrm{WD} = 1-10$ MG) or polar ($B_\textrm{WD} = 10-230$ MG). Once sufficient mass is depleted from the M dwarf and its mass drops below the hydrogen burning limit, it will effectively become a degenerate brown dwarf, expanding as further mass is lost. For that reason, the evolutionary tracks in Figure \ref{fig:mesa} turn around at a period of $\sim$70--80 min. Such systems are known as ``period bouncers'', and the evolution of CVs (once accretion begins) through this phase has been well-documented by seminal works in the field \citep[e.g.][ and references therein]{2011knigge}.

\section{Discussion}
\label{sec:population}

\subsection{Population Constraints and Detectability by Upcoming Surveys}

From two WD + M dwarf LPTs alone, we can calculate lower limits on the space density of systems in the solar neighborhood. We consider the effective stellar volume as a function of distance to be \citep[e.g.][]{2021elbadry_survey}:
\begin{gather}
    V_\mathrm{eff} (d_\mathrm{lim}) = 2\pi\int_0^{d_\mathrm{lim}} e^{-z/h} (d^2_\textrm{lim} - z^2) dz
    \label{eq:volume}
\end{gather}
where $V_\mathrm{eff}$ is the effective stellar volume, and $d_\mathrm{lim}$ is the limiting distance. $h$ is a characteristic height, which set to 1 kpc due to the thick disk nature of WD + M dwarf LPTs. Assuming 100 percent completeness out to 380 pc (the distance to GLEAM-X J0704), the space density is simply the total number of systems divided by $V_\mathrm{eff} (d_\mathrm{lim})$, which yields $\rho =1\times 10^{-8} \;\mathrm{pc}^{-3}$. If current radio surveys are only 10\% complete out to that distance, it follows that the true space density of WD + M dwarf LPTs is $\rho =1\times 10^{-7}\;\mathrm{pc}^{-3}$.

However, based on current radio surveys and the unknown duty cycle of WD + M dwarf LPTs, placing any stronger constraints on the space density is difficult. One possible way to constrain the total number of WD + M dwarf binaries that \textit{could undergo} LPT-like pulses would be to start with the observed space density of all WD + M dwarf close binaries. \cite{2021rebassa} found a space density of $\rho = 3\times10^{-5} \mathrm{pc}^{-3}$ from a volume-limited \textit{Gaia} sample (corrected to an upper limit of $4\times10^{-4} \mathrm{pc}^{-3}$ when accounting for all low-mass companions, not just M dwarfs), while \cite{2026shariat} found a space density of $\rho = 7\times10^{-5} \mathrm{pc}^{-3}$ from eclipsing sources discovered with the Zwicky Transient Facility (ZTF). A back-of-the envelope compromise would be $\sim10^{-4} \mathrm{pc}^{-3}$. Then, we can multiply by the fraction of systems that harbor WDs with $M\gtrsim0.8M_\odot$ (10\%, based on results from \cite{2021rebassa}), and by the fraction of systems with inclinations similar to GLEAM-X J0704 and lower (15\%). This leads to a space density of $\rho\sim10^{-6} \mathrm{pc}^{-3}$. The final step would be to multiply by the fraction of systems that harbor a magnetic WD. While it has been observed that over a third of accreting WDs are magnetic \citep{2020pala, 2024rodriguez_survey}, only a few percent of post common envelope binaries (PCEBs) have been observed to host a magnetic WD \citep[e.g.][]{2011zorotovic, 2021schreiber}. Because the rate of magnetism as a function of WD mass is unknown, we can choose the final factor to range from 1--10\%, leading to an estimated theoretical space density of WD + M dwarf LPTs between $\rho\sim10^{-8} - 10^{-7} \mathrm{pc}^{-3}$. 

\begin{figure}
    \centering
    \includegraphics[width=0.5\textwidth]{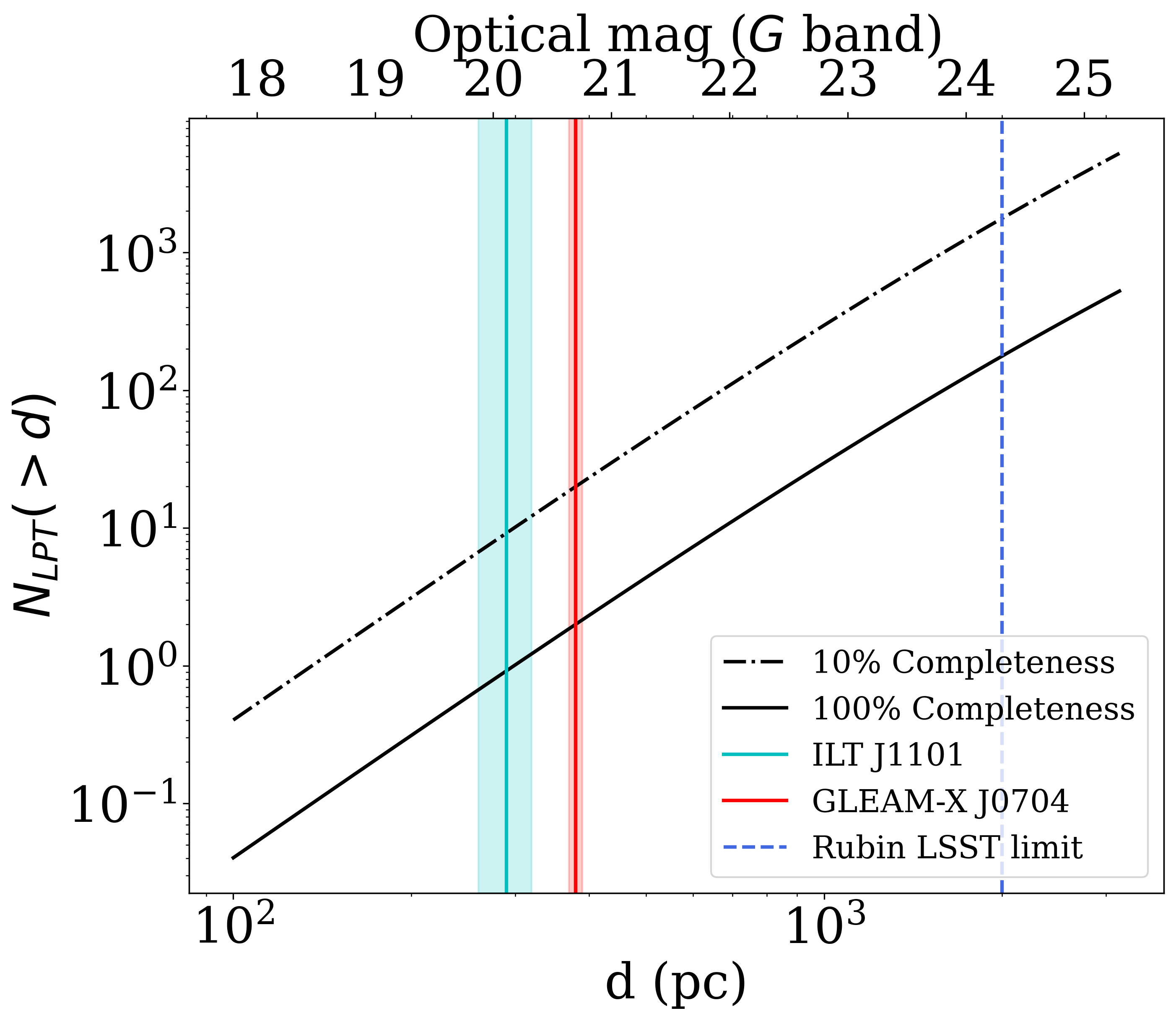}
    \caption{Based on these two systems alone, a lower limit on the space density of WD + M dwarf LPTs in the solar neighborhood can be placed ($\rho \gtrsim 10^{-8} \mathrm{pc}^{-3}$). If current radio surveys are 100 (10) percent complete, 100 (1000) similar systems should exist out to $\sim$2 kpc. Assuming a $\sim$3000 K M dwarf, like the one in ILT J1101 and GLEAM-X J0704, Rubin LSST will be able to detect optical counterparts to such systems out to $\sim$2 kpc. }
    \label{fig:density}
\end{figure}

Remarkably, the lower limits of this theoretical estimate agree with our observed lower limit, meaning that current radio surveys may be 10--100\% complete in finding WD + M dwarf LPTs out to 380 pc. In Figure \ref{fig:density}, we show the estimated total number of WD + M dwarf LPTs as a function of distance, assuming 10 and 100\% completeness of current radio surveys. We also show the number of systems as a function of optical magnitude of a 3000 K M dwarf, like the one in ILT J1101 and GLEAM-X J0704. 

While current all-sky surveys reach limits of 21--22 mag, the projected limit of the Rubin Observatory Legacy Survey of Space and Time is $\sim 24.3$ mag per visit in $g$ band \citep[LSST;][]{2019rubin}. Figure \ref{fig:density} shows that if new LPTs are discovered by radio facilities during the course of the LSST 10-year survey, optical counteraparts of M dwarf companions out to 2 kpc can be seen by LSST in real-time (assuming no extinction). Current large-scale efforts to find LPTs in radio surveys have placed strong upper limits on their occurrence rate, highlighting the need for deep optical surveys that can detect counterparts for more distant radio-identified LPTs over the next decade \citep{2025sherman, 2026lee_askap}. 

Moving forward, radio surveys will be of utmost importance for the discovery of new LPTs. The Coherent All Sky Monitor \citep[CASM; as proposed by][]{2023connor, 2026casm} will be sensitive to large fields of view, enabling discovery outside of the Galactic Plane, as is the case with ILT J1101 and GLEAM-X J0704. \textrm{Other upcoming radio surveys such as the Bustling Universe Radio Survey Telescope in Taiwan\citep[BURSTT;][]{2022burstt} and 
The Canadian Hydrogen Observatory and Radio-transient Detector \citep[CHORD;][]{2019chord} may also discover new LPTs due to their sensitivity to transient objects.} The Deep Synoptic Array \citep[DSA-2000;][]{2019hallinan} and Square Kilometre Array \citep[SKA;][]{2013ska} will also play important roles due to their increased sensitivity over current surveys.

\subsection{Open Questions}

We list below some of the major questions still surrounding WD + M dwarf LPTs and some suggestions for resolving them:
\begin{enumerate}
    \item \textit{Do WD + M dwarf LPTs represent the majority of the entire LPT population?} Here, we confirmed the nearby distance ($\sim$300 pc) of ILT J1101 found by \cite{2025deruiter}, and \cite{2025rodriguez_gleamx} established the distance of GLEAM-X J0704 to be $\sim$400 pc. This is in contrast to the majority of all LPTs, which have large radio DMs, placing them at 1--8 kpc distances. The only exception to date is CHIME J0630, whose DM places the source at $170^{+310}_{-100}$ pc \citep{2024dong}. Since that source is located towards the Galactic anticenter, where DM-inferred distances have at times proven to be unreliable, it is possible that the inferred distance is underestimated. Radio monitoring of nearby WD + M dwarf binaries identified through spectroscopic surveys (e.g. SDSS, DESI, LAMOST) and candidates identified through cuts in the \textit{Gaia} HR diagram for systems just below the main sequence would be most useful in establishing the fraction of such systems that exhibit the LPT phenomenon and their activity duty cycles.
    
    \item \textit{How are WD + M dwarf LPTs related to the evolutionary sequence of magnetic WDs in close binaries (e.g. WD propellers, pulsars, pre-polars, polars)?} The evolutionary sequence which ties the formation of strong WD magnetic fields in magnetic CVs presented by \cite{2021schreiber} requires an initial phase of accretion to spin-up the WD. Given the low temperature of the WD in both ILT J1101 and GLEAM-X J0704, it is unlikely that they are part of that evolutionary channel, as was proposed by \cite{2025yang}. Not only that, but the Roche lobe filling factor, $R_L$ of ILT J1101 and GLEAM-X J0704 is far below 0.94, which is the lowest value that $R_L$ takes on during the period gap when the WD and M dwarf detach, according to the models of \cite{2021schreiber}. Instead, it is likely that the WD in ILT J1101 and GLEAM-X J0704 is spinning at $\sim$hr long periods, not spun up by accretion (though the prospect of winds spinning up the WD should be explored). Due to their strongly crystallized cores, this would be consistent with forming $\sim$1 MG magnetic fields, as was shown by \cite{2022ginzburg}. Obtaining estimates of the initial mass ratio and orbital separation of WD + M dwarf binaries after common envelope evolution would be most helpful in determining which fraction of magnetic CVs undergo an initial accretion spin-up phase as outlined by \cite{2021schreiber} and which fraction do not come into contact until $\sim$1--2 hr long orbital periods like ILT J1101 and GLEAM-X J0704. 
    
    \item \textit{Is CHIME/ILT J1634+44 a WD binary LPT?} This enigmatic system has been associated with a very faint optical/UV counterpart and radio period derivative, leading \cite{2025bloot_1634} and \cite{2025dong_1634} to suggest that it could be a WD orbited by a very late-type M dwarf, brown dwarf or WD. Further UV/optical observations of this system are needed to establish the nature of this system. 
    
    \item \textit{What is the magnetic field strength of the WD in the WD + M dwarf LPTs?} We do not detect any clear Balmer lines from the WD itself in either ILT J1101 or GLEAM-X J0704, ruling out the possibility of measuring the magnetic field via Zeeman splitting. Furthermore, we do not see any cyclotron humps. \cite{2025vanroestel} found that magnetic WDs with companions that have similar masses Roche lobe filling factors to the ones in ILT J1101 and GLEAM-X J0704 are most likely to show strong cyclotron emission patterns only at specific magnetic pole and orbital inclination configurations, with low inclinations generally disfavoring the cyclotron beaming towards our line of sight. \textrm{Essentially, not only can we not measure the WD magnetic field strength, but we cannot determine if the WD is magnetic at all from our current observations.}

    \item \textit{Do the radio and orbital periods differ in the WD + M dwarf LPTs? What is the WD spin period?} These are the most important questions to answer in order to fully constrain the emission mechanisms in WD + M dwarf LPTs. Presently, phase-resolved observations of a single period constrain the orbital period to a precision of $\sim$few minutes, but multiple observations taken on near-simultaneous nights are needed to improve this to sub-second precision. 

    \item \textrm{\textit{How do ILT J1101 and GLEAM-X J0704 relate to the accreting WD, ASKAP J174508.9-505149, which shows LPT behavior? } During the review phase of this paper, ASKAP J174508.9-505149 was discovered as a cataclysmic variable (polar or asynchronous polar) showing LPT-type behavior \citep{2026rose}. This system shows radio pulsations on a period similar to the orbital period, both $\sim1.3$ hr. This discovery demonstrates that the LPT mechanism can operate during accretion, which implies that ILT J1101 and GLEAM-X J0704 may show LPT pulsations later in their evolution.   }
    
\end{enumerate}

\section{Conclusions}
We presented new, high signal-to-noise spectroscopy of the WD + M dwarf LPT, ILT J1101, over an entire radio period, and compared the binary properties to those of the other WD + M dwarf LPT, GLEAM-X J0704.  Our main conclusions are the following:
\begin{enumerate}
    \item ILT J1101 exhibits RV shifts ($96\pm6\;\mathrm{km s}^{-1}$) that show that the binary orbital period agrees with the radio pulse period to within five percent (Figure \ref{fig:rv}). The RV amplitude, along with most binary parameters agree with the findings of \cite{2025deruiter}, but have smaller error bars due to our higher signal-to-noise. The H$\alpha$ emission line RV curve is consistent with that of the Na I doublet, signaling an origin from the M dwarf, likely owing to chromospheric activity (Figure \ref{fig:halpha}).
    \item Radio pulses nearly coincide with the ascending node of the orbit when the M dwarf is at maximum redshift and WD at maximum blueshift (Figure \ref{fig:rv}). This differs from what was initially reported by \cite{2025deruiter}, and is similar to the behavior seen in GLEAM-X J0704 \citep{2025rodriguez_gleamx}. 
    \item ILT J1101 and GLEAM-X J0704 are old systems, whose orbital kinematics are consistent with being members of the Milky Way thick disk (Figure \ref{fig:toomre}). \textrm{If the more distant LPTs, which are more concentrated in the Galactic Plane, are established to be members of the thin disk, then} this establishes WD + MD LPTs as a unique sub-population of LPTs.
    \item The LPT phenomenon could favor low orbital inclinations, as well as cool, massive WDs that have undergone significant core crystallization (Figures \ref{fig:inclination} and \ref{fig:crystal}). This may also support theoretical work which has suggested that WDs in close binaries form strong magnetic fields through a crystallization-driven dynamo \citep[e.g.][]{2021schreiber, 2022ginzburg}.
    \item ILT J1101 and GLEAM-X J0704 will evolve into Roche lobe-filling binaries in 1 Gyr or less due to the gravitational wave-driven shrinkage of their orbits (Figures \ref{fig:mesa} and \ref{fig:rl}).
    \item The lower limit for the local space density of WD + M dwarf LPTs is $\rho \gtrsim 10^{-8} \mathrm{pc}^{-3}$. If current radio surveys are 100\% (10\%) complete, this means that $\sim$100 ($\sim$1000) optical counterparts to newly discovered WD + M dwarf LPTs will be detectable by the Rubin Observatory Legacy Survey of Space and Time (LSST) in the coming decade (Figure \ref{fig:density}).
\end{enumerate}

This work shows that LPTs are evolving into a true new class of astronomical objects, worthy of population studies and connection to previously known phenomena. In this case, only by jointly analyzing the binary properties of ILT J1101 and GLEAM-X J0704 could connections to WD magnetism have been made, as well as conclusions about their future evolution and underlying population. Upcoming radio surveys will play a major role in discovering more systems, and systematic surveys such as the work done by \cite{2025sherman} and \cite{2026lee_askap} will be vital in establishing the true space density of LPTs. Just as exciting is the role that multiwavelength facilities, such as the Rubin Observatory Legacy Survey of Space and Time (LSST) will play in establishing counterparts that will allow for detailed characterization of LPTs as has been done here.

\section{Acknowledgments}
\textrm{We thank the anonymous referees whose input led to an improved final manuscript.} We also thank Myles Sherman, Elias Most, Yici Zhong, and Ingrid Pelisoli for valuable discussions, as well as Shri Kulkarni for early encouragement to pursue this project. We thank Kaya Mori and Matthew Lundy for organizing the 2026 Columbia LPT workshop, and all attendees for insightful contributions. ACR acknowledges support from a Future Faculty Leader Fellowship and the Institute for Theory and Computation at the Center for Astrophysics $|$ Harvard \& Smithsonian. We are grateful to the staff of Keck Observatory for their support in carrying out the observations presented here.\\

Some of the data presented herein were obtained at Keck Observatory, which is a private 501(c)3 non-profit organization operated as a scientific partnership among the California Institute of Technology, the University of California, and the National Aeronautics and Space Administration. The Observatory was made possible by the generous financial support of the W. M. Keck Foundation. We wish to recognize and acknowledge the very significant cultural role and reverence that the summit of Maunakea has always had within the Native Hawaiian community. We are most fortunate to have the opportunity to conduct observations from this mountain.

This research was supported by NSF grants NSF-AST-2540180 and AST-2508988.

\bibliographystyle{apsrev4-1}

\bibliography{oja_template}

\begin{thebibliography}{93}%
\makeatletter
\providecommand \@ifxundefined [1]{%
 \@ifx{#1\undefined}
}%
\providecommand \@ifnum [1]{%
 \ifnum #1\expandafter \@firstoftwo
 \else \expandafter \@secondoftwo
 \fi
}%
\providecommand \@ifx [1]{%
 \ifx #1\expandafter \@firstoftwo
 \else \expandafter \@secondoftwo
 \fi
}%
\providecommand \natexlab [1]{#1}%
\providecommand \enquote  [1]{``#1''}%
\providecommand \bibnamefont  [1]{#1}%
\providecommand \bibfnamefont [1]{#1}%
\providecommand \citenamefont [1]{#1}%
\providecommand \href@noop [0]{\@secondoftwo}%
\providecommand \href [0]{\begingroup \@sanitize@url \@href}%
\providecommand \@href[1]{\@@startlink{#1}\@@href}%
\providecommand \@@href[1]{\endgroup#1\@@endlink}%
\providecommand \@sanitize@url [0]{\catcode `\\12\catcode `\$12\catcode `\&12\catcode `\#12\catcode `\^12\catcode `\_12\catcode `\%12\relax}%
\providecommand \@@startlink[1]{}%
\providecommand \@@endlink[0]{}%
\providecommand \url  [0]{\begingroup\@sanitize@url \@url }%
\providecommand \@url [1]{\endgroup\@href {#1}{\urlprefix }}%
\providecommand \urlprefix  [0]{URL }%
\providecommand \Eprint [0]{\href }%
\providecommand \doibase [0]{http://dx.doi.org/}%
\providecommand \selectlanguage [0]{\@gobble}%
\providecommand \bibinfo  [0]{\@secondoftwo}%
\providecommand \bibfield  [0]{\@secondoftwo}%
\providecommand \translation [1]{[#1]}%
\providecommand \BibitemOpen [0]{}%
\providecommand \bibitemStop [0]{}%
\providecommand \bibitemNoStop [0]{.\EOS\space}%
\providecommand \EOS [0]{\spacefactor3000\relax}%
\providecommand \BibitemShut  [1]{\csname bibitem#1\endcsname}%
\let\auto@bib@innerbib\@empty
\bibitem [{\citenamefont {{Rea}}\ \emph {et~al.}(2026)\citenamefont {{Rea}}, \citenamefont {{Hurley-Walker}},\ and\ \citenamefont {{Caleb}}}]{2026rea}%
  \BibitemOpen
  \bibfield  {author} {\bibinfo {author} {\bibfnamefont {N.}~\bibnamefont {{Rea}}}, \bibinfo {author} {\bibfnamefont {N.}~\bibnamefont {{Hurley-Walker}}}, \ and\ \bibinfo {author} {\bibfnamefont {M.}~\bibnamefont {{Caleb}}},\ }\href {\doibase 10.48550/arXiv.2601.10393} {\bibfield  {journal} {\bibinfo  {journal} {arXiv e-prints}\ ,\ \bibinfo {eid} {arXiv:2601.10393}} (\bibinfo {year} {2026})},\ \Eprint {http://arxiv.org/abs/2601.10393} {arXiv:2601.10393 [astro-ph.HE]} \BibitemShut {NoStop}%
\bibitem [{\citenamefont {{de Ruiter}}\ \emph {et~al.}(2025)\citenamefont {{de Ruiter}}, \citenamefont {{Rajwade}}, \citenamefont {{Bassa}}, \citenamefont {{Rowlinson}}, \citenamefont {{Wijers}}, \citenamefont {{Kilpatrick}}, \citenamefont {{Stefansson}}, \citenamefont {{Callingham}}, \citenamefont {{Hessels}}, \citenamefont {{Clarke}}, \citenamefont {{Peters}}, \citenamefont {{Wijnands}}, \citenamefont {{Shimwell}}, \citenamefont {{ter Veen}}, \citenamefont {{Morello}}, \citenamefont {{Zeimann}},\ and\ \citenamefont {{Mahadevan}}}]{2025deruiter}%
  \BibitemOpen
  \bibfield  {author} {\bibinfo {author} {\bibfnamefont {I.}~\bibnamefont {{de Ruiter}}}, \bibinfo {author} {\bibfnamefont {K.~M.}\ \bibnamefont {{Rajwade}}}, \bibinfo {author} {\bibfnamefont {C.~G.}\ \bibnamefont {{Bassa}}}, \bibinfo {author} {\bibfnamefont {A.}~\bibnamefont {{Rowlinson}}}, \bibinfo {author} {\bibfnamefont {R.~A.~M.~J.}\ \bibnamefont {{Wijers}}}, \bibinfo {author} {\bibfnamefont {C.~D.}\ \bibnamefont {{Kilpatrick}}}, \bibinfo {author} {\bibfnamefont {G.}~\bibnamefont {{Stefansson}}}, \bibinfo {author} {\bibfnamefont {J.~R.}\ \bibnamefont {{Callingham}}}, \bibinfo {author} {\bibfnamefont {J.~W.~T.}\ \bibnamefont {{Hessels}}}, \bibinfo {author} {\bibfnamefont {T.~E.}\ \bibnamefont {{Clarke}}}, \bibinfo {author} {\bibfnamefont {W.}~\bibnamefont {{Peters}}}, \bibinfo {author} {\bibfnamefont {R.~A.~D.}\ \bibnamefont {{Wijnands}}}, \bibinfo {author} {\bibfnamefont {T.~W.}\ \bibnamefont {{Shimwell}}}, \bibinfo {author} {\bibfnamefont {S.}~\bibnamefont {{ter Veen}}}, \bibinfo {author} {\bibfnamefont
  {V.}~\bibnamefont {{Morello}}}, \bibinfo {author} {\bibfnamefont {G.~R.}\ \bibnamefont {{Zeimann}}}, \ and\ \bibinfo {author} {\bibfnamefont {S.}~\bibnamefont {{Mahadevan}}},\ }\href {\doibase 10.1038/s41550-025-02491-0} {\bibfield  {journal} {\bibinfo  {journal} {Nature Astronomy}\ }\textbf {\bibinfo {volume} {9}},\ \bibinfo {pages} {672} (\bibinfo {year} {2025})},\ \Eprint {http://arxiv.org/abs/2408.11536} {arXiv:2408.11536 [astro-ph.HE]} \BibitemShut {NoStop}%
\bibitem [{\citenamefont {{Hurley-Walker}}\ \emph {et~al.}(2024)\citenamefont {{Hurley-Walker}}, \citenamefont {{McSweeney}}, \citenamefont {{Bahramian}}, \citenamefont {{Rea}}, \citenamefont {{Horv{\'a}th}}, \citenamefont {{Buchner}}, \citenamefont {{Williams}}, \citenamefont {{Meyers}}, \citenamefont {{Strader}}, \citenamefont {{Aydi}}, \citenamefont {{Urquhart}}, \citenamefont {{Chomiuk}}, \citenamefont {{Galvin}}, \citenamefont {{Coti Zelati}},\ and\ \citenamefont {{Bailes}}}]{2024discovery}%
  \BibitemOpen
  \bibfield  {author} {\bibinfo {author} {\bibfnamefont {N.}~\bibnamefont {{Hurley-Walker}}}, \bibinfo {author} {\bibfnamefont {S.~J.}\ \bibnamefont {{McSweeney}}}, \bibinfo {author} {\bibfnamefont {A.}~\bibnamefont {{Bahramian}}}, \bibinfo {author} {\bibfnamefont {N.}~\bibnamefont {{Rea}}}, \bibinfo {author} {\bibfnamefont {C.}~\bibnamefont {{Horv{\'a}th}}}, \bibinfo {author} {\bibfnamefont {S.}~\bibnamefont {{Buchner}}}, \bibinfo {author} {\bibfnamefont {A.}~\bibnamefont {{Williams}}}, \bibinfo {author} {\bibfnamefont {B.~W.}\ \bibnamefont {{Meyers}}}, \bibinfo {author} {\bibfnamefont {J.}~\bibnamefont {{Strader}}}, \bibinfo {author} {\bibfnamefont {E.}~\bibnamefont {{Aydi}}}, \bibinfo {author} {\bibfnamefont {R.}~\bibnamefont {{Urquhart}}}, \bibinfo {author} {\bibfnamefont {L.}~\bibnamefont {{Chomiuk}}}, \bibinfo {author} {\bibfnamefont {T.~J.}\ \bibnamefont {{Galvin}}}, \bibinfo {author} {\bibfnamefont {F.}~\bibnamefont {{Coti Zelati}}}, \ and\ \bibinfo {author} {\bibfnamefont {M.}~\bibnamefont
  {{Bailes}}},\ }\href {\doibase 10.3847/2041-8213/ad890e} {\bibfield  {journal} {\bibinfo  {journal} {\apjl}\ }\textbf {\bibinfo {volume} {976}},\ \bibinfo {eid} {L21} (\bibinfo {year} {2024})},\ \Eprint {http://arxiv.org/abs/2408.15757} {arXiv:2408.15757 [astro-ph.SR]} \BibitemShut {NoStop}%
\bibitem [{\citenamefont {{Rodriguez}}(2025)}]{2025rodriguez_gleamx}%
  \BibitemOpen
  \bibfield  {author} {\bibinfo {author} {\bibfnamefont {A.~C.}\ \bibnamefont {{Rodriguez}}},\ }\href {\doibase 10.1051/0004-6361/202553684} {\bibfield  {journal} {\bibinfo  {journal} {\aap}\ }\textbf {\bibinfo {volume} {695}},\ \bibinfo {eid} {L8} (\bibinfo {year} {2025})},\ \Eprint {http://arxiv.org/abs/2501.03315} {arXiv:2501.03315 [astro-ph.SR]} \BibitemShut {NoStop}%
\bibitem [{\citenamefont {{Pelisoli}}\ \emph {et~al.}(2025)\citenamefont {{Pelisoli}}, \citenamefont {{Brown}}, \citenamefont {{Castro Segura}}, \citenamefont {{Dhillon}}, \citenamefont {{Dyer}}, \citenamefont {{Garbutt}}, \citenamefont {{Green}}, \citenamefont {{Jarvis}}, \citenamefont {{Kennedy}}, \citenamefont {{Kerry}}, \citenamefont {{Littlefair}}, \citenamefont {{McCormac}}, \citenamefont {{Munday}}, \citenamefont {{Parsons}}, \citenamefont {{Pike}}, \citenamefont {{Sahman}},\ and\ \citenamefont {{Yates}}}]{2025pelisoli_limits}%
  \BibitemOpen
  \bibfield  {author} {\bibinfo {author} {\bibfnamefont {I.}~\bibnamefont {{Pelisoli}}}, \bibinfo {author} {\bibfnamefont {A.~J.}\ \bibnamefont {{Brown}}}, \bibinfo {author} {\bibfnamefont {N.}~\bibnamefont {{Castro Segura}}}, \bibinfo {author} {\bibfnamefont {V.~S.}\ \bibnamefont {{Dhillon}}}, \bibinfo {author} {\bibfnamefont {M.~J.}\ \bibnamefont {{Dyer}}}, \bibinfo {author} {\bibfnamefont {J.~A.}\ \bibnamefont {{Garbutt}}}, \bibinfo {author} {\bibfnamefont {M.~J.}\ \bibnamefont {{Green}}}, \bibinfo {author} {\bibfnamefont {D.}~\bibnamefont {{Jarvis}}}, \bibinfo {author} {\bibfnamefont {M.~R.}\ \bibnamefont {{Kennedy}}}, \bibinfo {author} {\bibfnamefont {P.}~\bibnamefont {{Kerry}}}, \bibinfo {author} {\bibfnamefont {S.~P.}\ \bibnamefont {{Littlefair}}}, \bibinfo {author} {\bibfnamefont {J.}~\bibnamefont {{McCormac}}}, \bibinfo {author} {\bibfnamefont {J.}~\bibnamefont {{Munday}}}, \bibinfo {author} {\bibfnamefont {S.~G.}\ \bibnamefont {{Parsons}}}, \bibinfo {author} {\bibfnamefont {E.}~\bibnamefont
  {{Pike}}}, \bibinfo {author} {\bibfnamefont {D.~I.}\ \bibnamefont {{Sahman}}}, \ and\ \bibinfo {author} {\bibfnamefont {A.}~\bibnamefont {{Yates}}},\ }\href {\doibase 10.1093/mnrasl/slaf101} {\bibfield  {journal} {\bibinfo  {journal} {\mnras}\ }\textbf {\bibinfo {volume} {544}},\ \bibinfo {pages} {L76} (\bibinfo {year} {2025})},\ \Eprint {http://arxiv.org/abs/2509.20438} {arXiv:2509.20438 [astro-ph.SR]} \BibitemShut {NoStop}%
\bibitem [{\citenamefont {{Gold}}(1968)}]{1968gold}%
  \BibitemOpen
  \bibfield  {author} {\bibinfo {author} {\bibfnamefont {T.}~\bibnamefont {{Gold}}},\ }\href {\doibase 10.1038/218731a0} {\bibfield  {journal} {\bibinfo  {journal} {\nat}\ }\textbf {\bibinfo {volume} {218}},\ \bibinfo {pages} {731} (\bibinfo {year} {1968})}\BibitemShut {NoStop}%
\bibitem [{\citenamefont {{Pacini}}(1968)}]{1968pacini}%
  \BibitemOpen
  \bibfield  {author} {\bibinfo {author} {\bibfnamefont {F.}~\bibnamefont {{Pacini}}},\ }\href {\doibase 10.1038/219145a0} {\bibfield  {journal} {\bibinfo  {journal} {\nat}\ }\textbf {\bibinfo {volume} {219}},\ \bibinfo {pages} {145} (\bibinfo {year} {1968})}\BibitemShut {NoStop}%
\bibitem [{\citenamefont {{Kulkarni}}(1986)}]{1986kulkarni}%
  \BibitemOpen
  \bibfield  {author} {\bibinfo {author} {\bibfnamefont {S.~R.}\ \bibnamefont {{Kulkarni}}},\ }\href {\doibase 10.1086/184711} {\bibfield  {journal} {\bibinfo  {journal} {\apjl}\ }\textbf {\bibinfo {volume} {306}},\ \bibinfo {pages} {L85} (\bibinfo {year} {1986})}\BibitemShut {NoStop}%
\bibitem [{\citenamefont {{Phinney}}\ and\ \citenamefont {{Kulkarni}}(1994)}]{1994pulsars}%
  \BibitemOpen
  \bibfield  {author} {\bibinfo {author} {\bibfnamefont {E.~S.}\ \bibnamefont {{Phinney}}}\ and\ \bibinfo {author} {\bibfnamefont {S.~R.}\ \bibnamefont {{Kulkarni}}},\ }\href {\doibase 10.1146/annurev.aa.32.090194.003111} {\bibfield  {journal} {\bibinfo  {journal} {\araa}\ }\textbf {\bibinfo {volume} {32}},\ \bibinfo {pages} {591} (\bibinfo {year} {1994})}\BibitemShut {NoStop}%
\bibitem [{\citenamefont {{Zhang}}\ and\ \citenamefont {{Gil}}(2005)}]{2005zhang}%
  \BibitemOpen
  \bibfield  {author} {\bibinfo {author} {\bibfnamefont {B.}~\bibnamefont {{Zhang}}}\ and\ \bibinfo {author} {\bibfnamefont {J.}~\bibnamefont {{Gil}}},\ }\href {\doibase 10.1086/497428} {\bibfield  {journal} {\bibinfo  {journal} {\apjl}\ }\textbf {\bibinfo {volume} {631}},\ \bibinfo {pages} {L143} (\bibinfo {year} {2005})},\ \Eprint {http://arxiv.org/abs/astro-ph/0508213} {arXiv:astro-ph/0508213 [astro-ph]} \BibitemShut {NoStop}%
\bibitem [{\citenamefont {{Katz}}(2022)}]{2022katz}%
  \BibitemOpen
  \bibfield  {author} {\bibinfo {author} {\bibfnamefont {J.~I.}\ \bibnamefont {{Katz}}},\ }\href {\doibase 10.1007/s10509-022-04146-2} {\bibfield  {journal} {\bibinfo  {journal} {\apss}\ }\textbf {\bibinfo {volume} {367}},\ \bibinfo {eid} {108} (\bibinfo {year} {2022})},\ \Eprint {http://arxiv.org/abs/2203.08112} {arXiv:2203.08112 [astro-ph.SR]} \BibitemShut {NoStop}%
\bibitem [{\citenamefont {{Loeb}}\ and\ \citenamefont {{Maoz}}(2022)}]{2022loeb}%
  \BibitemOpen
  \bibfield  {author} {\bibinfo {author} {\bibfnamefont {A.}~\bibnamefont {{Loeb}}}\ and\ \bibinfo {author} {\bibfnamefont {D.}~\bibnamefont {{Maoz}}},\ }\href {\doibase 10.3847/2515-5172/ac52f1} {\bibfield  {journal} {\bibinfo  {journal} {Research Notes of the American Astronomical Society}\ }\textbf {\bibinfo {volume} {6}},\ \bibinfo {eid} {27} (\bibinfo {year} {2022})},\ \Eprint {http://arxiv.org/abs/2202.04949} {arXiv:2202.04949 [astro-ph.SR]} \BibitemShut {NoStop}%
\bibitem [{\citenamefont {{Nathanail}}(2025)}]{2025nathanail}%
  \BibitemOpen
  \bibfield  {author} {\bibinfo {author} {\bibfnamefont {A.}~\bibnamefont {{Nathanail}}},\ }\href {\doibase 10.1051/0004-6361/202555993} {\bibfield  {journal} {\bibinfo  {journal} {\aap}\ }\textbf {\bibinfo {volume} {704}},\ \bibinfo {eid} {A123} (\bibinfo {year} {2025})},\ \Eprint {http://arxiv.org/abs/2506.17389} {arXiv:2506.17389 [astro-ph.HE]} \BibitemShut {NoStop}%
\bibitem [{\citenamefont {{Beniamini}}\ \emph {et~al.}(2023)\citenamefont {{Beniamini}}, \citenamefont {{Wadiasingh}}, \citenamefont {{Hare}}, \citenamefont {{Rajwade}}, \citenamefont {{Younes}},\ and\ \citenamefont {{van der Horst}}}]{2023beniamini}%
  \BibitemOpen
  \bibfield  {author} {\bibinfo {author} {\bibfnamefont {P.}~\bibnamefont {{Beniamini}}}, \bibinfo {author} {\bibfnamefont {Z.}~\bibnamefont {{Wadiasingh}}}, \bibinfo {author} {\bibfnamefont {J.}~\bibnamefont {{Hare}}}, \bibinfo {author} {\bibfnamefont {K.~M.}\ \bibnamefont {{Rajwade}}}, \bibinfo {author} {\bibfnamefont {G.}~\bibnamefont {{Younes}}}, \ and\ \bibinfo {author} {\bibfnamefont {A.~J.}\ \bibnamefont {{van der Horst}}},\ }\href {\doibase 10.1093/mnras/stad208} {\bibfield  {journal} {\bibinfo  {journal} {\mnras}\ }\textbf {\bibinfo {volume} {520}},\ \bibinfo {pages} {1872} (\bibinfo {year} {2023})},\ \Eprint {http://arxiv.org/abs/2210.09323} {arXiv:2210.09323 [astro-ph.HE]} \BibitemShut {NoStop}%
\bibitem [{\citenamefont {{Cooper}}\ and\ \citenamefont {{Wadiasingh}}(2024)}]{2024cooper}%
  \BibitemOpen
  \bibfield  {author} {\bibinfo {author} {\bibfnamefont {A.~J.}\ \bibnamefont {{Cooper}}}\ and\ \bibinfo {author} {\bibfnamefont {Z.}~\bibnamefont {{Wadiasingh}}},\ }\href {\doibase 10.1093/mnras/stae1813} {\bibfield  {journal} {\bibinfo  {journal} {\mnras}\ }\textbf {\bibinfo {volume} {533}},\ \bibinfo {pages} {2133} (\bibinfo {year} {2024})},\ \Eprint {http://arxiv.org/abs/2406.04135} {arXiv:2406.04135 [astro-ph.HE]} \BibitemShut {NoStop}%
\bibitem [{\citenamefont {{Cary}}\ \emph {et~al.}(2025)\citenamefont {{Cary}}, \citenamefont {{Lu}}, \citenamefont {{Leung}},\ and\ \citenamefont {{Wong}}}]{2025cary_lu}%
  \BibitemOpen
  \bibfield  {author} {\bibinfo {author} {\bibfnamefont {S.}~\bibnamefont {{Cary}}}, \bibinfo {author} {\bibfnamefont {W.}~\bibnamefont {{Lu}}}, \bibinfo {author} {\bibfnamefont {C.}~\bibnamefont {{Leung}}}, \ and\ \bibinfo {author} {\bibfnamefont {T.~L.~S.}\ \bibnamefont {{Wong}}},\ }\href {\doibase 10.48550/arXiv.2507.10682} {\bibfield  {journal} {\bibinfo  {journal} {arXiv e-prints}\ ,\ \bibinfo {eid} {arXiv:2507.10682}} (\bibinfo {year} {2025})},\ \Eprint {http://arxiv.org/abs/2507.10682} {arXiv:2507.10682 [astro-ph.HE]} \BibitemShut {NoStop}%
\bibitem [{\citenamefont {{Mao}}\ \emph {et~al.}(2025)\citenamefont {{Mao}}, \citenamefont {{Li}}, \citenamefont {{Lai}}, \citenamefont {{Deng}},\ and\ \citenamefont {{Yang}}}]{2025mao_magnetar}%
  \BibitemOpen
  \bibfield  {author} {\bibinfo {author} {\bibfnamefont {Y.-H.}\ \bibnamefont {{Mao}}}, \bibinfo {author} {\bibfnamefont {X.-D.}\ \bibnamefont {{Li}}}, \bibinfo {author} {\bibfnamefont {D.}~\bibnamefont {{Lai}}}, \bibinfo {author} {\bibfnamefont {Z.-L.}\ \bibnamefont {{Deng}}}, \ and\ \bibinfo {author} {\bibfnamefont {H.-R.}\ \bibnamefont {{Yang}}},\ }\href {\doibase 10.3847/2041-8213/ade80c} {\bibfield  {journal} {\bibinfo  {journal} {\apjl}\ }\textbf {\bibinfo {volume} {988}},\ \bibinfo {eid} {L11} (\bibinfo {year} {2025})},\ \Eprint {http://arxiv.org/abs/2507.00946} {arXiv:2507.00946 [astro-ph.HE]} \BibitemShut {NoStop}%
\bibitem [{\citenamefont {{Suvorov}}\ \emph {et~al.}(2025)\citenamefont {{Suvorov}}, \citenamefont {{Dehman}},\ and\ \citenamefont {{Pons}}}]{2025suvorov}%
  \BibitemOpen
  \bibfield  {author} {\bibinfo {author} {\bibfnamefont {A.~G.}\ \bibnamefont {{Suvorov}}}, \bibinfo {author} {\bibfnamefont {C.}~\bibnamefont {{Dehman}}}, \ and\ \bibinfo {author} {\bibfnamefont {J.~A.}\ \bibnamefont {{Pons}}},\ }\href {\doibase 10.3847/1538-4357/adfc55} {\bibfield  {journal} {\bibinfo  {journal} {\apj}\ }\textbf {\bibinfo {volume} {991}},\ \bibinfo {eid} {134} (\bibinfo {year} {2025})},\ \Eprint {http://arxiv.org/abs/2505.06125} {arXiv:2505.06125 [astro-ph.HE]} \BibitemShut {NoStop}%
\bibitem [{\citenamefont {{Qu}}\ and\ \citenamefont {{Zhang}}(2025)}]{2025qu_zhang}%
  \BibitemOpen
  \bibfield  {author} {\bibinfo {author} {\bibfnamefont {Y.}~\bibnamefont {{Qu}}}\ and\ \bibinfo {author} {\bibfnamefont {B.}~\bibnamefont {{Zhang}}},\ }\href {\doibase 10.3847/1538-4357/adb1b5} {\bibfield  {journal} {\bibinfo  {journal} {\apj}\ }\textbf {\bibinfo {volume} {981}},\ \bibinfo {eid} {34} (\bibinfo {year} {2025})},\ \Eprint {http://arxiv.org/abs/2409.05978} {arXiv:2409.05978 [astro-ph.HE]} \BibitemShut {NoStop}%
\bibitem [{\citenamefont {{Zhong}}\ and\ \citenamefont {{Most}}(2025)}]{2025zhong}%
  \BibitemOpen
  \bibfield  {author} {\bibinfo {author} {\bibfnamefont {Y.}~\bibnamefont {{Zhong}}}\ and\ \bibinfo {author} {\bibfnamefont {E.~R.}\ \bibnamefont {{Most}}},\ }\href {\doibase 10.48550/arXiv.2509.09057} {\bibfield  {journal} {\bibinfo  {journal} {arXiv e-prints}\ ,\ \bibinfo {eid} {arXiv:2509.09057}} (\bibinfo {year} {2025})},\ \Eprint {http://arxiv.org/abs/2509.09057} {arXiv:2509.09057 [astro-ph.HE]} \BibitemShut {NoStop}%
\bibitem [{\citenamefont {{Yang}}(2025)}]{2025yang}%
  \BibitemOpen
  \bibfield  {author} {\bibinfo {author} {\bibfnamefont {Y.-P.}\ \bibnamefont {{Yang}}},\ }\href {\doibase 10.48550/arXiv.2509.09224} {\bibfield  {journal} {\bibinfo  {journal} {arXiv e-prints}\ ,\ \bibinfo {eid} {arXiv:2509.09224}} (\bibinfo {year} {2025})},\ \Eprint {http://arxiv.org/abs/2509.09224} {arXiv:2509.09224 [astro-ph.HE]} \BibitemShut {NoStop}%
\bibitem [{\citenamefont {{Horv{\'a}th}}\ \emph {et~al.}(2025)\citenamefont {{Horv{\'a}th}}, \citenamefont {{Rea}}, \citenamefont {{Hurley-Walker}}, \citenamefont {{McSweeney}}, \citenamefont {{Perley}},\ and\ \citenamefont {{Lenc}}}]{2025horvath}%
  \BibitemOpen
  \bibfield  {author} {\bibinfo {author} {\bibfnamefont {C.}~\bibnamefont {{Horv{\'a}th}}}, \bibinfo {author} {\bibfnamefont {N.}~\bibnamefont {{Rea}}}, \bibinfo {author} {\bibfnamefont {N.}~\bibnamefont {{Hurley-Walker}}}, \bibinfo {author} {\bibfnamefont {S.~J.}\ \bibnamefont {{McSweeney}}}, \bibinfo {author} {\bibfnamefont {R.~A.}\ \bibnamefont {{Perley}}}, \ and\ \bibinfo {author} {\bibfnamefont {E.}~\bibnamefont {{Lenc}}},\ }\href {\doibase 10.48550/arXiv.2507.15352} {\bibfield  {journal} {\bibinfo  {journal} {arXiv e-prints}\ ,\ \bibinfo {eid} {arXiv:2507.15352}} (\bibinfo {year} {2025})},\ \Eprint {http://arxiv.org/abs/2507.15352} {arXiv:2507.15352 [astro-ph.HE]} \BibitemShut {NoStop}%
\bibitem [{\citenamefont {{Pala}}\ \emph {et~al.}(2020)\citenamefont {{Pala}}, \citenamefont {{G{\"a}nsicke}}, \citenamefont {{Breedt}}, \citenamefont {{Knigge}}, \citenamefont {{Hermes}}, \citenamefont {{Gentile Fusillo}}, \citenamefont {{Hollands}}, \citenamefont {{Naylor}}, \citenamefont {{Pelisoli}}, \citenamefont {{Schreiber}}, \citenamefont {{Toonen}}, \citenamefont {{Aungwerojwit}}, \citenamefont {{Cukanovaite}}, \citenamefont {{Dennihy}}, \citenamefont {{Manser}}, \citenamefont {{Pretorius}}, \citenamefont {{Scaringi}},\ and\ \citenamefont {{Toloza}}}]{2020pala}%
  \BibitemOpen
  \bibfield  {author} {\bibinfo {author} {\bibfnamefont {A.~F.}\ \bibnamefont {{Pala}}}, \bibinfo {author} {\bibfnamefont {B.~T.}\ \bibnamefont {{G{\"a}nsicke}}}, \bibinfo {author} {\bibfnamefont {E.}~\bibnamefont {{Breedt}}}, \bibinfo {author} {\bibfnamefont {C.}~\bibnamefont {{Knigge}}}, \bibinfo {author} {\bibfnamefont {J.~J.}\ \bibnamefont {{Hermes}}}, \bibinfo {author} {\bibfnamefont {N.~P.}\ \bibnamefont {{Gentile Fusillo}}}, \bibinfo {author} {\bibfnamefont {M.~A.}\ \bibnamefont {{Hollands}}}, \bibinfo {author} {\bibfnamefont {T.}~\bibnamefont {{Naylor}}}, \bibinfo {author} {\bibfnamefont {I.}~\bibnamefont {{Pelisoli}}}, \bibinfo {author} {\bibfnamefont {M.~R.}\ \bibnamefont {{Schreiber}}}, \bibinfo {author} {\bibfnamefont {S.}~\bibnamefont {{Toonen}}}, \bibinfo {author} {\bibfnamefont {A.}~\bibnamefont {{Aungwerojwit}}}, \bibinfo {author} {\bibfnamefont {E.}~\bibnamefont {{Cukanovaite}}}, \bibinfo {author} {\bibfnamefont {E.}~\bibnamefont {{Dennihy}}}, \bibinfo {author} {\bibfnamefont {C.~J.}\
  \bibnamefont {{Manser}}}, \bibinfo {author} {\bibfnamefont {M.~L.}\ \bibnamefont {{Pretorius}}}, \bibinfo {author} {\bibfnamefont {S.}~\bibnamefont {{Scaringi}}}, \ and\ \bibinfo {author} {\bibfnamefont {O.}~\bibnamefont {{Toloza}}},\ }\href {\doibase 10.1093/mnras/staa764} {\bibfield  {journal} {\bibinfo  {journal} {\mnras}\ }\textbf {\bibinfo {volume} {494}},\ \bibinfo {pages} {3799} (\bibinfo {year} {2020})},\ \Eprint {http://arxiv.org/abs/1907.13152} {arXiv:1907.13152 [astro-ph.SR]} \BibitemShut {NoStop}%
\bibitem [{\citenamefont {{Rodriguez}}\ \emph {et~al.}(2025)\citenamefont {{Rodriguez}}, \citenamefont {{El-Badry}}, \citenamefont {{Suleimanov}}, \citenamefont {{Pala}}, \citenamefont {{Kulkarni}}, \citenamefont {{Gaensicke}}, \citenamefont {{Mori}}, \citenamefont {{Rich}}, \citenamefont {{Sarkar}}, \citenamefont {{Bao}}, \citenamefont {{Lopes de Oliveira}}, \citenamefont {{Ramsay}}, \citenamefont {{Szkody}}, \citenamefont {{Graham}}, \citenamefont {{Prince}}, \citenamefont {{Caiazzo}}, \citenamefont {{Vanderbosch}}, \citenamefont {{van Roestel}}, \citenamefont {{Das}}, \citenamefont {{Qin}}, \citenamefont {{Kasliwal}}, \citenamefont {{Wold}}, \citenamefont {{Groom}}, \citenamefont {{Reiley}},\ and\ \citenamefont {{Riddle}}}]{2024rodriguez_survey}%
  \BibitemOpen
  \bibfield  {author} {\bibinfo {author} {\bibfnamefont {A.~C.}\ \bibnamefont {{Rodriguez}}}, \bibinfo {author} {\bibfnamefont {K.}~\bibnamefont {{El-Badry}}}, \bibinfo {author} {\bibfnamefont {V.}~\bibnamefont {{Suleimanov}}}, \bibinfo {author} {\bibfnamefont {A.~F.}\ \bibnamefont {{Pala}}}, \bibinfo {author} {\bibfnamefont {S.~R.}\ \bibnamefont {{Kulkarni}}}, \bibinfo {author} {\bibfnamefont {B.}~\bibnamefont {{Gaensicke}}}, \bibinfo {author} {\bibfnamefont {K.}~\bibnamefont {{Mori}}}, \bibinfo {author} {\bibfnamefont {R.~M.}\ \bibnamefont {{Rich}}}, \bibinfo {author} {\bibfnamefont {A.}~\bibnamefont {{Sarkar}}}, \bibinfo {author} {\bibfnamefont {T.}~\bibnamefont {{Bao}}}, \bibinfo {author} {\bibfnamefont {R.}~\bibnamefont {{Lopes de Oliveira}}}, \bibinfo {author} {\bibfnamefont {G.}~\bibnamefont {{Ramsay}}}, \bibinfo {author} {\bibfnamefont {P.}~\bibnamefont {{Szkody}}}, \bibinfo {author} {\bibfnamefont {M.}~\bibnamefont {{Graham}}}, \bibinfo {author} {\bibfnamefont {T.~A.}\ \bibnamefont {{Prince}}},
  \bibinfo {author} {\bibfnamefont {I.}~\bibnamefont {{Caiazzo}}}, \bibinfo {author} {\bibfnamefont {Z.~P.}\ \bibnamefont {{Vanderbosch}}}, \bibinfo {author} {\bibfnamefont {J.}~\bibnamefont {{van Roestel}}}, \bibinfo {author} {\bibfnamefont {K.~K.}\ \bibnamefont {{Das}}}, \bibinfo {author} {\bibfnamefont {Y.-J.}\ \bibnamefont {{Qin}}}, \bibinfo {author} {\bibfnamefont {M.~M.}\ \bibnamefont {{Kasliwal}}}, \bibinfo {author} {\bibfnamefont {A.}~\bibnamefont {{Wold}}}, \bibinfo {author} {\bibfnamefont {S.~L.}\ \bibnamefont {{Groom}}}, \bibinfo {author} {\bibfnamefont {D.}~\bibnamefont {{Reiley}}}, \ and\ \bibinfo {author} {\bibfnamefont {R.}~\bibnamefont {{Riddle}}},\ }\href {\doibase 10.1088/1538-3873/ada185} {\bibfield  {journal} {\bibinfo  {journal} {\pasp}\ }\textbf {\bibinfo {volume} {137}},\ \bibinfo {eid} {014201} (\bibinfo {year} {2025})},\ \Eprint {http://arxiv.org/abs/2408.16053} {arXiv:2408.16053 [astro-ph.HE]} \BibitemShut {NoStop}%
\bibitem [{\citenamefont {{Zorotovic}}\ \emph {et~al.}(2011)\citenamefont {{Zorotovic}}, \citenamefont {{Schreiber}},\ and\ \citenamefont {{G{\"a}nsicke}}}]{2011zorotovic}%
  \BibitemOpen
  \bibfield  {author} {\bibinfo {author} {\bibfnamefont {M.}~\bibnamefont {{Zorotovic}}}, \bibinfo {author} {\bibfnamefont {M.~R.}\ \bibnamefont {{Schreiber}}}, \ and\ \bibinfo {author} {\bibfnamefont {B.~T.}\ \bibnamefont {{G{\"a}nsicke}}},\ }\href {\doibase 10.1051/0004-6361/201116626} {\bibfield  {journal} {\bibinfo  {journal} {\aap}\ }\textbf {\bibinfo {volume} {536}},\ \bibinfo {eid} {A42} (\bibinfo {year} {2011})},\ \Eprint {http://arxiv.org/abs/1108.4600} {arXiv:1108.4600 [astro-ph.SR]} \BibitemShut {NoStop}%
\bibitem [{\citenamefont {{Schreiber}}\ \emph {et~al.}(2021)\citenamefont {{Schreiber}}, \citenamefont {{Belloni}}, \citenamefont {{G{\"a}nsicke}}, \citenamefont {{Parsons}},\ and\ \citenamefont {{Zorotovic}}}]{2021schreiber}%
  \BibitemOpen
  \bibfield  {author} {\bibinfo {author} {\bibfnamefont {M.~R.}\ \bibnamefont {{Schreiber}}}, \bibinfo {author} {\bibfnamefont {D.}~\bibnamefont {{Belloni}}}, \bibinfo {author} {\bibfnamefont {B.~T.}\ \bibnamefont {{G{\"a}nsicke}}}, \bibinfo {author} {\bibfnamefont {S.~G.}\ \bibnamefont {{Parsons}}}, \ and\ \bibinfo {author} {\bibfnamefont {M.}~\bibnamefont {{Zorotovic}}},\ }\href {\doibase 10.1038/s41550-021-01346-8} {\bibfield  {journal} {\bibinfo  {journal} {Nature Astronomy}\ }\textbf {\bibinfo {volume} {5}},\ \bibinfo {pages} {648} (\bibinfo {year} {2021})},\ \Eprint {http://arxiv.org/abs/2104.14607} {arXiv:2104.14607 [astro-ph.SR]} \BibitemShut {NoStop}%
\bibitem [{\citenamefont {{Isern}}\ \emph {et~al.}(2017)\citenamefont {{Isern}}, \citenamefont {{Garc{\'\i}a-Berro}}, \citenamefont {{K{\"u}lebi}},\ and\ \citenamefont {{Lor{\'e}n-Aguilar}}}]{2017isern}%
  \BibitemOpen
  \bibfield  {author} {\bibinfo {author} {\bibfnamefont {J.}~\bibnamefont {{Isern}}}, \bibinfo {author} {\bibfnamefont {E.}~\bibnamefont {{Garc{\'\i}a-Berro}}}, \bibinfo {author} {\bibfnamefont {B.}~\bibnamefont {{K{\"u}lebi}}}, \ and\ \bibinfo {author} {\bibfnamefont {P.}~\bibnamefont {{Lor{\'e}n-Aguilar}}},\ }\href {\doibase 10.3847/2041-8213/aa5eae} {\bibfield  {journal} {\bibinfo  {journal} {\apjl}\ }\textbf {\bibinfo {volume} {836}},\ \bibinfo {eid} {L28} (\bibinfo {year} {2017})},\ \Eprint {http://arxiv.org/abs/1702.01813} {arXiv:1702.01813 [astro-ph.SR]} \BibitemShut {NoStop}%
\bibitem [{\citenamefont {{Ginzburg}}\ \emph {et~al.}(2022)\citenamefont {{Ginzburg}}, \citenamefont {{Fuller}}, \citenamefont {{Kawka}},\ and\ \citenamefont {{Caiazzo}}}]{2022ginzburg}%
  \BibitemOpen
  \bibfield  {author} {\bibinfo {author} {\bibfnamefont {S.}~\bibnamefont {{Ginzburg}}}, \bibinfo {author} {\bibfnamefont {J.}~\bibnamefont {{Fuller}}}, \bibinfo {author} {\bibfnamefont {A.}~\bibnamefont {{Kawka}}}, \ and\ \bibinfo {author} {\bibfnamefont {I.}~\bibnamefont {{Caiazzo}}},\ }\href {\doibase 10.1093/mnras/stac1363} {\bibfield  {journal} {\bibinfo  {journal} {\mnras}\ }\textbf {\bibinfo {volume} {514}},\ \bibinfo {pages} {4111} (\bibinfo {year} {2022})},\ \Eprint {http://arxiv.org/abs/2202.12902} {arXiv:2202.12902 [astro-ph.SR]} \BibitemShut {NoStop}%
\bibitem [{\citenamefont {{Rose}}\ \emph {et~al.}(2026)\citenamefont {{Rose}}, \citenamefont {{Pritchard}}, \citenamefont {{Murphy}}, \citenamefont {{Driessen}}, \citenamefont {{Kaplan}}, \citenamefont {{Caleb}}, \citenamefont {{Wang}}, \citenamefont {{Zic}}, \citenamefont {{Andreoni}}, \citenamefont {{Carney}}, \citenamefont {{Barlow}}, \citenamefont {{Dobie}}, \citenamefont {{Gu}}, \citenamefont {{Heald}}, \citenamefont {{Huber}}, \citenamefont {{Lenc}}, \citenamefont {{Leung}}, \citenamefont {{Lu}}, \citenamefont {{Momose}}, \citenamefont {{Pedersen}}, \citenamefont {{Qu}}, \citenamefont {{Rea}}, \citenamefont {{de Ruiter}}, \citenamefont {{Shaji}}, \citenamefont {{Sivakoff}}, \citenamefont {{Thomson}}, \citenamefont {{Wang}}, \citenamefont {{Yang}},\ and\ \citenamefont {{Zahedy}}}]{2026rose}%
  \BibitemOpen
  \bibfield  {author} {\bibinfo {author} {\bibfnamefont {K.}~\bibnamefont {{Rose}}}, \bibinfo {author} {\bibfnamefont {J.}~\bibnamefont {{Pritchard}}}, \bibinfo {author} {\bibfnamefont {T.}~\bibnamefont {{Murphy}}}, \bibinfo {author} {\bibfnamefont {L.~N.}\ \bibnamefont {{Driessen}}}, \bibinfo {author} {\bibfnamefont {D.~L.}\ \bibnamefont {{Kaplan}}}, \bibinfo {author} {\bibfnamefont {M.}~\bibnamefont {{Caleb}}}, \bibinfo {author} {\bibfnamefont {Z.}~\bibnamefont {{Wang}}}, \bibinfo {author} {\bibfnamefont {A.}~\bibnamefont {{Zic}}}, \bibinfo {author} {\bibfnamefont {I.}~\bibnamefont {{Andreoni}}}, \bibinfo {author} {\bibfnamefont {J.}~\bibnamefont {{Carney}}}, \bibinfo {author} {\bibfnamefont {B.~N.}\ \bibnamefont {{Barlow}}}, \bibinfo {author} {\bibfnamefont {D.}~\bibnamefont {{Dobie}}}, \bibinfo {author} {\bibfnamefont {M.}~\bibnamefont {{Gu}}}, \bibinfo {author} {\bibfnamefont {G.}~\bibnamefont {{Heald}}}, \bibinfo {author} {\bibfnamefont {D.}~\bibnamefont {{Huber}}}, \bibinfo {author} {\bibfnamefont
  {E.}~\bibnamefont {{Lenc}}}, \bibinfo {author} {\bibfnamefont {J.~K.}\ \bibnamefont {{Leung}}}, \bibinfo {author} {\bibfnamefont {W.}~\bibnamefont {{Lu}}}, \bibinfo {author} {\bibfnamefont {R.}~\bibnamefont {{Momose}}}, \bibinfo {author} {\bibfnamefont {M.~G.}\ \bibnamefont {{Pedersen}}}, \bibinfo {author} {\bibfnamefont {Y.}~\bibnamefont {{Qu}}}, \bibinfo {author} {\bibfnamefont {N.}~\bibnamefont {{Rea}}}, \bibinfo {author} {\bibfnamefont {I.}~\bibnamefont {{de Ruiter}}}, \bibinfo {author} {\bibfnamefont {K.}~\bibnamefont {{Shaji}}}, \bibinfo {author} {\bibfnamefont {G.~R.}\ \bibnamefont {{Sivakoff}}}, \bibinfo {author} {\bibfnamefont {A.~J.~M.}\ \bibnamefont {{Thomson}}}, \bibinfo {author} {\bibfnamefont {Y.~L.}\ \bibnamefont {{Wang}}}, \bibinfo {author} {\bibfnamefont {G.~J.}\ \bibnamefont {{Yang}}}, \ and\ \bibinfo {author} {\bibfnamefont {F.}~\bibnamefont {{Zahedy}}},\ }\href {\doibase 10.1038/s41550-026-02882-x} {\bibfield  {journal} {\bibinfo  {journal} {Nature Astronomy}\ } (\bibinfo {year}
  {2026}),\ 10.1038/s41550-026-02882-x}\BibitemShut {NoStop}%
\bibitem [{\citenamefont {{Hyman}}\ \emph {et~al.}(2005)\citenamefont {{Hyman}}, \citenamefont {{Lazio}}, \citenamefont {{Kassim}}, \citenamefont {{Ray}}, \citenamefont {{Markwardt}},\ and\ \citenamefont {{Yusef-Zadeh}}}]{2005hyman}%
  \BibitemOpen
  \bibfield  {author} {\bibinfo {author} {\bibfnamefont {S.~D.}\ \bibnamefont {{Hyman}}}, \bibinfo {author} {\bibfnamefont {T.~J.~W.}\ \bibnamefont {{Lazio}}}, \bibinfo {author} {\bibfnamefont {N.~E.}\ \bibnamefont {{Kassim}}}, \bibinfo {author} {\bibfnamefont {P.~S.}\ \bibnamefont {{Ray}}}, \bibinfo {author} {\bibfnamefont {C.~B.}\ \bibnamefont {{Markwardt}}}, \ and\ \bibinfo {author} {\bibfnamefont {F.}~\bibnamefont {{Yusef-Zadeh}}},\ }\href {\doibase 10.1038/nature03400} {\bibfield  {journal} {\bibinfo  {journal} {\nat}\ }\textbf {\bibinfo {volume} {434}},\ \bibinfo {pages} {50} (\bibinfo {year} {2005})},\ \Eprint {http://arxiv.org/abs/astro-ph/0503052} {arXiv:astro-ph/0503052 [astro-ph]} \BibitemShut {NoStop}%
\bibitem [{\citenamefont {{Hurley-Walker}}\ \emph {et~al.}(2022)\citenamefont {{Hurley-Walker}}, \citenamefont {{Zhang}}, \citenamefont {{Bahramian}}, \citenamefont {{McSweeney}}, \citenamefont {{O'Doherty}}, \citenamefont {{Hancock}}, \citenamefont {{Morgan}}, \citenamefont {{Anderson}}, \citenamefont {{Heald}},\ and\ \citenamefont {{Galvin}}}]{2022hurley-walker}%
  \BibitemOpen
  \bibfield  {author} {\bibinfo {author} {\bibfnamefont {N.}~\bibnamefont {{Hurley-Walker}}}, \bibinfo {author} {\bibfnamefont {X.}~\bibnamefont {{Zhang}}}, \bibinfo {author} {\bibfnamefont {A.}~\bibnamefont {{Bahramian}}}, \bibinfo {author} {\bibfnamefont {S.~J.}\ \bibnamefont {{McSweeney}}}, \bibinfo {author} {\bibfnamefont {T.~N.}\ \bibnamefont {{O'Doherty}}}, \bibinfo {author} {\bibfnamefont {P.~J.}\ \bibnamefont {{Hancock}}}, \bibinfo {author} {\bibfnamefont {J.~S.}\ \bibnamefont {{Morgan}}}, \bibinfo {author} {\bibfnamefont {G.~E.}\ \bibnamefont {{Anderson}}}, \bibinfo {author} {\bibfnamefont {G.~H.}\ \bibnamefont {{Heald}}}, \ and\ \bibinfo {author} {\bibfnamefont {T.~J.}\ \bibnamefont {{Galvin}}},\ }\href {\doibase 10.1038/s41586-021-04272-x} {\bibfield  {journal} {\bibinfo  {journal} {\nat}\ }\textbf {\bibinfo {volume} {601}},\ \bibinfo {pages} {526} (\bibinfo {year} {2022})}\BibitemShut {NoStop}%
\bibitem [{\citenamefont {{Hurley-Walker}}\ \emph {et~al.}(2023)\citenamefont {{Hurley-Walker}}, \citenamefont {{Rea}}, \citenamefont {{McSweeney}}, \citenamefont {{Meyers}}, \citenamefont {{Lenc}}, \citenamefont {{Heywood}}, \citenamefont {{Hyman}}, \citenamefont {{Men}}, \citenamefont {{Clarke}}, \citenamefont {{Coti Zelati}}, \citenamefont {{Price}}, \citenamefont {{Horv{\'a}th}}, \citenamefont {{Galvin}}, \citenamefont {{Anderson}}, \citenamefont {{Bahramian}}, \citenamefont {{Barr}}, \citenamefont {{Bhat}}, \citenamefont {{Caleb}}, \citenamefont {{Dall'Ora}}, \citenamefont {{de Martino}}, \citenamefont {{Giacintucci}}, \citenamefont {{Morgan}}, \citenamefont {{Rajwade}}, \citenamefont {{Stappers}},\ and\ \citenamefont {{Williams}}}]{2023hurley-walker}%
  \BibitemOpen
  \bibfield  {author} {\bibinfo {author} {\bibfnamefont {N.}~\bibnamefont {{Hurley-Walker}}}, \bibinfo {author} {\bibfnamefont {N.}~\bibnamefont {{Rea}}}, \bibinfo {author} {\bibfnamefont {S.~J.}\ \bibnamefont {{McSweeney}}}, \bibinfo {author} {\bibfnamefont {B.~W.}\ \bibnamefont {{Meyers}}}, \bibinfo {author} {\bibfnamefont {E.}~\bibnamefont {{Lenc}}}, \bibinfo {author} {\bibfnamefont {I.}~\bibnamefont {{Heywood}}}, \bibinfo {author} {\bibfnamefont {S.~D.}\ \bibnamefont {{Hyman}}}, \bibinfo {author} {\bibfnamefont {Y.~P.}\ \bibnamefont {{Men}}}, \bibinfo {author} {\bibfnamefont {T.~E.}\ \bibnamefont {{Clarke}}}, \bibinfo {author} {\bibfnamefont {F.}~\bibnamefont {{Coti Zelati}}}, \bibinfo {author} {\bibfnamefont {D.~C.}\ \bibnamefont {{Price}}}, \bibinfo {author} {\bibfnamefont {C.}~\bibnamefont {{Horv{\'a}th}}}, \bibinfo {author} {\bibfnamefont {T.~J.}\ \bibnamefont {{Galvin}}}, \bibinfo {author} {\bibfnamefont {G.~E.}\ \bibnamefont {{Anderson}}}, \bibinfo {author} {\bibfnamefont {A.}~\bibnamefont
  {{Bahramian}}}, \bibinfo {author} {\bibfnamefont {E.~D.}\ \bibnamefont {{Barr}}}, \bibinfo {author} {\bibfnamefont {N.~D.~R.}\ \bibnamefont {{Bhat}}}, \bibinfo {author} {\bibfnamefont {M.}~\bibnamefont {{Caleb}}}, \bibinfo {author} {\bibfnamefont {M.}~\bibnamefont {{Dall'Ora}}}, \bibinfo {author} {\bibfnamefont {D.}~\bibnamefont {{de Martino}}}, \bibinfo {author} {\bibfnamefont {S.}~\bibnamefont {{Giacintucci}}}, \bibinfo {author} {\bibfnamefont {J.~S.}\ \bibnamefont {{Morgan}}}, \bibinfo {author} {\bibfnamefont {K.~M.}\ \bibnamefont {{Rajwade}}}, \bibinfo {author} {\bibfnamefont {B.}~\bibnamefont {{Stappers}}}, \ and\ \bibinfo {author} {\bibfnamefont {A.}~\bibnamefont {{Williams}}},\ }\href {\doibase 10.1038/s41586-023-06202-5} {\bibfield  {journal} {\bibinfo  {journal} {\nat}\ }\textbf {\bibinfo {volume} {619}},\ \bibinfo {pages} {487} (\bibinfo {year} {2023})}\BibitemShut {NoStop}%
\bibitem [{\citenamefont {{Caleb}}\ \emph {et~al.}(2024)\citenamefont {{Caleb}}, \citenamefont {{Lenc}}, \citenamefont {{Kaplan}}, \citenamefont {{Murphy}}, \citenamefont {{Men}}, \citenamefont {{Shannon}}, \citenamefont {{Ferrario}}, \citenamefont {{Rajwade}}, \citenamefont {{Clarke}}, \citenamefont {{Giacintucci}}, \citenamefont {{Hurley-Walker}}, \citenamefont {{Hyman}}, \citenamefont {{Lower}}, \citenamefont {{McSweeney}}, \citenamefont {{Ravi}}, \citenamefont {{Barr}}, \citenamefont {{Buchner}}, \citenamefont {{Flynn}}, \citenamefont {{Hessels}}, \citenamefont {{Kramer}}, \citenamefont {{Pritchard}},\ and\ \citenamefont {{Stappers}}}]{2024caleb}%
  \BibitemOpen
  \bibfield  {author} {\bibinfo {author} {\bibfnamefont {M.}~\bibnamefont {{Caleb}}}, \bibinfo {author} {\bibfnamefont {E.}~\bibnamefont {{Lenc}}}, \bibinfo {author} {\bibfnamefont {D.~L.}\ \bibnamefont {{Kaplan}}}, \bibinfo {author} {\bibfnamefont {T.}~\bibnamefont {{Murphy}}}, \bibinfo {author} {\bibfnamefont {Y.~P.}\ \bibnamefont {{Men}}}, \bibinfo {author} {\bibfnamefont {R.~M.}\ \bibnamefont {{Shannon}}}, \bibinfo {author} {\bibfnamefont {L.}~\bibnamefont {{Ferrario}}}, \bibinfo {author} {\bibfnamefont {K.~M.}\ \bibnamefont {{Rajwade}}}, \bibinfo {author} {\bibfnamefont {T.~E.}\ \bibnamefont {{Clarke}}}, \bibinfo {author} {\bibfnamefont {S.}~\bibnamefont {{Giacintucci}}}, \bibinfo {author} {\bibfnamefont {N.}~\bibnamefont {{Hurley-Walker}}}, \bibinfo {author} {\bibfnamefont {S.~D.}\ \bibnamefont {{Hyman}}}, \bibinfo {author} {\bibfnamefont {M.~E.}\ \bibnamefont {{Lower}}}, \bibinfo {author} {\bibfnamefont {S.}~\bibnamefont {{McSweeney}}}, \bibinfo {author} {\bibfnamefont {V.}~\bibnamefont {{Ravi}}},
  \bibinfo {author} {\bibfnamefont {E.~D.}\ \bibnamefont {{Barr}}}, \bibinfo {author} {\bibfnamefont {S.}~\bibnamefont {{Buchner}}}, \bibinfo {author} {\bibfnamefont {C.~M.~L.}\ \bibnamefont {{Flynn}}}, \bibinfo {author} {\bibfnamefont {J.~W.~T.}\ \bibnamefont {{Hessels}}}, \bibinfo {author} {\bibfnamefont {M.}~\bibnamefont {{Kramer}}}, \bibinfo {author} {\bibfnamefont {J.}~\bibnamefont {{Pritchard}}}, \ and\ \bibinfo {author} {\bibfnamefont {B.~W.}\ \bibnamefont {{Stappers}}},\ }\href {\doibase 10.1038/s41550-024-02277-w} {\bibfield  {journal} {\bibinfo  {journal} {Nature Astronomy}\ }\textbf {\bibinfo {volume} {8}},\ \bibinfo {pages} {1159} (\bibinfo {year} {2024})},\ \Eprint {http://arxiv.org/abs/2407.12266} {arXiv:2407.12266 [astro-ph.HE]} \BibitemShut {NoStop}%
\bibitem [{\citenamefont {{Dong}}\ \emph {et~al.}(2024)\citenamefont {{Dong}}, \citenamefont {{Clarke}}, \citenamefont {{Curtin}}, \citenamefont {{Kumar}}, \citenamefont {{Stairs}}, \citenamefont {{Chatterjee}}, \citenamefont {{Cook}}, \citenamefont {{Fonseca}}, \citenamefont {{Gaensler}}, \citenamefont {{Hessels}}, \citenamefont {{Kaspi}}, \citenamefont {{Lazda}}, \citenamefont {{Masui}}, \citenamefont {{McKee}}, \citenamefont {{Meyers}}, \citenamefont {{Pearlman}}, \citenamefont {{Ransom}}, \citenamefont {{Scholz}}, \citenamefont {{Shin}}, \citenamefont {{Smith}},\ and\ \citenamefont {{Tan}}}]{2024dong}%
  \BibitemOpen
  \bibfield  {author} {\bibinfo {author} {\bibfnamefont {F.~A.}\ \bibnamefont {{Dong}}}, \bibinfo {author} {\bibfnamefont {T.}~\bibnamefont {{Clarke}}}, \bibinfo {author} {\bibfnamefont {A.~P.}\ \bibnamefont {{Curtin}}}, \bibinfo {author} {\bibfnamefont {A.}~\bibnamefont {{Kumar}}}, \bibinfo {author} {\bibfnamefont {I.}~\bibnamefont {{Stairs}}}, \bibinfo {author} {\bibfnamefont {S.}~\bibnamefont {{Chatterjee}}}, \bibinfo {author} {\bibfnamefont {A.~M.}\ \bibnamefont {{Cook}}}, \bibinfo {author} {\bibfnamefont {E.}~\bibnamefont {{Fonseca}}}, \bibinfo {author} {\bibfnamefont {B.~M.}\ \bibnamefont {{Gaensler}}}, \bibinfo {author} {\bibfnamefont {J.~W.~T.}\ \bibnamefont {{Hessels}}}, \bibinfo {author} {\bibfnamefont {V.~M.}\ \bibnamefont {{Kaspi}}}, \bibinfo {author} {\bibfnamefont {M.}~\bibnamefont {{Lazda}}}, \bibinfo {author} {\bibfnamefont {K.~W.}\ \bibnamefont {{Masui}}}, \bibinfo {author} {\bibfnamefont {J.~W.}\ \bibnamefont {{McKee}}}, \bibinfo {author} {\bibfnamefont {B.~W.}\ \bibnamefont {{Meyers}}},
  \bibinfo {author} {\bibfnamefont {A.~B.}\ \bibnamefont {{Pearlman}}}, \bibinfo {author} {\bibfnamefont {S.~M.}\ \bibnamefont {{Ransom}}}, \bibinfo {author} {\bibfnamefont {P.}~\bibnamefont {{Scholz}}}, \bibinfo {author} {\bibfnamefont {K.}~\bibnamefont {{Shin}}}, \bibinfo {author} {\bibfnamefont {K.~M.}\ \bibnamefont {{Smith}}}, \ and\ \bibinfo {author} {\bibfnamefont {C.~M.}\ \bibnamefont {{Tan}}},\ }\href {\doibase 10.48550/arXiv.2407.07480} {\bibfield  {journal} {\bibinfo  {journal} {arXiv e-prints}\ ,\ \bibinfo {eid} {arXiv:2407.07480}} (\bibinfo {year} {2024})},\ \Eprint {http://arxiv.org/abs/2407.07480} {arXiv:2407.07480 [astro-ph.HE]} \BibitemShut {NoStop}%
\bibitem [{\citenamefont {{Wang}}\ \emph {et~al.}(2024)\citenamefont {{Wang}}, \citenamefont {{Rea}}, \citenamefont {{Bao}}, \citenamefont {{Kaplan}}, \citenamefont {{Lenc}}, \citenamefont {{Wadiasingh}}, \citenamefont {{Hare}}, \citenamefont {{Zic}}, \citenamefont {{Anumarlapudi}}, \citenamefont {{Bera}}, \citenamefont {{Beniamini}}, \citenamefont {{Cooper}}, \citenamefont {{Clarke}}, \citenamefont {{Deller}}, \citenamefont {{Dawson}}, \citenamefont {{Glowacki}}, \citenamefont {{Hurley-Walker}}, \citenamefont {{McSweeney}}, \citenamefont {{Polisensky}}, \citenamefont {{Peters}}, \citenamefont {{Younes}}, \citenamefont {{Bannister}}, \citenamefont {{Caleb}}, \citenamefont {{Dage}}, \citenamefont {{James}}, \citenamefont {{Kasliwal}}, \citenamefont {{Karambelkar}}, \citenamefont {{Lower}}, \citenamefont {{Mori}}, \citenamefont {{Ocker}}, \citenamefont {{P{\'e}rez-Torres}}, \citenamefont {{Qiu}}, \citenamefont {{Rose}}, \citenamefont {{Shannon}}, \citenamefont {{Taub}}, \citenamefont {{Wang}}, \citenamefont
  {{Wang}}, \citenamefont {{Zhao}}, \citenamefont {{Bhat}}, \citenamefont {{Dobie}}, \citenamefont {{Driessen}}, \citenamefont {{Murphy}}, \citenamefont {{Jaini}}, \citenamefont {{Deng}}, \citenamefont {{Jahns-Schindler}}, \citenamefont {{Lee}}, \citenamefont {{Pritchard}}, \citenamefont {{Tuthill}},\ and\ \citenamefont {{Thyagarajan}}}]{2024wang}%
  \BibitemOpen
  \bibfield  {author} {\bibinfo {author} {\bibfnamefont {Z.}~\bibnamefont {{Wang}}}, \bibinfo {author} {\bibfnamefont {N.}~\bibnamefont {{Rea}}}, \bibinfo {author} {\bibfnamefont {T.}~\bibnamefont {{Bao}}}, \bibinfo {author} {\bibfnamefont {D.~L.}\ \bibnamefont {{Kaplan}}}, \bibinfo {author} {\bibfnamefont {E.}~\bibnamefont {{Lenc}}}, \bibinfo {author} {\bibfnamefont {Z.}~\bibnamefont {{Wadiasingh}}}, \bibinfo {author} {\bibfnamefont {J.}~\bibnamefont {{Hare}}}, \bibinfo {author} {\bibfnamefont {A.}~\bibnamefont {{Zic}}}, \bibinfo {author} {\bibfnamefont {A.}~\bibnamefont {{Anumarlapudi}}}, \bibinfo {author} {\bibfnamefont {A.}~\bibnamefont {{Bera}}}, \bibinfo {author} {\bibfnamefont {P.}~\bibnamefont {{Beniamini}}}, \bibinfo {author} {\bibfnamefont {A.~J.}\ \bibnamefont {{Cooper}}}, \bibinfo {author} {\bibfnamefont {T.~E.}\ \bibnamefont {{Clarke}}}, \bibinfo {author} {\bibfnamefont {A.~T.}\ \bibnamefont {{Deller}}}, \bibinfo {author} {\bibfnamefont {J.~R.}\ \bibnamefont {{Dawson}}}, \bibinfo {author}
  {\bibfnamefont {M.}~\bibnamefont {{Glowacki}}}, \bibinfo {author} {\bibfnamefont {N.}~\bibnamefont {{Hurley-Walker}}}, \bibinfo {author} {\bibfnamefont {S.~J.}\ \bibnamefont {{McSweeney}}}, \bibinfo {author} {\bibfnamefont {E.~J.}\ \bibnamefont {{Polisensky}}}, \bibinfo {author} {\bibfnamefont {W.~M.}\ \bibnamefont {{Peters}}}, \bibinfo {author} {\bibfnamefont {G.}~\bibnamefont {{Younes}}}, \bibinfo {author} {\bibfnamefont {K.~W.}\ \bibnamefont {{Bannister}}}, \bibinfo {author} {\bibfnamefont {M.}~\bibnamefont {{Caleb}}}, \bibinfo {author} {\bibfnamefont {K.~C.}\ \bibnamefont {{Dage}}}, \bibinfo {author} {\bibfnamefont {C.~W.}\ \bibnamefont {{James}}}, \bibinfo {author} {\bibfnamefont {M.~M.}\ \bibnamefont {{Kasliwal}}}, \bibinfo {author} {\bibfnamefont {V.}~\bibnamefont {{Karambelkar}}}, \bibinfo {author} {\bibfnamefont {M.~E.}\ \bibnamefont {{Lower}}}, \bibinfo {author} {\bibfnamefont {K.}~\bibnamefont {{Mori}}}, \bibinfo {author} {\bibfnamefont {S.~K.}\ \bibnamefont {{Ocker}}}, \bibinfo {author}
  {\bibfnamefont {M.}~\bibnamefont {{P{\'e}rez-Torres}}}, \bibinfo {author} {\bibfnamefont {H.}~\bibnamefont {{Qiu}}}, \bibinfo {author} {\bibfnamefont {K.}~\bibnamefont {{Rose}}}, \bibinfo {author} {\bibfnamefont {R.~M.}\ \bibnamefont {{Shannon}}}, \bibinfo {author} {\bibfnamefont {R.}~\bibnamefont {{Taub}}}, \bibinfo {author} {\bibfnamefont {F.}~\bibnamefont {{Wang}}}, \bibinfo {author} {\bibfnamefont {Y.}~\bibnamefont {{Wang}}}, \bibinfo {author} {\bibfnamefont {Z.}~\bibnamefont {{Zhao}}}, \bibinfo {author} {\bibfnamefont {N.~D.~R.}\ \bibnamefont {{Bhat}}}, \bibinfo {author} {\bibfnamefont {D.}~\bibnamefont {{Dobie}}}, \bibinfo {author} {\bibfnamefont {L.~N.}\ \bibnamefont {{Driessen}}}, \bibinfo {author} {\bibfnamefont {T.}~\bibnamefont {{Murphy}}}, \bibinfo {author} {\bibfnamefont {A.}~\bibnamefont {{Jaini}}}, \bibinfo {author} {\bibfnamefont {X.}~\bibnamefont {{Deng}}}, \bibinfo {author} {\bibfnamefont {J.~N.}\ \bibnamefont {{Jahns-Schindler}}}, \bibinfo {author} {\bibfnamefont {Y.~W.~J.}\ \bibnamefont
  {{Lee}}}, \bibinfo {author} {\bibfnamefont {J.}~\bibnamefont {{Pritchard}}}, \bibinfo {author} {\bibfnamefont {J.}~\bibnamefont {{Tuthill}}}, \ and\ \bibinfo {author} {\bibfnamefont {N.}~\bibnamefont {{Thyagarajan}}},\ }\href {\doibase 10.48550/arXiv.2411.16606} {\bibfield  {journal} {\bibinfo  {journal} {arXiv e-prints}\ ,\ \bibinfo {eid} {arXiv:2411.16606}} (\bibinfo {year} {2024})},\ \Eprint {http://arxiv.org/abs/2411.16606} {arXiv:2411.16606 [astro-ph.HE]} \BibitemShut {NoStop}%
\bibitem [{\citenamefont {{Li}}\ \emph {et~al.}(2024)\citenamefont {{Li}}, \citenamefont {{Yuan}}, \citenamefont {{Wu}}, \citenamefont {{Yan}}, \citenamefont {{Lv}}, \citenamefont {{Tsai}}, \citenamefont {{Wang}}, \citenamefont {{Zhu}}, \citenamefont {{Deng}}, \citenamefont {{Lan}}, \citenamefont {{Xu}}, \citenamefont {{Chen}}, \citenamefont {{Meng}}, \citenamefont {{Li}}, \citenamefont {{Li}}, \citenamefont {{Zhou}}, \citenamefont {{Yang}}, \citenamefont {{Xue}}, \citenamefont {{Lu}}, \citenamefont {{Miao}}, \citenamefont {{Wang}}, \citenamefont {{Niu}}, \citenamefont {{Fang}}, \citenamefont {{Fu}}, \citenamefont {{Feng}}, \citenamefont {{Zhang}}, \citenamefont {{Jiang}}, \citenamefont {{Miao}}, \citenamefont {{Chen}}, \citenamefont {{Sun}}, \citenamefont {{Yang}}, \citenamefont {{Deng}}, \citenamefont {{Dai}}, \citenamefont {{Chen}}, \citenamefont {{Yao}}, \citenamefont {{Liu}}, \citenamefont {{Li}}, \citenamefont {{Zhang}}, \citenamefont {{Yang}}, \citenamefont {{Zhou}}, \citenamefont {{Yiyizhou}},
  \citenamefont {{Zhang}}, \citenamefont {{Niu}}, \citenamefont {{Zhao}}, \citenamefont {{Zhang}}, \citenamefont {{Peng}}, \citenamefont {{Wu}},\ and\ \citenamefont {{Wang}}}]{2024li}%
  \BibitemOpen
  \bibfield  {author} {\bibinfo {author} {\bibfnamefont {D.}~\bibnamefont {{Li}}}, \bibinfo {author} {\bibfnamefont {M.}~\bibnamefont {{Yuan}}}, \bibinfo {author} {\bibfnamefont {L.}~\bibnamefont {{Wu}}}, \bibinfo {author} {\bibfnamefont {J.}~\bibnamefont {{Yan}}}, \bibinfo {author} {\bibfnamefont {X.}~\bibnamefont {{Lv}}}, \bibinfo {author} {\bibfnamefont {C.-W.}\ \bibnamefont {{Tsai}}}, \bibinfo {author} {\bibfnamefont {P.}~\bibnamefont {{Wang}}}, \bibinfo {author} {\bibfnamefont {W.}~\bibnamefont {{Zhu}}}, \bibinfo {author} {\bibfnamefont {L.}~\bibnamefont {{Deng}}}, \bibinfo {author} {\bibfnamefont {A.}~\bibnamefont {{Lan}}}, \bibinfo {author} {\bibfnamefont {R.}~\bibnamefont {{Xu}}}, \bibinfo {author} {\bibfnamefont {X.}~\bibnamefont {{Chen}}}, \bibinfo {author} {\bibfnamefont {L.}~\bibnamefont {{Meng}}}, \bibinfo {author} {\bibfnamefont {J.}~\bibnamefont {{Li}}}, \bibinfo {author} {\bibfnamefont {X.}~\bibnamefont {{Li}}}, \bibinfo {author} {\bibfnamefont {P.}~\bibnamefont {{Zhou}}}, \bibinfo {author}
  {\bibfnamefont {H.}~\bibnamefont {{Yang}}}, \bibinfo {author} {\bibfnamefont {M.}~\bibnamefont {{Xue}}}, \bibinfo {author} {\bibfnamefont {J.}~\bibnamefont {{Lu}}}, \bibinfo {author} {\bibfnamefont {C.}~\bibnamefont {{Miao}}}, \bibinfo {author} {\bibfnamefont {W.}~\bibnamefont {{Wang}}}, \bibinfo {author} {\bibfnamefont {J.}~\bibnamefont {{Niu}}}, \bibinfo {author} {\bibfnamefont {Z.}~\bibnamefont {{Fang}}}, \bibinfo {author} {\bibfnamefont {Q.}~\bibnamefont {{Fu}}}, \bibinfo {author} {\bibfnamefont {Y.}~\bibnamefont {{Feng}}}, \bibinfo {author} {\bibfnamefont {P.}~\bibnamefont {{Zhang}}}, \bibinfo {author} {\bibfnamefont {J.}~\bibnamefont {{Jiang}}}, \bibinfo {author} {\bibfnamefont {X.}~\bibnamefont {{Miao}}}, \bibinfo {author} {\bibfnamefont {Y.}~\bibnamefont {{Chen}}}, \bibinfo {author} {\bibfnamefont {L.}~\bibnamefont {{Sun}}}, \bibinfo {author} {\bibfnamefont {Y.}~\bibnamefont {{Yang}}}, \bibinfo {author} {\bibfnamefont {X.}~\bibnamefont {{Deng}}}, \bibinfo {author} {\bibfnamefont {S.}~\bibnamefont
  {{Dai}}}, \bibinfo {author} {\bibfnamefont {X.}~\bibnamefont {{Chen}}}, \bibinfo {author} {\bibfnamefont {J.}~\bibnamefont {{Yao}}}, \bibinfo {author} {\bibfnamefont {Y.}~\bibnamefont {{Liu}}}, \bibinfo {author} {\bibfnamefont {C.}~\bibnamefont {{Li}}}, \bibinfo {author} {\bibfnamefont {M.}~\bibnamefont {{Zhang}}}, \bibinfo {author} {\bibfnamefont {Y.}~\bibnamefont {{Yang}}}, \bibinfo {author} {\bibfnamefont {Y.}~\bibnamefont {{Zhou}}}, \bibinfo {author} {\bibnamefont {{Yiyizhou}}}, \bibinfo {author} {\bibfnamefont {Y.}~\bibnamefont {{Zhang}}}, \bibinfo {author} {\bibfnamefont {C.}~\bibnamefont {{Niu}}}, \bibinfo {author} {\bibfnamefont {R.}~\bibnamefont {{Zhao}}}, \bibinfo {author} {\bibfnamefont {L.}~\bibnamefont {{Zhang}}}, \bibinfo {author} {\bibfnamefont {B.}~\bibnamefont {{Peng}}}, \bibinfo {author} {\bibfnamefont {J.}~\bibnamefont {{Wu}}}, \ and\ \bibinfo {author} {\bibfnamefont {C.}~\bibnamefont {{Wang}}},\ }\href {\doibase 10.48550/arXiv.2411.15739} {\bibfield  {journal} {\bibinfo  {journal} {arXiv
  e-prints}\ ,\ \bibinfo {eid} {arXiv:2411.15739}} (\bibinfo {year} {2024})},\ \Eprint {http://arxiv.org/abs/2411.15739} {arXiv:2411.15739 [astro-ph.HE]} \BibitemShut {NoStop}%
\bibitem [{\citenamefont {Lee}\ \emph {et~al.}(2025)\citenamefont {Lee}, \citenamefont {Caleb}, \citenamefont {Murphy}, \citenamefont {Lenc}, \citenamefont {Kaplan}, \citenamefont {Ferrario}, \citenamefont {Wadiasingh}, \citenamefont {Anumarlapudi}, \citenamefont {Hurley-Walker}, \citenamefont {Karambelkar}, \citenamefont {Ocker}, \citenamefont {McSweeney}, \citenamefont {Qiu}, \citenamefont {Rajwade}, \citenamefont {Zic}, \citenamefont {Bannister}, \citenamefont {Bhat}, \citenamefont {Deller}, \citenamefont {Dobie}, \citenamefont {Driessen}, \citenamefont {Gendreau}, \citenamefont {Glowacki}, \citenamefont {Gupta}, \citenamefont {Jahns-Schindler}, \citenamefont {Jaini}, \citenamefont {James}, \citenamefont {Kasliwal}, \citenamefont {Lower}, \citenamefont {Shannon}, \citenamefont {Uttarkar}, \citenamefont {Wang},\ and\ \citenamefont {Wang}}]{2025lee}%
  \BibitemOpen
  \bibfield  {author} {\bibinfo {author} {\bibfnamefont {Y.~W.~J.}\ \bibnamefont {Lee}}, \bibinfo {author} {\bibfnamefont {M.}~\bibnamefont {Caleb}}, \bibinfo {author} {\bibfnamefont {T.}~\bibnamefont {Murphy}}, \bibinfo {author} {\bibfnamefont {E.}~\bibnamefont {Lenc}}, \bibinfo {author} {\bibfnamefont {D.~L.}\ \bibnamefont {Kaplan}}, \bibinfo {author} {\bibfnamefont {L.}~\bibnamefont {Ferrario}}, \bibinfo {author} {\bibfnamefont {Z.}~\bibnamefont {Wadiasingh}}, \bibinfo {author} {\bibfnamefont {A.}~\bibnamefont {Anumarlapudi}}, \bibinfo {author} {\bibfnamefont {N.}~\bibnamefont {Hurley-Walker}}, \bibinfo {author} {\bibfnamefont {V.}~\bibnamefont {Karambelkar}}, \bibinfo {author} {\bibfnamefont {S.~K.}\ \bibnamefont {Ocker}}, \bibinfo {author} {\bibfnamefont {S.}~\bibnamefont {McSweeney}}, \bibinfo {author} {\bibfnamefont {H.}~\bibnamefont {Qiu}}, \bibinfo {author} {\bibfnamefont {K.~M.}\ \bibnamefont {Rajwade}}, \bibinfo {author} {\bibfnamefont {A.}~\bibnamefont {Zic}}, \bibinfo {author} {\bibfnamefont
  {K.~W.}\ \bibnamefont {Bannister}}, \bibinfo {author} {\bibfnamefont {N.~D.~R.}\ \bibnamefont {Bhat}}, \bibinfo {author} {\bibfnamefont {A.}~\bibnamefont {Deller}}, \bibinfo {author} {\bibfnamefont {D.}~\bibnamefont {Dobie}}, \bibinfo {author} {\bibfnamefont {L.~N.}\ \bibnamefont {Driessen}}, \bibinfo {author} {\bibfnamefont {K.}~\bibnamefont {Gendreau}}, \bibinfo {author} {\bibfnamefont {M.}~\bibnamefont {Glowacki}}, \bibinfo {author} {\bibfnamefont {V.}~\bibnamefont {Gupta}}, \bibinfo {author} {\bibfnamefont {J.~N.}\ \bibnamefont {Jahns-Schindler}}, \bibinfo {author} {\bibfnamefont {A.}~\bibnamefont {Jaini}}, \bibinfo {author} {\bibfnamefont {C.~W.}\ \bibnamefont {James}}, \bibinfo {author} {\bibfnamefont {M.~M.}\ \bibnamefont {Kasliwal}}, \bibinfo {author} {\bibfnamefont {M.~E.}\ \bibnamefont {Lower}}, \bibinfo {author} {\bibfnamefont {R.~M.}\ \bibnamefont {Shannon}}, \bibinfo {author} {\bibfnamefont {P.~A.}\ \bibnamefont {Uttarkar}}, \bibinfo {author} {\bibfnamefont {Y.}~\bibnamefont {Wang}}, \ and\
  \bibinfo {author} {\bibfnamefont {Z.}~\bibnamefont {Wang}},\ }\href {\doibase 10.1038/s41550-024-02452-z} {\bibfield  {journal} {\bibinfo  {journal} {Nature Astronomy}\ } (\bibinfo {year} {2025}),\ 10.1038/s41550-024-02452-z}\BibitemShut {NoStop}%
\bibitem [{\citenamefont {{Dong}}\ \emph {et~al.}(2025)\citenamefont {{Dong}}, \citenamefont {{Shin}}, \citenamefont {{Law}}, \citenamefont {{Ng}}, \citenamefont {{Stairs}}, \citenamefont {{Bower}}, \citenamefont {{Cassity}}, \citenamefont {{Fonseca}}, \citenamefont {{Gaensler}}, \citenamefont {{Hessels}}, \citenamefont {{Kaspi}}, \citenamefont {{Kharel}}, \citenamefont {{Leung}}, \citenamefont {{Main}}, \citenamefont {{Masui}}, \citenamefont {{McKee}}, \citenamefont {{Meyers}}, \citenamefont {{Modilim}}, \citenamefont {{Pandhi}}, \citenamefont {{Pearlman}}, \citenamefont {{Ransom}}, \citenamefont {{Scholz}},\ and\ \citenamefont {{Smith}}}]{2025dong_1634}%
  \BibitemOpen
  \bibfield  {author} {\bibinfo {author} {\bibfnamefont {F.~A.}\ \bibnamefont {{Dong}}}, \bibinfo {author} {\bibfnamefont {K.}~\bibnamefont {{Shin}}}, \bibinfo {author} {\bibfnamefont {C.}~\bibnamefont {{Law}}}, \bibinfo {author} {\bibfnamefont {M.}~\bibnamefont {{Ng}}}, \bibinfo {author} {\bibfnamefont {I.}~\bibnamefont {{Stairs}}}, \bibinfo {author} {\bibfnamefont {G.}~\bibnamefont {{Bower}}}, \bibinfo {author} {\bibfnamefont {A.}~\bibnamefont {{Cassity}}}, \bibinfo {author} {\bibfnamefont {E.}~\bibnamefont {{Fonseca}}}, \bibinfo {author} {\bibfnamefont {B.~M.}\ \bibnamefont {{Gaensler}}}, \bibinfo {author} {\bibfnamefont {J.~W.~T.}\ \bibnamefont {{Hessels}}}, \bibinfo {author} {\bibfnamefont {V.~M.}\ \bibnamefont {{Kaspi}}}, \bibinfo {author} {\bibfnamefont {B.}~\bibnamefont {{Kharel}}}, \bibinfo {author} {\bibfnamefont {C.}~\bibnamefont {{Leung}}}, \bibinfo {author} {\bibfnamefont {R.~A.}\ \bibnamefont {{Main}}}, \bibinfo {author} {\bibfnamefont {K.~W.}\ \bibnamefont {{Masui}}}, \bibinfo {author}
  {\bibfnamefont {J.~W.}\ \bibnamefont {{McKee}}}, \bibinfo {author} {\bibfnamefont {B.~W.}\ \bibnamefont {{Meyers}}}, \bibinfo {author} {\bibfnamefont {O.}~\bibnamefont {{Modilim}}}, \bibinfo {author} {\bibfnamefont {A.}~\bibnamefont {{Pandhi}}}, \bibinfo {author} {\bibfnamefont {A.~B.}\ \bibnamefont {{Pearlman}}}, \bibinfo {author} {\bibfnamefont {S.~M.}\ \bibnamefont {{Ransom}}}, \bibinfo {author} {\bibfnamefont {P.}~\bibnamefont {{Scholz}}}, \ and\ \bibinfo {author} {\bibfnamefont {K.}~\bibnamefont {{Smith}}},\ }\href {\doibase 10.3847/2041-8213/adeaab} {\bibfield  {journal} {\bibinfo  {journal} {\apjl}\ }\textbf {\bibinfo {volume} {988}},\ \bibinfo {eid} {L29} (\bibinfo {year} {2025})},\ \Eprint {http://arxiv.org/abs/2507.05139} {arXiv:2507.05139 [astro-ph.HE]} \BibitemShut {NoStop}%
\bibitem [{\citenamefont {{Bloot}}\ \emph {et~al.}(2025)\citenamefont {{Bloot}}, \citenamefont {{Vedantham}}, \citenamefont {{Bassa}}, \citenamefont {{Callingham}}, \citenamefont {{Best}}, \citenamefont {{Liu}}, \citenamefont {{Magnier}}, \citenamefont {{Shimwell}},\ and\ \citenamefont {{Dupuy}}}]{2025bloot_1634}%
  \BibitemOpen
  \bibfield  {author} {\bibinfo {author} {\bibfnamefont {S.}~\bibnamefont {{Bloot}}}, \bibinfo {author} {\bibfnamefont {H.~K.}\ \bibnamefont {{Vedantham}}}, \bibinfo {author} {\bibfnamefont {C.~G.}\ \bibnamefont {{Bassa}}}, \bibinfo {author} {\bibfnamefont {J.~R.}\ \bibnamefont {{Callingham}}}, \bibinfo {author} {\bibfnamefont {W.~M.~J.}\ \bibnamefont {{Best}}}, \bibinfo {author} {\bibfnamefont {M.~C.}\ \bibnamefont {{Liu}}}, \bibinfo {author} {\bibfnamefont {E.~A.}\ \bibnamefont {{Magnier}}}, \bibinfo {author} {\bibfnamefont {T.~W.}\ \bibnamefont {{Shimwell}}}, \ and\ \bibinfo {author} {\bibfnamefont {T.~J.}\ \bibnamefont {{Dupuy}}},\ }\href {\doibase 10.1051/0004-6361/202555131} {\bibfield  {journal} {\bibinfo  {journal} {\aap}\ }\textbf {\bibinfo {volume} {699}},\ \bibinfo {eid} {A341} (\bibinfo {year} {2025})},\ \Eprint {http://arxiv.org/abs/2507.05078} {arXiv:2507.05078 [astro-ph.HE]} \BibitemShut {NoStop}%
\bibitem [{\citenamefont {{McSweeney}}\ \emph {et~al.}(2025)\citenamefont {{McSweeney}}, \citenamefont {{Hurley-Walker}}, \citenamefont {{Horv{\'a}th}}, \citenamefont {{Anumarlapudi}}, \citenamefont {{Waszewski}}, \citenamefont {{Dobie}}, \citenamefont {{Kaplan}}, \citenamefont {{Morgan}}, \citenamefont {{Rose}},\ and\ \citenamefont {{Wang}}}]{2025mcsweeney}%
  \BibitemOpen
  \bibfield  {author} {\bibinfo {author} {\bibfnamefont {S.~J.}\ \bibnamefont {{McSweeney}}}, \bibinfo {author} {\bibfnamefont {N.}~\bibnamefont {{Hurley-Walker}}}, \bibinfo {author} {\bibfnamefont {C.}~\bibnamefont {{Horv{\'a}th}}}, \bibinfo {author} {\bibfnamefont {A.}~\bibnamefont {{Anumarlapudi}}}, \bibinfo {author} {\bibfnamefont {A.}~\bibnamefont {{Waszewski}}}, \bibinfo {author} {\bibfnamefont {D.}~\bibnamefont {{Dobie}}}, \bibinfo {author} {\bibfnamefont {D.~L.}\ \bibnamefont {{Kaplan}}}, \bibinfo {author} {\bibfnamefont {J.}~\bibnamefont {{Morgan}}}, \bibinfo {author} {\bibfnamefont {K.}~\bibnamefont {{Rose}}}, \ and\ \bibinfo {author} {\bibfnamefont {Z.}~\bibnamefont {{Wang}}},\ }\href {\doibase 10.1093/mnras/staf1203} {\bibfield  {journal} {\bibinfo  {journal} {\mnras}\ }\textbf {\bibinfo {volume} {542}},\ \bibinfo {pages} {203} (\bibinfo {year} {2025})},\ \Eprint {http://arxiv.org/abs/2507.14448} {arXiv:2507.14448 [astro-ph.HE]} \BibitemShut {NoStop}%
\bibitem [{\citenamefont {{Anumarlapudi}}\ \emph {et~al.}(2025)\citenamefont {{Anumarlapudi}}, \citenamefont {{Kaplan}}, \citenamefont {{Rea}}, \citenamefont {{Erasmus}}, \citenamefont {{Kelson}}, \citenamefont {{Ocker}}, \citenamefont {{Lenc}}, \citenamefont {{Dobie}}, \citenamefont {{Hurley-Walker}}, \citenamefont {{Sivakoff}}, \citenamefont {{Buckley}}, \citenamefont {{Murphy}}, \citenamefont {{Pritchard}}, \citenamefont {{Driessen}}, \citenamefont {{Rose}},\ and\ \citenamefont {{Zic}}}]{2025Anumarlapudi}%
  \BibitemOpen
  \bibfield  {author} {\bibinfo {author} {\bibfnamefont {A.}~\bibnamefont {{Anumarlapudi}}}, \bibinfo {author} {\bibfnamefont {D.~L.}\ \bibnamefont {{Kaplan}}}, \bibinfo {author} {\bibfnamefont {N.}~\bibnamefont {{Rea}}}, \bibinfo {author} {\bibfnamefont {N.}~\bibnamefont {{Erasmus}}}, \bibinfo {author} {\bibfnamefont {D.}~\bibnamefont {{Kelson}}}, \bibinfo {author} {\bibfnamefont {S.~K.}\ \bibnamefont {{Ocker}}}, \bibinfo {author} {\bibfnamefont {E.}~\bibnamefont {{Lenc}}}, \bibinfo {author} {\bibfnamefont {D.}~\bibnamefont {{Dobie}}}, \bibinfo {author} {\bibfnamefont {N.}~\bibnamefont {{Hurley-Walker}}}, \bibinfo {author} {\bibfnamefont {G.}~\bibnamefont {{Sivakoff}}}, \bibinfo {author} {\bibfnamefont {D.~A.~H.}\ \bibnamefont {{Buckley}}}, \bibinfo {author} {\bibfnamefont {T.}~\bibnamefont {{Murphy}}}, \bibinfo {author} {\bibfnamefont {J.}~\bibnamefont {{Pritchard}}}, \bibinfo {author} {\bibfnamefont {L.}~\bibnamefont {{Driessen}}}, \bibinfo {author} {\bibfnamefont {K.}~\bibnamefont {{Rose}}}, \ and\
  \bibinfo {author} {\bibfnamefont {A.}~\bibnamefont {{Zic}}},\ }\href {\doibase 10.1093/mnras/staf1227} {\bibfield  {journal} {\bibinfo  {journal} {\mnras}\ }\textbf {\bibinfo {volume} {542}},\ \bibinfo {pages} {1208} (\bibinfo {year} {2025})},\ \Eprint {http://arxiv.org/abs/2507.13453} {arXiv:2507.13453 [astro-ph.HE]} \BibitemShut {NoStop}%
\bibitem [{\citenamefont {{Pritchard}}\ \emph {et~al.}(2026)\citenamefont {{Pritchard}}, \citenamefont {{Murphy}}, \citenamefont {{Dobie}}, \citenamefont {{Lenc}}, \citenamefont {{Anumarlapudi}}, \citenamefont {{Caleb}}, \citenamefont {{Grainger}}, \citenamefont {{Hurley-Walker}}, \citenamefont {{Kaplan}}, \citenamefont {{McSweeney}}, \citenamefont {{Mitchell-Bolton}}, \citenamefont {{Rose}}, \citenamefont {{Sengar}}, \citenamefont {{Wang}}, \citenamefont {{Willingham}},\ and\ \citenamefont {{Zic}}}]{2026Pritchard}%
  \BibitemOpen
  \bibfield  {author} {\bibinfo {author} {\bibfnamefont {J.}~\bibnamefont {{Pritchard}}}, \bibinfo {author} {\bibfnamefont {T.}~\bibnamefont {{Murphy}}}, \bibinfo {author} {\bibfnamefont {D.}~\bibnamefont {{Dobie}}}, \bibinfo {author} {\bibfnamefont {E.}~\bibnamefont {{Lenc}}}, \bibinfo {author} {\bibfnamefont {A.}~\bibnamefont {{Anumarlapudi}}}, \bibinfo {author} {\bibfnamefont {M.}~\bibnamefont {{Caleb}}}, \bibinfo {author} {\bibfnamefont {S.}~\bibnamefont {{Grainger}}}, \bibinfo {author} {\bibfnamefont {N.}~\bibnamefont {{Hurley-Walker}}}, \bibinfo {author} {\bibfnamefont {D.~L.}\ \bibnamefont {{Kaplan}}}, \bibinfo {author} {\bibfnamefont {S.~J.}\ \bibnamefont {{McSweeney}}}, \bibinfo {author} {\bibfnamefont {J.}~\bibnamefont {{Mitchell-Bolton}}}, \bibinfo {author} {\bibfnamefont {K.}~\bibnamefont {{Rose}}}, \bibinfo {author} {\bibfnamefont {R.}~\bibnamefont {{Sengar}}}, \bibinfo {author} {\bibfnamefont {Z.}~\bibnamefont {{Wang}}}, \bibinfo {author} {\bibfnamefont {J.}~\bibnamefont {{Willingham}}}, \ and\
  \bibinfo {author} {\bibfnamefont {A.}~\bibnamefont {{Zic}}},\ }\href {\doibase 10.48550/arXiv.2603.07857} {\bibfield  {journal} {\bibinfo  {journal} {arXiv e-prints}\ ,\ \bibinfo {eid} {arXiv:2603.07857}} (\bibinfo {year} {2026})},\ \Eprint {http://arxiv.org/abs/2603.07857} {arXiv:2603.07857 [astro-ph.HE]} \BibitemShut {NoStop}%
\bibitem [{\citenamefont {{Oke}}\ \emph {et~al.}(1995)\citenamefont {{Oke}}, \citenamefont {{Cohen}}, \citenamefont {{Carr}}, \citenamefont {{Cromer}}, \citenamefont {{Dingizian}}, \citenamefont {{Harris}}, \citenamefont {{Labrecque}}, \citenamefont {{Lucinio}}, \citenamefont {{Schaal}}, \citenamefont {{Epps}},\ and\ \citenamefont {{Miller}}}]{lris}%
  \BibitemOpen
  \bibfield  {author} {\bibinfo {author} {\bibfnamefont {J.~B.}\ \bibnamefont {{Oke}}}, \bibinfo {author} {\bibfnamefont {J.~G.}\ \bibnamefont {{Cohen}}}, \bibinfo {author} {\bibfnamefont {M.}~\bibnamefont {{Carr}}}, \bibinfo {author} {\bibfnamefont {J.}~\bibnamefont {{Cromer}}}, \bibinfo {author} {\bibfnamefont {A.}~\bibnamefont {{Dingizian}}}, \bibinfo {author} {\bibfnamefont {F.~H.}\ \bibnamefont {{Harris}}}, \bibinfo {author} {\bibfnamefont {S.}~\bibnamefont {{Labrecque}}}, \bibinfo {author} {\bibfnamefont {R.}~\bibnamefont {{Lucinio}}}, \bibinfo {author} {\bibfnamefont {W.}~\bibnamefont {{Schaal}}}, \bibinfo {author} {\bibfnamefont {H.}~\bibnamefont {{Epps}}}, \ and\ \bibinfo {author} {\bibfnamefont {J.}~\bibnamefont {{Miller}}},\ }\href {\doibase 10.1086/133562} {\bibfield  {journal} {\bibinfo  {journal} {\pasp}\ }\textbf {\bibinfo {volume} {107}},\ \bibinfo {pages} {375} (\bibinfo {year} {1995})}\BibitemShut {NoStop}%
\bibitem [{\citenamefont {{Perley}}(2019)}]{2019perley_lpipe}%
  \BibitemOpen
  \bibfield  {author} {\bibinfo {author} {\bibfnamefont {D.~A.}\ \bibnamefont {{Perley}}},\ }\href {\doibase 10.1088/1538-3873/ab215d} {\bibfield  {journal} {\bibinfo  {journal} {\pasp}\ }\textbf {\bibinfo {volume} {131}},\ \bibinfo {pages} {084503} (\bibinfo {year} {2019})},\ \Eprint {http://arxiv.org/abs/1903.07629} {arXiv:1903.07629 [astro-ph.IM]} \BibitemShut {NoStop}%
\bibitem [{\citenamefont {{Allard}}\ \emph {et~al.}(2011)\citenamefont {{Allard}}, \citenamefont {{Homeier}},\ and\ \citenamefont {{Freytag}}}]{2011allard}%
  \BibitemOpen
  \bibfield  {author} {\bibinfo {author} {\bibfnamefont {F.}~\bibnamefont {{Allard}}}, \bibinfo {author} {\bibfnamefont {D.}~\bibnamefont {{Homeier}}}, \ and\ \bibinfo {author} {\bibfnamefont {B.}~\bibnamefont {{Freytag}}},\ }in\ \href {\doibase 10.48550/arXiv.1011.5405} {\emph {\bibinfo {booktitle} {16th Cambridge Workshop on Cool Stars, Stellar Systems, and the Sun}}},\ \bibinfo {series} {Astronomical Society of the Pacific Conference Series}, Vol.\ \bibinfo {volume} {448},\ \bibinfo {editor} {edited by\ \bibinfo {editor} {\bibfnamefont {C.}~\bibnamefont {{Johns-Krull}}}, \bibinfo {editor} {\bibfnamefont {M.~K.}\ \bibnamefont {{Browning}}}, \ and\ \bibinfo {editor} {\bibfnamefont {A.~A.}\ \bibnamefont {{West}}}}\ (\bibinfo {year} {2011})\ p.~\bibinfo {pages} {91},\ \Eprint {http://arxiv.org/abs/1011.5405} {arXiv:1011.5405 [astro-ph.SR]} \BibitemShut {NoStop}%
\bibitem [{\citenamefont {{Hastings}}(1970)}]{1970mcmc}%
  \BibitemOpen
  \bibfield  {author} {\bibinfo {author} {\bibfnamefont {W.~K.}\ \bibnamefont {{Hastings}}},\ }\href {\doibase 10.1093/biomet/57.1.97} {\bibfield  {journal} {\bibinfo  {journal} {Biometrika}\ }\textbf {\bibinfo {volume} {57}},\ \bibinfo {pages} {97} (\bibinfo {year} {1970})}\BibitemShut {NoStop}%
\bibitem [{\citenamefont {{Foreman-Mackey}}\ \emph {et~al.}(2019)\citenamefont {{Foreman-Mackey}}, \citenamefont {{Farr}}, \citenamefont {{Sinha}}, \citenamefont {{Archibald}}, \citenamefont {{Hogg}}, \citenamefont {{Sanders}}, \citenamefont {{Zuntz}}, \citenamefont {{Williams}}, \citenamefont {{Nelson}}, \citenamefont {{de Val-Borro}}, \citenamefont {{Erhardt}}, \citenamefont {{Pashchenko}},\ and\ \citenamefont {{Pla}}}]{2019emcee}%
  \BibitemOpen
  \bibfield  {author} {\bibinfo {author} {\bibfnamefont {D.}~\bibnamefont {{Foreman-Mackey}}}, \bibinfo {author} {\bibfnamefont {W.}~\bibnamefont {{Farr}}}, \bibinfo {author} {\bibfnamefont {M.}~\bibnamefont {{Sinha}}}, \bibinfo {author} {\bibfnamefont {A.}~\bibnamefont {{Archibald}}}, \bibinfo {author} {\bibfnamefont {D.}~\bibnamefont {{Hogg}}}, \bibinfo {author} {\bibfnamefont {J.}~\bibnamefont {{Sanders}}}, \bibinfo {author} {\bibfnamefont {J.}~\bibnamefont {{Zuntz}}}, \bibinfo {author} {\bibfnamefont {P.}~\bibnamefont {{Williams}}}, \bibinfo {author} {\bibfnamefont {A.}~\bibnamefont {{Nelson}}}, \bibinfo {author} {\bibfnamefont {M.}~\bibnamefont {{de Val-Borro}}}, \bibinfo {author} {\bibfnamefont {T.}~\bibnamefont {{Erhardt}}}, \bibinfo {author} {\bibfnamefont {I.}~\bibnamefont {{Pashchenko}}}, \ and\ \bibinfo {author} {\bibfnamefont {O.}~\bibnamefont {{Pla}}},\ }\href {\doibase 10.21105/joss.01864} {\bibfield  {journal} {\bibinfo  {journal} {The Journal of Open Source Software}\ }\textbf {\bibinfo
  {volume} {4}},\ \bibinfo {eid} {1864} (\bibinfo {year} {2019})},\ \Eprint {http://arxiv.org/abs/1911.07688} {arXiv:1911.07688 [astro-ph.IM]} \BibitemShut {NoStop}%
\bibitem [{\citenamefont {{Pelisoli}}\ \emph {et~al.}(2023)\citenamefont {{Pelisoli}}, \citenamefont {{Marsh}}, \citenamefont {{Buckley}}, \citenamefont {{Heywood}}, \citenamefont {{Potter}}, \citenamefont {{Schwope}}, \citenamefont {{Brink}}, \citenamefont {{Standke}}, \citenamefont {{Woudt}}, \citenamefont {{Parsons}}, \citenamefont {{Green}}, \citenamefont {{Kepler}}, \citenamefont {{Munday}}, \citenamefont {{Romero}}, \citenamefont {{Breedt}}, \citenamefont {{Brown}}, \citenamefont {{Dhillon}}, \citenamefont {{Dyer}}, \citenamefont {{Kerry}}, \citenamefont {{Littlefair}}, \citenamefont {{Sahman}},\ and\ \citenamefont {{Wild}}}]{2023j1912}%
  \BibitemOpen
  \bibfield  {author} {\bibinfo {author} {\bibfnamefont {I.}~\bibnamefont {{Pelisoli}}}, \bibinfo {author} {\bibfnamefont {T.~R.}\ \bibnamefont {{Marsh}}}, \bibinfo {author} {\bibfnamefont {D.~A.~H.}\ \bibnamefont {{Buckley}}}, \bibinfo {author} {\bibfnamefont {I.}~\bibnamefont {{Heywood}}}, \bibinfo {author} {\bibfnamefont {S.~B.}\ \bibnamefont {{Potter}}}, \bibinfo {author} {\bibfnamefont {A.}~\bibnamefont {{Schwope}}}, \bibinfo {author} {\bibfnamefont {J.}~\bibnamefont {{Brink}}}, \bibinfo {author} {\bibfnamefont {A.}~\bibnamefont {{Standke}}}, \bibinfo {author} {\bibfnamefont {P.~A.}\ \bibnamefont {{Woudt}}}, \bibinfo {author} {\bibfnamefont {S.~G.}\ \bibnamefont {{Parsons}}}, \bibinfo {author} {\bibfnamefont {M.~J.}\ \bibnamefont {{Green}}}, \bibinfo {author} {\bibfnamefont {S.~O.}\ \bibnamefont {{Kepler}}}, \bibinfo {author} {\bibfnamefont {J.}~\bibnamefont {{Munday}}}, \bibinfo {author} {\bibfnamefont {A.~D.}\ \bibnamefont {{Romero}}}, \bibinfo {author} {\bibfnamefont {E.}~\bibnamefont {{Breedt}}},
  \bibinfo {author} {\bibfnamefont {A.~J.}\ \bibnamefont {{Brown}}}, \bibinfo {author} {\bibfnamefont {V.~S.}\ \bibnamefont {{Dhillon}}}, \bibinfo {author} {\bibfnamefont {M.~J.}\ \bibnamefont {{Dyer}}}, \bibinfo {author} {\bibfnamefont {P.}~\bibnamefont {{Kerry}}}, \bibinfo {author} {\bibfnamefont {S.~P.}\ \bibnamefont {{Littlefair}}}, \bibinfo {author} {\bibfnamefont {D.~I.}\ \bibnamefont {{Sahman}}}, \ and\ \bibinfo {author} {\bibfnamefont {J.~F.}\ \bibnamefont {{Wild}}},\ }\href {\doibase 10.1038/s41550-023-01995-x} {\bibfield  {journal} {\bibinfo  {journal} {Nature Astronomy}\ }\textbf {\bibinfo {volume} {7}},\ \bibinfo {pages} {931} (\bibinfo {year} {2023})},\ \Eprint {http://arxiv.org/abs/2306.09272} {arXiv:2306.09272 [astro-ph.SR]} \BibitemShut {NoStop}%
\bibitem [{\citenamefont {{Gaia Collaboration}}\ \emph {et~al.}(2023)\citenamefont {{Gaia Collaboration}}, \citenamefont {{Vallenari}}, \citenamefont {{Brown}}, \citenamefont {{Prusti}}, \citenamefont {{de Bruijne}}, \citenamefont {{Arenou}}, \citenamefont {{Babusiaux}}, \citenamefont {{Biermann}}, \citenamefont {{Creevey}}, \citenamefont {{Ducourant}}, \citenamefont {{Evans}}, \citenamefont {{Eyer}}, \citenamefont {{Guerra}}, \citenamefont {{Hutton}}, \citenamefont {{Jordi}}, \citenamefont {{Klioner}}, \citenamefont {{Lammers}}, \citenamefont {{Lindegren}}, \citenamefont {{Luri}}, \citenamefont {{Mignard}}, \citenamefont {{Panem}}, \citenamefont {{Pourbaix}}, \citenamefont {{Randich}}, \citenamefont {{Sartoretti}}, \citenamefont {{Soubiran}}, \citenamefont {{Tanga}}, \citenamefont {{Walton}}, \citenamefont {{Bailer-Jones}}, \citenamefont {{Bastian}}, \citenamefont {{Drimmel}}, \citenamefont {{Jansen}}, \citenamefont {{Katz}}, \citenamefont {{Lattanzi}}, \citenamefont {{van Leeuwen}}, \citenamefont
  {{Bakker}}, \citenamefont {{Cacciari}}, \citenamefont {{Casta{\~n}eda}}, \citenamefont {{De Angeli}}, \citenamefont {{Fabricius}}, \citenamefont {{Fouesneau}}, \citenamefont {{Fr{\'e}mat}}, \citenamefont {{Galluccio}}, \citenamefont {{Guerrier}}, \citenamefont {{Heiter}}, \citenamefont {{Masana}}, \citenamefont {{Messineo}}, \citenamefont {{Mowlavi}}, \citenamefont {{Nicolas}}, \citenamefont {{Nienartowicz}}, \citenamefont {{Pailler}}, \citenamefont {{Panuzzo}}, \citenamefont {{Riclet}}, \citenamefont {{Roux}}, \citenamefont {{Seabroke}}, \citenamefont {{Sordo}}, \citenamefont {{Th{\'e}venin}}, \citenamefont {{Gracia-Abril}}, \citenamefont {{Portell}}, \citenamefont {{Teyssier}}, \citenamefont {{Altmann}}, \citenamefont {{Andrae}}, \citenamefont {{Audard}}, \citenamefont {{Bellas-Velidis}}, \citenamefont {{Benson}}, \citenamefont {{Berthier}}, \citenamefont {{Blomme}}, \citenamefont {{Burgess}}, \citenamefont {{Busonero}}, \citenamefont {{Busso}}, \citenamefont {{C{\'a}novas}}, \citenamefont {{Carry}},
  \citenamefont {{Cellino}}, \citenamefont {{Cheek}}, \citenamefont {{Clementini}}, \citenamefont {{Damerdji}}, \citenamefont {{Davidson}}, \citenamefont {{de Teodoro}}, \citenamefont {{Nu{\~n}ez Campos}}, \citenamefont {{Delchambre}}, \citenamefont {{Dell'Oro}}, \citenamefont {{Esquej}}, \citenamefont {{Fern{\'a}ndez-Hern{\'a}ndez}}, \citenamefont {{Fraile}}, \citenamefont {{Garabato}}, \citenamefont {{Garc{\'\i}a-Lario}}, \citenamefont {{Gosset}}, \citenamefont {{Haigron}}, \citenamefont {{Halbwachs}}, \citenamefont {{Hambly}}, \citenamefont {{Harrison}}, \citenamefont {{Hern{\'a}ndez}}, \citenamefont {{Hestroffer}}, \citenamefont {{Hodgkin}}, \citenamefont {{Holl}}, \citenamefont {{Jan{\ss}en}}, \citenamefont {{Jevardat de Fombelle}}, \citenamefont {{Jordan}}, \citenamefont {{Krone-Martins}}, \citenamefont {{Lanzafame}}, \citenamefont {{L{\"o}ffler}}, \citenamefont {{Marchal}}, \citenamefont {{Marrese}}, \citenamefont {{Moitinho}}, \citenamefont {{Muinonen}}, \citenamefont {{Osborne}}, \citenamefont
  {{Pancino}}, \citenamefont {{Pauwels}}, \citenamefont {{Recio-Blanco}}, \citenamefont {{Reyl{\'e}}}, \citenamefont {{Riello}}, \citenamefont {{Rimoldini}}, \citenamefont {{Roegiers}}, \citenamefont {{Rybizki}}, \citenamefont {{Sarro}}, \citenamefont {{Siopis}}, \citenamefont {{Smith}}, \citenamefont {{Sozzetti}}, \citenamefont {{Utrilla}}, \citenamefont {{van Leeuwen}}, \citenamefont {{Abbas}}, \citenamefont {{{\'A}brah{\'a}m}}, \citenamefont {{Abreu Aramburu}}, \citenamefont {{Aerts}}, \citenamefont {{Aguado}}, \citenamefont {{Ajaj}}, \citenamefont {{Aldea-Montero}}, \citenamefont {{Altavilla}}, \citenamefont {{{\'A}lvarez}}, \citenamefont {{Alves}}, \citenamefont {{Anders}}, \citenamefont {{Anderson}}, \citenamefont {{Anglada Varela}}, \citenamefont {{Antoja}}, \citenamefont {{Baines}}, \citenamefont {{Baker}}, \citenamefont {{Balaguer-N{\'u}{\~n}ez}}, \citenamefont {{Balbinot}}, \citenamefont {{Balog}}, \citenamefont {{Barache}}, \citenamefont {{Barbato}}, \citenamefont {{Barros}}, \citenamefont
  {{Barstow}}, \citenamefont {{Bartolom{\'e}}}, \citenamefont {{Bassilana}}, \citenamefont {{Bauchet}}, \citenamefont {{Becciani}}, \citenamefont {{Bellazzini}}, \citenamefont {{Berihuete}}, \citenamefont {{Bernet}}, \citenamefont {{Bertone}}, \citenamefont {{Bianchi}}, \citenamefont {{Binnenfeld}}, \citenamefont {{Blanco-Cuaresma}}, \citenamefont {{Blazere}}, \citenamefont {{Boch}}, \citenamefont {{Bombrun}}, \citenamefont {{Bossini}}, \citenamefont {{Bouquillon}}, \citenamefont {{Bragaglia}}, \citenamefont {{Bramante}}, \citenamefont {{Breedt}}, \citenamefont {{Bressan}}, \citenamefont {{Brouillet}}, \citenamefont {{Brugaletta}}, \citenamefont {{Bucciarelli}}, \citenamefont {{Burlacu}}, \citenamefont {{Butkevich}}, \citenamefont {{Buzzi}}, \citenamefont {{Caffau}}, \citenamefont {{Cancelliere}}, \citenamefont {{Cantat-Gaudin}}, \citenamefont {{Carballo}}, \citenamefont {{Carlucci}}, \citenamefont {{Carnerero}}, \citenamefont {{Carrasco}}, \citenamefont {{Casamiquela}}, \citenamefont {{Castellani}},
  \citenamefont {{Castro-Ginard}}, \citenamefont {{Chaoul}}, \citenamefont {{Charlot}}, \citenamefont {{Chemin}}, \citenamefont {{Chiaramida}}, \citenamefont {{Chiavassa}}, \citenamefont {{Chornay}}, \citenamefont {{Comoretto}}, \citenamefont {{Contursi}}, \citenamefont {{Cooper}}, \citenamefont {{Cornez}}, \citenamefont {{Cowell}}, \citenamefont {{Crifo}}, \citenamefont {{Cropper}}, \citenamefont {{Crosta}}, \citenamefont {{Crowley}}, \citenamefont {{Dafonte}}, \citenamefont {{Dapergolas}}, \citenamefont {{David}}, \citenamefont {{David}}, \citenamefont {{de Laverny}}, \citenamefont {{De Luise}},\ and\ \citenamefont {{De March}}}]{2023gaiadr3}%
  \BibitemOpen
  \bibfield  {author} {\bibinfo {author} {\bibnamefont {{Gaia Collaboration}}}, \bibinfo {author} {\bibfnamefont {A.}~\bibnamefont {{Vallenari}}}, \bibinfo {author} {\bibfnamefont {A.~G.~A.}\ \bibnamefont {{Brown}}}, \bibinfo {author} {\bibfnamefont {T.}~\bibnamefont {{Prusti}}}, \bibinfo {author} {\bibfnamefont {J.~H.~J.}\ \bibnamefont {{de Bruijne}}}, \bibinfo {author} {\bibfnamefont {F.}~\bibnamefont {{Arenou}}}, \bibinfo {author} {\bibfnamefont {C.}~\bibnamefont {{Babusiaux}}}, \bibinfo {author} {\bibfnamefont {M.}~\bibnamefont {{Biermann}}}, \bibinfo {author} {\bibfnamefont {O.~L.}\ \bibnamefont {{Creevey}}}, \bibinfo {author} {\bibfnamefont {C.}~\bibnamefont {{Ducourant}}}, \bibinfo {author} {\bibfnamefont {D.~W.}\ \bibnamefont {{Evans}}}, \bibinfo {author} {\bibfnamefont {L.}~\bibnamefont {{Eyer}}}, \bibinfo {author} {\bibfnamefont {R.}~\bibnamefont {{Guerra}}}, \bibinfo {author} {\bibfnamefont {A.}~\bibnamefont {{Hutton}}}, \bibinfo {author} {\bibfnamefont {C.}~\bibnamefont {{Jordi}}}, \bibinfo
  {author} {\bibfnamefont {S.~A.}\ \bibnamefont {{Klioner}}}, \bibinfo {author} {\bibfnamefont {U.~L.}\ \bibnamefont {{Lammers}}}, \bibinfo {author} {\bibfnamefont {L.}~\bibnamefont {{Lindegren}}}, \bibinfo {author} {\bibfnamefont {X.}~\bibnamefont {{Luri}}}, \bibinfo {author} {\bibfnamefont {F.}~\bibnamefont {{Mignard}}}, \bibinfo {author} {\bibfnamefont {C.}~\bibnamefont {{Panem}}}, \bibinfo {author} {\bibfnamefont {D.}~\bibnamefont {{Pourbaix}}}, \bibinfo {author} {\bibfnamefont {S.}~\bibnamefont {{Randich}}}, \bibinfo {author} {\bibfnamefont {P.}~\bibnamefont {{Sartoretti}}}, \bibinfo {author} {\bibfnamefont {C.}~\bibnamefont {{Soubiran}}}, \bibinfo {author} {\bibfnamefont {P.}~\bibnamefont {{Tanga}}}, \bibinfo {author} {\bibfnamefont {N.~A.}\ \bibnamefont {{Walton}}}, \bibinfo {author} {\bibfnamefont {C.~A.~L.}\ \bibnamefont {{Bailer-Jones}}}, \bibinfo {author} {\bibfnamefont {U.}~\bibnamefont {{Bastian}}}, \bibinfo {author} {\bibfnamefont {R.}~\bibnamefont {{Drimmel}}}, \bibinfo {author} {\bibfnamefont
  {F.}~\bibnamefont {{Jansen}}}, \bibinfo {author} {\bibfnamefont {D.}~\bibnamefont {{Katz}}}, \bibinfo {author} {\bibfnamefont {M.~G.}\ \bibnamefont {{Lattanzi}}}, \bibinfo {author} {\bibfnamefont {F.}~\bibnamefont {{van Leeuwen}}}, \bibinfo {author} {\bibfnamefont {J.}~\bibnamefont {{Bakker}}}, \bibinfo {author} {\bibfnamefont {C.}~\bibnamefont {{Cacciari}}}, \bibinfo {author} {\bibfnamefont {J.}~\bibnamefont {{Casta{\~n}eda}}}, \bibinfo {author} {\bibfnamefont {F.}~\bibnamefont {{De Angeli}}}, \bibinfo {author} {\bibfnamefont {C.}~\bibnamefont {{Fabricius}}}, \bibinfo {author} {\bibfnamefont {M.}~\bibnamefont {{Fouesneau}}}, \bibinfo {author} {\bibfnamefont {Y.}~\bibnamefont {{Fr{\'e}mat}}}, \bibinfo {author} {\bibfnamefont {L.}~\bibnamefont {{Galluccio}}}, \bibinfo {author} {\bibfnamefont {A.}~\bibnamefont {{Guerrier}}}, \bibinfo {author} {\bibfnamefont {U.}~\bibnamefont {{Heiter}}}, \bibinfo {author} {\bibfnamefont {E.}~\bibnamefont {{Masana}}}, \bibinfo {author} {\bibfnamefont {R.}~\bibnamefont
  {{Messineo}}}, \bibinfo {author} {\bibfnamefont {N.}~\bibnamefont {{Mowlavi}}}, \bibinfo {author} {\bibfnamefont {C.}~\bibnamefont {{Nicolas}}}, \bibinfo {author} {\bibfnamefont {K.}~\bibnamefont {{Nienartowicz}}}, \bibinfo {author} {\bibfnamefont {F.}~\bibnamefont {{Pailler}}}, \bibinfo {author} {\bibfnamefont {P.}~\bibnamefont {{Panuzzo}}}, \bibinfo {author} {\bibfnamefont {F.}~\bibnamefont {{Riclet}}}, \bibinfo {author} {\bibfnamefont {W.}~\bibnamefont {{Roux}}}, \bibinfo {author} {\bibfnamefont {G.~M.}\ \bibnamefont {{Seabroke}}}, \bibinfo {author} {\bibfnamefont {R.}~\bibnamefont {{Sordo}}}, \bibinfo {author} {\bibfnamefont {F.}~\bibnamefont {{Th{\'e}venin}}}, \bibinfo {author} {\bibfnamefont {G.}~\bibnamefont {{Gracia-Abril}}}, \bibinfo {author} {\bibfnamefont {J.}~\bibnamefont {{Portell}}}, \bibinfo {author} {\bibfnamefont {D.}~\bibnamefont {{Teyssier}}}, \bibinfo {author} {\bibfnamefont {M.}~\bibnamefont {{Altmann}}}, \bibinfo {author} {\bibfnamefont {R.}~\bibnamefont {{Andrae}}}, \bibinfo {author}
  {\bibfnamefont {M.}~\bibnamefont {{Audard}}}, \bibinfo {author} {\bibfnamefont {I.}~\bibnamefont {{Bellas-Velidis}}}, \bibinfo {author} {\bibfnamefont {K.}~\bibnamefont {{Benson}}}, \bibinfo {author} {\bibfnamefont {J.}~\bibnamefont {{Berthier}}}, \bibinfo {author} {\bibfnamefont {R.}~\bibnamefont {{Blomme}}}, \bibinfo {author} {\bibfnamefont {P.~W.}\ \bibnamefont {{Burgess}}}, \bibinfo {author} {\bibfnamefont {D.}~\bibnamefont {{Busonero}}}, \bibinfo {author} {\bibfnamefont {G.}~\bibnamefont {{Busso}}}, \bibinfo {author} {\bibfnamefont {H.}~\bibnamefont {{C{\'a}novas}}}, \bibinfo {author} {\bibfnamefont {B.}~\bibnamefont {{Carry}}}, \bibinfo {author} {\bibfnamefont {A.}~\bibnamefont {{Cellino}}}, \bibinfo {author} {\bibfnamefont {N.}~\bibnamefont {{Cheek}}}, \bibinfo {author} {\bibfnamefont {G.}~\bibnamefont {{Clementini}}}, \bibinfo {author} {\bibfnamefont {Y.}~\bibnamefont {{Damerdji}}}, \bibinfo {author} {\bibfnamefont {M.}~\bibnamefont {{Davidson}}}, \bibinfo {author} {\bibfnamefont {P.}~\bibnamefont
  {{de Teodoro}}}, \bibinfo {author} {\bibfnamefont {M.}~\bibnamefont {{Nu{\~n}ez Campos}}}, \bibinfo {author} {\bibfnamefont {L.}~\bibnamefont {{Delchambre}}}, \bibinfo {author} {\bibfnamefont {A.}~\bibnamefont {{Dell'Oro}}}, \bibinfo {author} {\bibfnamefont {P.}~\bibnamefont {{Esquej}}}, \bibinfo {author} {\bibfnamefont {J.}~\bibnamefont {{Fern{\'a}ndez-Hern{\'a}ndez}}}, \bibinfo {author} {\bibfnamefont {E.}~\bibnamefont {{Fraile}}}, \bibinfo {author} {\bibfnamefont {D.}~\bibnamefont {{Garabato}}}, \bibinfo {author} {\bibfnamefont {P.}~\bibnamefont {{Garc{\'\i}a-Lario}}}, \bibinfo {author} {\bibfnamefont {E.}~\bibnamefont {{Gosset}}}, \bibinfo {author} {\bibfnamefont {R.}~\bibnamefont {{Haigron}}}, \bibinfo {author} {\bibfnamefont {J.~L.}\ \bibnamefont {{Halbwachs}}}, \bibinfo {author} {\bibfnamefont {N.~C.}\ \bibnamefont {{Hambly}}}, \bibinfo {author} {\bibfnamefont {D.~L.}\ \bibnamefont {{Harrison}}}, \bibinfo {author} {\bibfnamefont {J.}~\bibnamefont {{Hern{\'a}ndez}}}, \bibinfo {author} {\bibfnamefont
  {D.}~\bibnamefont {{Hestroffer}}}, \bibinfo {author} {\bibfnamefont {S.~T.}\ \bibnamefont {{Hodgkin}}}, \bibinfo {author} {\bibfnamefont {B.}~\bibnamefont {{Holl}}}, \bibinfo {author} {\bibfnamefont {K.}~\bibnamefont {{Jan{\ss}en}}}, \bibinfo {author} {\bibfnamefont {G.}~\bibnamefont {{Jevardat de Fombelle}}}, \bibinfo {author} {\bibfnamefont {S.}~\bibnamefont {{Jordan}}}, \bibinfo {author} {\bibfnamefont {A.}~\bibnamefont {{Krone-Martins}}}, \bibinfo {author} {\bibfnamefont {A.~C.}\ \bibnamefont {{Lanzafame}}}, \bibinfo {author} {\bibfnamefont {W.}~\bibnamefont {{L{\"o}ffler}}}, \bibinfo {author} {\bibfnamefont {O.}~\bibnamefont {{Marchal}}}, \bibinfo {author} {\bibfnamefont {P.~M.}\ \bibnamefont {{Marrese}}}, \bibinfo {author} {\bibfnamefont {A.}~\bibnamefont {{Moitinho}}}, \bibinfo {author} {\bibfnamefont {K.}~\bibnamefont {{Muinonen}}}, \bibinfo {author} {\bibfnamefont {P.}~\bibnamefont {{Osborne}}}, \bibinfo {author} {\bibfnamefont {E.}~\bibnamefont {{Pancino}}}, \bibinfo {author} {\bibfnamefont
  {T.}~\bibnamefont {{Pauwels}}}, \bibinfo {author} {\bibfnamefont {A.}~\bibnamefont {{Recio-Blanco}}}, \bibinfo {author} {\bibfnamefont {C.}~\bibnamefont {{Reyl{\'e}}}}, \bibinfo {author} {\bibfnamefont {M.}~\bibnamefont {{Riello}}}, \bibinfo {author} {\bibfnamefont {L.}~\bibnamefont {{Rimoldini}}}, \bibinfo {author} {\bibfnamefont {T.}~\bibnamefont {{Roegiers}}}, \bibinfo {author} {\bibfnamefont {J.}~\bibnamefont {{Rybizki}}}, \bibinfo {author} {\bibfnamefont {L.~M.}\ \bibnamefont {{Sarro}}}, \bibinfo {author} {\bibfnamefont {C.}~\bibnamefont {{Siopis}}}, \bibinfo {author} {\bibfnamefont {M.}~\bibnamefont {{Smith}}}, \bibinfo {author} {\bibfnamefont {A.}~\bibnamefont {{Sozzetti}}}, \bibinfo {author} {\bibfnamefont {E.}~\bibnamefont {{Utrilla}}}, \bibinfo {author} {\bibfnamefont {M.}~\bibnamefont {{van Leeuwen}}}, \bibinfo {author} {\bibfnamefont {U.}~\bibnamefont {{Abbas}}}, \bibinfo {author} {\bibfnamefont {P.}~\bibnamefont {{{\'A}brah{\'a}m}}}, \bibinfo {author} {\bibfnamefont {A.}~\bibnamefont {{Abreu
  Aramburu}}}, \bibinfo {author} {\bibfnamefont {C.}~\bibnamefont {{Aerts}}}, \bibinfo {author} {\bibfnamefont {J.~J.}\ \bibnamefont {{Aguado}}}, \bibinfo {author} {\bibfnamefont {M.}~\bibnamefont {{Ajaj}}}, \bibinfo {author} {\bibfnamefont {F.}~\bibnamefont {{Aldea-Montero}}}, \bibinfo {author} {\bibfnamefont {G.}~\bibnamefont {{Altavilla}}}, \bibinfo {author} {\bibfnamefont {M.~A.}\ \bibnamefont {{{\'A}lvarez}}}, \bibinfo {author} {\bibfnamefont {J.}~\bibnamefont {{Alves}}}, \bibinfo {author} {\bibfnamefont {F.}~\bibnamefont {{Anders}}}, \bibinfo {author} {\bibfnamefont {R.~I.}\ \bibnamefont {{Anderson}}}, \bibinfo {author} {\bibfnamefont {E.}~\bibnamefont {{Anglada Varela}}}, \bibinfo {author} {\bibfnamefont {T.}~\bibnamefont {{Antoja}}}, \bibinfo {author} {\bibfnamefont {D.}~\bibnamefont {{Baines}}}, \bibinfo {author} {\bibfnamefont {S.~G.}\ \bibnamefont {{Baker}}}, \bibinfo {author} {\bibfnamefont {L.}~\bibnamefont {{Balaguer-N{\'u}{\~n}ez}}}, \bibinfo {author} {\bibfnamefont {E.}~\bibnamefont
  {{Balbinot}}}, \bibinfo {author} {\bibfnamefont {Z.}~\bibnamefont {{Balog}}}, \bibinfo {author} {\bibfnamefont {C.}~\bibnamefont {{Barache}}}, \bibinfo {author} {\bibfnamefont {D.}~\bibnamefont {{Barbato}}}, \bibinfo {author} {\bibfnamefont {M.}~\bibnamefont {{Barros}}}, \bibinfo {author} {\bibfnamefont {M.~A.}\ \bibnamefont {{Barstow}}}, \bibinfo {author} {\bibfnamefont {S.}~\bibnamefont {{Bartolom{\'e}}}}, \bibinfo {author} {\bibfnamefont {J.~L.}\ \bibnamefont {{Bassilana}}}, \bibinfo {author} {\bibfnamefont {N.}~\bibnamefont {{Bauchet}}}, \bibinfo {author} {\bibfnamefont {U.}~\bibnamefont {{Becciani}}}, \bibinfo {author} {\bibfnamefont {M.}~\bibnamefont {{Bellazzini}}}, \bibinfo {author} {\bibfnamefont {A.}~\bibnamefont {{Berihuete}}}, \bibinfo {author} {\bibfnamefont {M.}~\bibnamefont {{Bernet}}}, \bibinfo {author} {\bibfnamefont {S.}~\bibnamefont {{Bertone}}}, \bibinfo {author} {\bibfnamefont {L.}~\bibnamefont {{Bianchi}}}, \bibinfo {author} {\bibfnamefont {A.}~\bibnamefont {{Binnenfeld}}}, \bibinfo
  {author} {\bibfnamefont {S.}~\bibnamefont {{Blanco-Cuaresma}}}, \bibinfo {author} {\bibfnamefont {A.}~\bibnamefont {{Blazere}}}, \bibinfo {author} {\bibfnamefont {T.}~\bibnamefont {{Boch}}}, \bibinfo {author} {\bibfnamefont {A.}~\bibnamefont {{Bombrun}}}, \bibinfo {author} {\bibfnamefont {D.}~\bibnamefont {{Bossini}}}, \bibinfo {author} {\bibfnamefont {S.}~\bibnamefont {{Bouquillon}}}, \bibinfo {author} {\bibfnamefont {A.}~\bibnamefont {{Bragaglia}}}, \bibinfo {author} {\bibfnamefont {L.}~\bibnamefont {{Bramante}}}, \bibinfo {author} {\bibfnamefont {E.}~\bibnamefont {{Breedt}}}, \bibinfo {author} {\bibfnamefont {A.}~\bibnamefont {{Bressan}}}, \bibinfo {author} {\bibfnamefont {N.}~\bibnamefont {{Brouillet}}}, \bibinfo {author} {\bibfnamefont {E.}~\bibnamefont {{Brugaletta}}}, \bibinfo {author} {\bibfnamefont {B.}~\bibnamefont {{Bucciarelli}}}, \bibinfo {author} {\bibfnamefont {A.}~\bibnamefont {{Burlacu}}}, \bibinfo {author} {\bibfnamefont {A.~G.}\ \bibnamefont {{Butkevich}}}, \bibinfo {author}
  {\bibfnamefont {R.}~\bibnamefont {{Buzzi}}}, \bibinfo {author} {\bibfnamefont {E.}~\bibnamefont {{Caffau}}}, \bibinfo {author} {\bibfnamefont {R.}~\bibnamefont {{Cancelliere}}}, \bibinfo {author} {\bibfnamefont {T.}~\bibnamefont {{Cantat-Gaudin}}}, \bibinfo {author} {\bibfnamefont {R.}~\bibnamefont {{Carballo}}}, \bibinfo {author} {\bibfnamefont {T.}~\bibnamefont {{Carlucci}}}, \bibinfo {author} {\bibfnamefont {M.~I.}\ \bibnamefont {{Carnerero}}}, \bibinfo {author} {\bibfnamefont {J.~M.}\ \bibnamefont {{Carrasco}}}, \bibinfo {author} {\bibfnamefont {L.}~\bibnamefont {{Casamiquela}}}, \bibinfo {author} {\bibfnamefont {M.}~\bibnamefont {{Castellani}}}, \bibinfo {author} {\bibfnamefont {A.}~\bibnamefont {{Castro-Ginard}}}, \bibinfo {author} {\bibfnamefont {L.}~\bibnamefont {{Chaoul}}}, \bibinfo {author} {\bibfnamefont {P.}~\bibnamefont {{Charlot}}}, \bibinfo {author} {\bibfnamefont {L.}~\bibnamefont {{Chemin}}}, \bibinfo {author} {\bibfnamefont {V.}~\bibnamefont {{Chiaramida}}}, \bibinfo {author}
  {\bibfnamefont {A.}~\bibnamefont {{Chiavassa}}}, \bibinfo {author} {\bibfnamefont {N.}~\bibnamefont {{Chornay}}}, \bibinfo {author} {\bibfnamefont {G.}~\bibnamefont {{Comoretto}}}, \bibinfo {author} {\bibfnamefont {G.}~\bibnamefont {{Contursi}}}, \bibinfo {author} {\bibfnamefont {W.~J.}\ \bibnamefont {{Cooper}}}, \bibinfo {author} {\bibfnamefont {T.}~\bibnamefont {{Cornez}}}, \bibinfo {author} {\bibfnamefont {S.}~\bibnamefont {{Cowell}}}, \bibinfo {author} {\bibfnamefont {F.}~\bibnamefont {{Crifo}}}, \bibinfo {author} {\bibfnamefont {M.}~\bibnamefont {{Cropper}}}, \bibinfo {author} {\bibfnamefont {M.}~\bibnamefont {{Crosta}}}, \bibinfo {author} {\bibfnamefont {C.}~\bibnamefont {{Crowley}}}, \bibinfo {author} {\bibfnamefont {C.}~\bibnamefont {{Dafonte}}}, \bibinfo {author} {\bibfnamefont {A.}~\bibnamefont {{Dapergolas}}}, \bibinfo {author} {\bibfnamefont {M.}~\bibnamefont {{David}}}, \bibinfo {author} {\bibfnamefont {P.}~\bibnamefont {{David}}}, \bibinfo {author} {\bibfnamefont {P.}~\bibnamefont {{de
  Laverny}}}, \bibinfo {author} {\bibfnamefont {F.}~\bibnamefont {{De Luise}}}, \ and\ \bibinfo {author} {\bibfnamefont {R.}~\bibnamefont {{De March}}},\ }\href {\doibase 10.1051/0004-6361/202243940} {\bibfield  {journal} {\bibinfo  {journal} {\aap}\ }\textbf {\bibinfo {volume} {674}},\ \bibinfo {eid} {A1} (\bibinfo {year} {2023})},\ \Eprint {http://arxiv.org/abs/2208.00211} {arXiv:2208.00211 [astro-ph.GA]} \BibitemShut {NoStop}%
\bibitem [{\citenamefont {{Toomre}}(1964)}]{1964toomre}%
  \BibitemOpen
  \bibfield  {author} {\bibinfo {author} {\bibfnamefont {A.}~\bibnamefont {{Toomre}}},\ }\href {\doibase 10.1086/147861} {\bibfield  {journal} {\bibinfo  {journal} {\apj}\ }\textbf {\bibinfo {volume} {139}},\ \bibinfo {pages} {1217} (\bibinfo {year} {1964})}\BibitemShut {NoStop}%
\bibitem [{\citenamefont {{Bensby}}\ \emph {et~al.}(2003)\citenamefont {{Bensby}}, \citenamefont {{Feltzing}},\ and\ \citenamefont {{Lundstr{\"o}m}}}]{2003bensby}%
  \BibitemOpen
  \bibfield  {author} {\bibinfo {author} {\bibfnamefont {T.}~\bibnamefont {{Bensby}}}, \bibinfo {author} {\bibfnamefont {S.}~\bibnamefont {{Feltzing}}}, \ and\ \bibinfo {author} {\bibfnamefont {I.}~\bibnamefont {{Lundstr{\"o}m}}},\ }\href {\doibase 10.1051/0004-6361:20031213} {\bibfield  {journal} {\bibinfo  {journal} {\aap}\ }\textbf {\bibinfo {volume} {410}},\ \bibinfo {pages} {527} (\bibinfo {year} {2003})}\BibitemShut {NoStop}%
\bibitem [{\citenamefont {{Zorotovic}}\ \emph {et~al.}(2016)\citenamefont {{Zorotovic}}, \citenamefont {{Schreiber}}, \citenamefont {{Parsons}}, \citenamefont {{G{\"a}nsicke}}, \citenamefont {{Hardy}}, \citenamefont {{Agurto-Gangas}}, \citenamefont {{Nebot G{\'o}mez-Mor{\'a}n}}, \citenamefont {{Rebassa-Mansergas}},\ and\ \citenamefont {{Schwope}}}]{2016zorotovic}%
  \BibitemOpen
  \bibfield  {author} {\bibinfo {author} {\bibfnamefont {M.}~\bibnamefont {{Zorotovic}}}, \bibinfo {author} {\bibfnamefont {M.~R.}\ \bibnamefont {{Schreiber}}}, \bibinfo {author} {\bibfnamefont {S.~G.}\ \bibnamefont {{Parsons}}}, \bibinfo {author} {\bibfnamefont {B.~T.}\ \bibnamefont {{G{\"a}nsicke}}}, \bibinfo {author} {\bibfnamefont {A.}~\bibnamefont {{Hardy}}}, \bibinfo {author} {\bibfnamefont {C.}~\bibnamefont {{Agurto-Gangas}}}, \bibinfo {author} {\bibfnamefont {A.}~\bibnamefont {{Nebot G{\'o}mez-Mor{\'a}n}}}, \bibinfo {author} {\bibfnamefont {A.}~\bibnamefont {{Rebassa-Mansergas}}}, \ and\ \bibinfo {author} {\bibfnamefont {A.~D.}\ \bibnamefont {{Schwope}}},\ }\href {\doibase 10.1093/mnras/stw246} {\bibfield  {journal} {\bibinfo  {journal} {\mnras}\ }\textbf {\bibinfo {volume} {457}},\ \bibinfo {pages} {3867} (\bibinfo {year} {2016})},\ \Eprint {http://arxiv.org/abs/1601.07785} {arXiv:1601.07785 [astro-ph.SR]} \BibitemShut {NoStop}%
\bibitem [{\citenamefont {{Kilic}}\ \emph {et~al.}(2017)\citenamefont {{Kilic}}, \citenamefont {{Munn}}, \citenamefont {{Harris}}, \citenamefont {{von Hippel}}, \citenamefont {{Liebert}}, \citenamefont {{Williams}}, \citenamefont {{Jeffery}},\ and\ \citenamefont {{DeGennaro}}}]{2017kilic}%
  \BibitemOpen
  \bibfield  {author} {\bibinfo {author} {\bibfnamefont {M.}~\bibnamefont {{Kilic}}}, \bibinfo {author} {\bibfnamefont {J.~A.}\ \bibnamefont {{Munn}}}, \bibinfo {author} {\bibfnamefont {H.~C.}\ \bibnamefont {{Harris}}}, \bibinfo {author} {\bibfnamefont {T.}~\bibnamefont {{von Hippel}}}, \bibinfo {author} {\bibfnamefont {J.~W.}\ \bibnamefont {{Liebert}}}, \bibinfo {author} {\bibfnamefont {K.~A.}\ \bibnamefont {{Williams}}}, \bibinfo {author} {\bibfnamefont {E.}~\bibnamefont {{Jeffery}}}, \ and\ \bibinfo {author} {\bibfnamefont {S.}~\bibnamefont {{DeGennaro}}},\ }\href {\doibase 10.3847/1538-4357/aa62a5} {\bibfield  {journal} {\bibinfo  {journal} {\apj}\ }\textbf {\bibinfo {volume} {837}},\ \bibinfo {eid} {162} (\bibinfo {year} {2017})},\ \Eprint {http://arxiv.org/abs/1702.06984} {arXiv:1702.06984 [astro-ph.SR]} \BibitemShut {NoStop}%
\bibitem [{\citenamefont {{Edenhofer}}\ \emph {et~al.}(2024)\citenamefont {{Edenhofer}}, \citenamefont {{Zucker}}, \citenamefont {{Frank}}, \citenamefont {{Saydjari}}, \citenamefont {{Speagle}}, \citenamefont {{Finkbeiner}},\ and\ \citenamefont {{En{\ss}lin}}}]{2024edenhofer}%
  \BibitemOpen
  \bibfield  {author} {\bibinfo {author} {\bibfnamefont {G.}~\bibnamefont {{Edenhofer}}}, \bibinfo {author} {\bibfnamefont {C.}~\bibnamefont {{Zucker}}}, \bibinfo {author} {\bibfnamefont {P.}~\bibnamefont {{Frank}}}, \bibinfo {author} {\bibfnamefont {A.~K.}\ \bibnamefont {{Saydjari}}}, \bibinfo {author} {\bibfnamefont {J.~S.}\ \bibnamefont {{Speagle}}}, \bibinfo {author} {\bibfnamefont {D.}~\bibnamefont {{Finkbeiner}}}, \ and\ \bibinfo {author} {\bibfnamefont {T.~A.}\ \bibnamefont {{En{\ss}lin}}},\ }\href {\doibase 10.1051/0004-6361/202347628} {\bibfield  {journal} {\bibinfo  {journal} {\aap}\ }\textbf {\bibinfo {volume} {685}},\ \bibinfo {eid} {A82} (\bibinfo {year} {2024})},\ \Eprint {http://arxiv.org/abs/2308.01295} {arXiv:2308.01295 [astro-ph.GA]} \BibitemShut {NoStop}%
\bibitem [{\citenamefont {{Baraffe}}\ \emph {et~al.}(2015)\citenamefont {{Baraffe}}, \citenamefont {{Homeier}}, \citenamefont {{Allard}},\ and\ \citenamefont {{Chabrier}}}]{2015baraffe}%
  \BibitemOpen
  \bibfield  {author} {\bibinfo {author} {\bibfnamefont {I.}~\bibnamefont {{Baraffe}}}, \bibinfo {author} {\bibfnamefont {D.}~\bibnamefont {{Homeier}}}, \bibinfo {author} {\bibfnamefont {F.}~\bibnamefont {{Allard}}}, \ and\ \bibinfo {author} {\bibfnamefont {G.}~\bibnamefont {{Chabrier}}},\ }\href {\doibase 10.1051/0004-6361/201425481} {\bibfield  {journal} {\bibinfo  {journal} {\aap}\ }\textbf {\bibinfo {volume} {577}},\ \bibinfo {eid} {A42} (\bibinfo {year} {2015})},\ \Eprint {http://arxiv.org/abs/1503.04107} {arXiv:1503.04107 [astro-ph.SR]} \BibitemShut {NoStop}%
\bibitem [{\citenamefont {{Parsons}}\ \emph {et~al.}(2018)\citenamefont {{Parsons}}, \citenamefont {{G{\"a}nsicke}}, \citenamefont {{Marsh}}, \citenamefont {{Ashley}}, \citenamefont {{Breedt}}, \citenamefont {{Burleigh}}, \citenamefont {{Copperwheat}}, \citenamefont {{Dhillon}}, \citenamefont {{Green}}, \citenamefont {{Hermes}}, \citenamefont {{Irawati}}, \citenamefont {{Kerry}}, \citenamefont {{Littlefair}}, \citenamefont {{Rebassa-Mansergas}}, \citenamefont {{Sahman}}, \citenamefont {{Schreiber}},\ and\ \citenamefont {{Zorotovic}}}]{2018parsons}%
  \BibitemOpen
  \bibfield  {author} {\bibinfo {author} {\bibfnamefont {S.~G.}\ \bibnamefont {{Parsons}}}, \bibinfo {author} {\bibfnamefont {B.~T.}\ \bibnamefont {{G{\"a}nsicke}}}, \bibinfo {author} {\bibfnamefont {T.~R.}\ \bibnamefont {{Marsh}}}, \bibinfo {author} {\bibfnamefont {R.~P.}\ \bibnamefont {{Ashley}}}, \bibinfo {author} {\bibfnamefont {E.}~\bibnamefont {{Breedt}}}, \bibinfo {author} {\bibfnamefont {M.~R.}\ \bibnamefont {{Burleigh}}}, \bibinfo {author} {\bibfnamefont {C.~M.}\ \bibnamefont {{Copperwheat}}}, \bibinfo {author} {\bibfnamefont {V.~S.}\ \bibnamefont {{Dhillon}}}, \bibinfo {author} {\bibfnamefont {M.~J.}\ \bibnamefont {{Green}}}, \bibinfo {author} {\bibfnamefont {J.~J.}\ \bibnamefont {{Hermes}}}, \bibinfo {author} {\bibfnamefont {P.}~\bibnamefont {{Irawati}}}, \bibinfo {author} {\bibfnamefont {P.}~\bibnamefont {{Kerry}}}, \bibinfo {author} {\bibfnamefont {S.~P.}\ \bibnamefont {{Littlefair}}}, \bibinfo {author} {\bibfnamefont {A.}~\bibnamefont {{Rebassa-Mansergas}}}, \bibinfo {author} {\bibfnamefont
  {D.~I.}\ \bibnamefont {{Sahman}}}, \bibinfo {author} {\bibfnamefont {M.~R.}\ \bibnamefont {{Schreiber}}}, \ and\ \bibinfo {author} {\bibfnamefont {M.}~\bibnamefont {{Zorotovic}}},\ }\href {\doibase 10.1093/mnras/sty2345} {\bibfield  {journal} {\bibinfo  {journal} {\mnras}\ }\textbf {\bibinfo {volume} {481}},\ \bibinfo {pages} {1083} (\bibinfo {year} {2018})},\ \Eprint {http://arxiv.org/abs/1808.07780} {arXiv:1808.07780 [astro-ph.SR]} \BibitemShut {NoStop}%
\bibitem [{\citenamefont {{Koester}}(2010)}]{2010koester}%
  \BibitemOpen
  \bibfield  {author} {\bibinfo {author} {\bibfnamefont {D.}~\bibnamefont {{Koester}}},\ }\href@noop {} {\bibfield  {journal} {\bibinfo  {journal} {\memsai}\ }\textbf {\bibinfo {volume} {81}},\ \bibinfo {pages} {921} (\bibinfo {year} {2010})}\BibitemShut {NoStop}%
\bibitem [{\citenamefont {{Gelman}}\ and\ \citenamefont {{Rubin}}(1992)}]{1992gelman}%
  \BibitemOpen
  \bibfield  {author} {\bibinfo {author} {\bibfnamefont {A.}~\bibnamefont {{Gelman}}}\ and\ \bibinfo {author} {\bibfnamefont {D.~B.}\ \bibnamefont {{Rubin}}},\ }\href {\doibase 10.1214/ss/1177011136} {\bibfield  {journal} {\bibinfo  {journal} {Statistical Science}\ }\textbf {\bibinfo {volume} {7}},\ \bibinfo {pages} {457} (\bibinfo {year} {1992})}\BibitemShut {NoStop}%
\bibitem [{\citenamefont {{B{\'e}dard}}\ \emph {et~al.}(2020)\citenamefont {{B{\'e}dard}}, \citenamefont {{Bergeron}}, \citenamefont {{Brassard}},\ and\ \citenamefont {{Fontaine}}}]{2020bedard}%
  \BibitemOpen
  \bibfield  {author} {\bibinfo {author} {\bibfnamefont {A.}~\bibnamefont {{B{\'e}dard}}}, \bibinfo {author} {\bibfnamefont {P.}~\bibnamefont {{Bergeron}}}, \bibinfo {author} {\bibfnamefont {P.}~\bibnamefont {{Brassard}}}, \ and\ \bibinfo {author} {\bibfnamefont {G.}~\bibnamefont {{Fontaine}}},\ }\href {\doibase 10.3847/1538-4357/abafbe} {\bibfield  {journal} {\bibinfo  {journal} {\apj}\ }\textbf {\bibinfo {volume} {901}},\ \bibinfo {eid} {93} (\bibinfo {year} {2020})},\ \Eprint {http://arxiv.org/abs/2008.07469} {arXiv:2008.07469 [astro-ph.SR]} \BibitemShut {NoStop}%
\bibitem [{\citenamefont {{Eggleton}}(1983)}]{1983eggleton}%
  \BibitemOpen
  \bibfield  {author} {\bibinfo {author} {\bibfnamefont {P.~P.}\ \bibnamefont {{Eggleton}}},\ }\href {\doibase 10.1086/160960} {\bibfield  {journal} {\bibinfo  {journal} {\apj}\ }\textbf {\bibinfo {volume} {268}},\ \bibinfo {pages} {368} (\bibinfo {year} {1983})}\BibitemShut {NoStop}%
\bibitem [{\citenamefont {{Kotze}}\ \emph {et~al.}(2015)\citenamefont {{Kotze}}, \citenamefont {{Potter}},\ and\ \citenamefont {{McBride}}}]{2015doptomog}%
  \BibitemOpen
  \bibfield  {author} {\bibinfo {author} {\bibfnamefont {E.~J.}\ \bibnamefont {{Kotze}}}, \bibinfo {author} {\bibfnamefont {S.~B.}\ \bibnamefont {{Potter}}}, \ and\ \bibinfo {author} {\bibfnamefont {V.~A.}\ \bibnamefont {{McBride}}},\ }\href {\doibase 10.1051/0004-6361/201526381} {\bibfield  {journal} {\bibinfo  {journal} {\aap}\ }\textbf {\bibinfo {volume} {579}},\ \bibinfo {eid} {A77} (\bibinfo {year} {2015})},\ \Eprint {http://arxiv.org/abs/1507.05213} {arXiv:1507.05213 [astro-ph.SR]} \BibitemShut {NoStop}%
\bibitem [{\citenamefont {{Marsh}}(2001)}]{2001marsh}%
  \BibitemOpen
  \bibfield  {author} {\bibinfo {author} {\bibfnamefont {T.~R.}\ \bibnamefont {{Marsh}}},\ }in\ \href {\doibase 10.48550/arXiv.astro-ph/0011020} {\emph {\bibinfo {booktitle} {Astrotomography, Indirect Imaging Methods in Observational Astronomy}}},\ Vol.\ \bibinfo {volume} {573},\ \bibinfo {editor} {edited by\ \bibinfo {editor} {\bibfnamefont {H.~M.~J.}\ \bibnamefont {{Boffin}}}, \bibinfo {editor} {\bibfnamefont {D.}~\bibnamefont {{Steeghs}}}, \ and\ \bibinfo {editor} {\bibfnamefont {J.}~\bibnamefont {{Cuypers}}}}\ (\bibinfo {year} {2001})\ p.~\bibinfo {pages} {1}\BibitemShut {NoStop}%
\bibitem [{\citenamefont {{Delfosse}}\ \emph {et~al.}(1998)\citenamefont {{Delfosse}}, \citenamefont {{Forveille}}, \citenamefont {{Perrier}},\ and\ \citenamefont {{Mayor}}}]{1998delfosse}%
  \BibitemOpen
  \bibfield  {author} {\bibinfo {author} {\bibfnamefont {X.}~\bibnamefont {{Delfosse}}}, \bibinfo {author} {\bibfnamefont {T.}~\bibnamefont {{Forveille}}}, \bibinfo {author} {\bibfnamefont {C.}~\bibnamefont {{Perrier}}}, \ and\ \bibinfo {author} {\bibfnamefont {M.}~\bibnamefont {{Mayor}}},\ }\href@noop {} {\bibfield  {journal} {\bibinfo  {journal} {\aap}\ }\textbf {\bibinfo {volume} {331}},\ \bibinfo {pages} {581} (\bibinfo {year} {1998})}\BibitemShut {NoStop}%
\bibitem [{\citenamefont {{Skinner}}\ \emph {et~al.}(2017)\citenamefont {{Skinner}}, \citenamefont {{Morgan}}, \citenamefont {{West}}, \citenamefont {{L{\'e}pine}},\ and\ \citenamefont {{Thorstensen}}}]{2017skinner}%
  \BibitemOpen
  \bibfield  {author} {\bibinfo {author} {\bibfnamefont {J.~N.}\ \bibnamefont {{Skinner}}}, \bibinfo {author} {\bibfnamefont {D.~P.}\ \bibnamefont {{Morgan}}}, \bibinfo {author} {\bibfnamefont {A.~A.}\ \bibnamefont {{West}}}, \bibinfo {author} {\bibfnamefont {S.}~\bibnamefont {{L{\'e}pine}}}, \ and\ \bibinfo {author} {\bibfnamefont {J.~R.}\ \bibnamefont {{Thorstensen}}},\ }\href {\doibase 10.3847/1538-3881/aa83b5} {\bibfield  {journal} {\bibinfo  {journal} {\aj}\ }\textbf {\bibinfo {volume} {154}},\ \bibinfo {eid} {118} (\bibinfo {year} {2017})},\ \Eprint {http://arxiv.org/abs/1707.08576} {arXiv:1707.08576 [astro-ph.SR]} \BibitemShut {NoStop}%
\bibitem [{\citenamefont {{El-Badry}}\ \emph {et~al.}(2023)\citenamefont {{El-Badry}}, \citenamefont {{Burdge}}, \citenamefont {{van Roestel}},\ and\ \citenamefont {{Rodriguez}}}]{2023elbadry_bd}%
  \BibitemOpen
  \bibfield  {author} {\bibinfo {author} {\bibfnamefont {K.}~\bibnamefont {{El-Badry}}}, \bibinfo {author} {\bibfnamefont {K.~B.}\ \bibnamefont {{Burdge}}}, \bibinfo {author} {\bibfnamefont {J.}~\bibnamefont {{van Roestel}}}, \ and\ \bibinfo {author} {\bibfnamefont {A.~C.}\ \bibnamefont {{Rodriguez}}},\ }\href {\doibase 10.21105/astro.2307.15729} {\bibfield  {journal} {\bibinfo  {journal} {The Open Journal of Astrophysics}\ }\textbf {\bibinfo {volume} {6}},\ \bibinfo {eid} {33} (\bibinfo {year} {2023})},\ \Eprint {http://arxiv.org/abs/2307.15729} {arXiv:2307.15729 [astro-ph.SR]} \BibitemShut {NoStop}%
\bibitem [{\citenamefont {{Marsh}}\ \emph {et~al.}(2016)\citenamefont {{Marsh}}, \citenamefont {{G{\"a}nsicke}}, \citenamefont {{H{\"u}mmerich}}, \citenamefont {{Hambsch}}, \citenamefont {{Bernhard}}, \citenamefont {{Lloyd}}, \citenamefont {{Breedt}}, \citenamefont {{Stanway}}, \citenamefont {{Steeghs}}, \citenamefont {{Parsons}}, \citenamefont {{Toloza}}, \citenamefont {{Schreiber}}, \citenamefont {{Jonker}}, \citenamefont {{van Roestel}}, \citenamefont {{Kupfer}}, \citenamefont {{Pala}}, \citenamefont {{Dhillon}}, \citenamefont {{Hardy}}, \citenamefont {{Littlefair}}, \citenamefont {{Aungwerojwit}}, \citenamefont {{Arjyotha}}, \citenamefont {{Koester}}, \citenamefont {{Bochinski}}, \citenamefont {{Haswell}}, \citenamefont {{Frank}},\ and\ \citenamefont {{Wheatley}}}]{2016marsh}%
  \BibitemOpen
  \bibfield  {author} {\bibinfo {author} {\bibfnamefont {T.~R.}\ \bibnamefont {{Marsh}}}, \bibinfo {author} {\bibfnamefont {B.~T.}\ \bibnamefont {{G{\"a}nsicke}}}, \bibinfo {author} {\bibfnamefont {S.}~\bibnamefont {{H{\"u}mmerich}}}, \bibinfo {author} {\bibfnamefont {F.~J.}\ \bibnamefont {{Hambsch}}}, \bibinfo {author} {\bibfnamefont {K.}~\bibnamefont {{Bernhard}}}, \bibinfo {author} {\bibfnamefont {C.}~\bibnamefont {{Lloyd}}}, \bibinfo {author} {\bibfnamefont {E.}~\bibnamefont {{Breedt}}}, \bibinfo {author} {\bibfnamefont {E.~R.}\ \bibnamefont {{Stanway}}}, \bibinfo {author} {\bibfnamefont {D.~T.}\ \bibnamefont {{Steeghs}}}, \bibinfo {author} {\bibfnamefont {S.~G.}\ \bibnamefont {{Parsons}}}, \bibinfo {author} {\bibfnamefont {O.}~\bibnamefont {{Toloza}}}, \bibinfo {author} {\bibfnamefont {M.~R.}\ \bibnamefont {{Schreiber}}}, \bibinfo {author} {\bibfnamefont {P.~G.}\ \bibnamefont {{Jonker}}}, \bibinfo {author} {\bibfnamefont {J.}~\bibnamefont {{van Roestel}}}, \bibinfo {author} {\bibfnamefont
  {T.}~\bibnamefont {{Kupfer}}}, \bibinfo {author} {\bibfnamefont {A.~F.}\ \bibnamefont {{Pala}}}, \bibinfo {author} {\bibfnamefont {V.~S.}\ \bibnamefont {{Dhillon}}}, \bibinfo {author} {\bibfnamefont {L.~K.}\ \bibnamefont {{Hardy}}}, \bibinfo {author} {\bibfnamefont {S.~P.}\ \bibnamefont {{Littlefair}}}, \bibinfo {author} {\bibfnamefont {A.}~\bibnamefont {{Aungwerojwit}}}, \bibinfo {author} {\bibfnamefont {S.}~\bibnamefont {{Arjyotha}}}, \bibinfo {author} {\bibfnamefont {D.}~\bibnamefont {{Koester}}}, \bibinfo {author} {\bibfnamefont {J.~J.}\ \bibnamefont {{Bochinski}}}, \bibinfo {author} {\bibfnamefont {C.~A.}\ \bibnamefont {{Haswell}}}, \bibinfo {author} {\bibfnamefont {P.}~\bibnamefont {{Frank}}}, \ and\ \bibinfo {author} {\bibfnamefont {P.~J.}\ \bibnamefont {{Wheatley}}},\ }\href {\doibase 10.1038/nature18620} {\bibfield  {journal} {\bibinfo  {journal} {\nat}\ }\textbf {\bibinfo {volume} {537}},\ \bibinfo {pages} {374} (\bibinfo {year} {2016})},\ \Eprint {http://arxiv.org/abs/1607.08265}
  {arXiv:1607.08265 [astro-ph.SR]} \BibitemShut {NoStop}%
\bibitem [{\citenamefont {{Kepler}}\ \emph {et~al.}(2007)\citenamefont {{Kepler}}, \citenamefont {{Kleinman}}, \citenamefont {{Nitta}}, \citenamefont {{Koester}}, \citenamefont {{Castanheira}}, \citenamefont {{Giovannini}}, \citenamefont {{Costa}},\ and\ \citenamefont {{Althaus}}}]{2007wd}%
  \BibitemOpen
  \bibfield  {author} {\bibinfo {author} {\bibfnamefont {S.~O.}\ \bibnamefont {{Kepler}}}, \bibinfo {author} {\bibfnamefont {S.~J.}\ \bibnamefont {{Kleinman}}}, \bibinfo {author} {\bibfnamefont {A.}~\bibnamefont {{Nitta}}}, \bibinfo {author} {\bibfnamefont {D.}~\bibnamefont {{Koester}}}, \bibinfo {author} {\bibfnamefont {B.~G.}\ \bibnamefont {{Castanheira}}}, \bibinfo {author} {\bibfnamefont {O.}~\bibnamefont {{Giovannini}}}, \bibinfo {author} {\bibfnamefont {A.~F.~M.}\ \bibnamefont {{Costa}}}, \ and\ \bibinfo {author} {\bibfnamefont {L.}~\bibnamefont {{Althaus}}},\ }\href {\doibase 10.1111/j.1365-2966.2006.11388.x} {\bibfield  {journal} {\bibinfo  {journal} {\mnras}\ }\textbf {\bibinfo {volume} {375}},\ \bibinfo {pages} {1315} (\bibinfo {year} {2007})},\ \Eprint {http://arxiv.org/abs/astro-ph/0612277} {arXiv:astro-ph/0612277 [astro-ph]} \BibitemShut {NoStop}%
\bibitem [{\citenamefont {{Pala}}\ \emph {et~al.}(2022)\citenamefont {{Pala}}, \citenamefont {{G{\"a}nsicke}}, \citenamefont {{Belloni}}, \citenamefont {{Parsons}}, \citenamefont {{Marsh}}, \citenamefont {{Schreiber}}, \citenamefont {{Breedt}}, \citenamefont {{Knigge}}, \citenamefont {{Sion}}, \citenamefont {{Szkody}}, \citenamefont {{Townsley}}, \citenamefont {{Bildsten}}, \citenamefont {{Boyd}}, \citenamefont {{Cook}}, \citenamefont {{De Martino}}, \citenamefont {{Godon}}, \citenamefont {{Kafka}}, \citenamefont {{Kouprianov}}, \citenamefont {{Long}}, \citenamefont {{Monard}}, \citenamefont {{Myers}}, \citenamefont {{Nelson}}, \citenamefont {{Nogami}}, \citenamefont {{Oksanen}}, \citenamefont {{Pickard}}, \citenamefont {{Poyner}}, \citenamefont {{Reichart}}, \citenamefont {{Rodriguez Perez}}, \citenamefont {{Shears}}, \citenamefont {{Stubbings}},\ and\ \citenamefont {{Toloza}}}]{2022pala}%
  \BibitemOpen
  \bibfield  {author} {\bibinfo {author} {\bibfnamefont {A.~F.}\ \bibnamefont {{Pala}}}, \bibinfo {author} {\bibfnamefont {B.~T.}\ \bibnamefont {{G{\"a}nsicke}}}, \bibinfo {author} {\bibfnamefont {D.}~\bibnamefont {{Belloni}}}, \bibinfo {author} {\bibfnamefont {S.~G.}\ \bibnamefont {{Parsons}}}, \bibinfo {author} {\bibfnamefont {T.~R.}\ \bibnamefont {{Marsh}}}, \bibinfo {author} {\bibfnamefont {M.~R.}\ \bibnamefont {{Schreiber}}}, \bibinfo {author} {\bibfnamefont {E.}~\bibnamefont {{Breedt}}}, \bibinfo {author} {\bibfnamefont {C.}~\bibnamefont {{Knigge}}}, \bibinfo {author} {\bibfnamefont {E.~M.}\ \bibnamefont {{Sion}}}, \bibinfo {author} {\bibfnamefont {P.}~\bibnamefont {{Szkody}}}, \bibinfo {author} {\bibfnamefont {D.}~\bibnamefont {{Townsley}}}, \bibinfo {author} {\bibfnamefont {L.}~\bibnamefont {{Bildsten}}}, \bibinfo {author} {\bibfnamefont {D.}~\bibnamefont {{Boyd}}}, \bibinfo {author} {\bibfnamefont {M.~J.}\ \bibnamefont {{Cook}}}, \bibinfo {author} {\bibfnamefont {D.}~\bibnamefont {{De Martino}}},
  \bibinfo {author} {\bibfnamefont {P.}~\bibnamefont {{Godon}}}, \bibinfo {author} {\bibfnamefont {S.}~\bibnamefont {{Kafka}}}, \bibinfo {author} {\bibfnamefont {V.}~\bibnamefont {{Kouprianov}}}, \bibinfo {author} {\bibfnamefont {K.~S.}\ \bibnamefont {{Long}}}, \bibinfo {author} {\bibfnamefont {B.}~\bibnamefont {{Monard}}}, \bibinfo {author} {\bibfnamefont {G.}~\bibnamefont {{Myers}}}, \bibinfo {author} {\bibfnamefont {P.}~\bibnamefont {{Nelson}}}, \bibinfo {author} {\bibfnamefont {D.}~\bibnamefont {{Nogami}}}, \bibinfo {author} {\bibfnamefont {A.}~\bibnamefont {{Oksanen}}}, \bibinfo {author} {\bibfnamefont {R.}~\bibnamefont {{Pickard}}}, \bibinfo {author} {\bibfnamefont {G.}~\bibnamefont {{Poyner}}}, \bibinfo {author} {\bibfnamefont {D.~E.}\ \bibnamefont {{Reichart}}}, \bibinfo {author} {\bibfnamefont {D.}~\bibnamefont {{Rodriguez Perez}}}, \bibinfo {author} {\bibfnamefont {J.}~\bibnamefont {{Shears}}}, \bibinfo {author} {\bibfnamefont {R.}~\bibnamefont {{Stubbings}}}, \ and\ \bibinfo {author} {\bibfnamefont
  {O.}~\bibnamefont {{Toloza}}},\ }\href {\doibase 10.1093/mnras/stab3449} {\bibfield  {journal} {\bibinfo  {journal} {\mnras}\ }\textbf {\bibinfo {volume} {510}},\ \bibinfo {pages} {6110} (\bibinfo {year} {2022})},\ \Eprint {http://arxiv.org/abs/2111.13706} {arXiv:2111.13706 [astro-ph.SR]} \BibitemShut {NoStop}%
\bibitem [{\citenamefont {{Paczynski}}(1976)}]{1976paczynski}%
  \BibitemOpen
  \bibfield  {author} {\bibinfo {author} {\bibfnamefont {B.}~\bibnamefont {{Paczynski}}},\ }in\ \href@noop {} {\emph {\bibinfo {booktitle} {Structure and Evolution of Close Binary Systems}}},\ \bibinfo {series} {IAU Symposium}, Vol.~\bibinfo {volume} {73},\ \bibinfo {editor} {edited by\ \bibinfo {editor} {\bibfnamefont {P.}~\bibnamefont {{Eggleton}}}, \bibinfo {editor} {\bibfnamefont {S.}~\bibnamefont {{Mitton}}}, \ and\ \bibinfo {editor} {\bibfnamefont {J.}~\bibnamefont {{Whelan}}}}\ (\bibinfo {year} {1976})\ p.~\bibinfo {pages} {75}\BibitemShut {NoStop}%
\bibitem [{\citenamefont {{Ivanova}}\ \emph {et~al.}(2013)\citenamefont {{Ivanova}}, \citenamefont {{Justham}}, \citenamefont {{Chen}}, \citenamefont {{De Marco}}, \citenamefont {{Fryer}}, \citenamefont {{Gaburov}}, \citenamefont {{Ge}}, \citenamefont {{Glebbeek}}, \citenamefont {{Han}}, \citenamefont {{Li}}, \citenamefont {{Lu}}, \citenamefont {{Marsh}}, \citenamefont {{Podsiadlowski}}, \citenamefont {{Potter}}, \citenamefont {{Soker}}, \citenamefont {{Taam}}, \citenamefont {{Tauris}}, \citenamefont {{van den Heuvel}},\ and\ \citenamefont {{Webbink}}}]{2013ivanova}%
  \BibitemOpen
  \bibfield  {author} {\bibinfo {author} {\bibfnamefont {N.}~\bibnamefont {{Ivanova}}}, \bibinfo {author} {\bibfnamefont {S.}~\bibnamefont {{Justham}}}, \bibinfo {author} {\bibfnamefont {X.}~\bibnamefont {{Chen}}}, \bibinfo {author} {\bibfnamefont {O.}~\bibnamefont {{De Marco}}}, \bibinfo {author} {\bibfnamefont {C.~L.}\ \bibnamefont {{Fryer}}}, \bibinfo {author} {\bibfnamefont {E.}~\bibnamefont {{Gaburov}}}, \bibinfo {author} {\bibfnamefont {H.}~\bibnamefont {{Ge}}}, \bibinfo {author} {\bibfnamefont {E.}~\bibnamefont {{Glebbeek}}}, \bibinfo {author} {\bibfnamefont {Z.}~\bibnamefont {{Han}}}, \bibinfo {author} {\bibfnamefont {X.-D.}\ \bibnamefont {{Li}}}, \bibinfo {author} {\bibfnamefont {G.}~\bibnamefont {{Lu}}}, \bibinfo {author} {\bibfnamefont {T.}~\bibnamefont {{Marsh}}}, \bibinfo {author} {\bibfnamefont {P.}~\bibnamefont {{Podsiadlowski}}}, \bibinfo {author} {\bibfnamefont {A.}~\bibnamefont {{Potter}}}, \bibinfo {author} {\bibfnamefont {N.}~\bibnamefont {{Soker}}}, \bibinfo {author} {\bibfnamefont
  {R.}~\bibnamefont {{Taam}}}, \bibinfo {author} {\bibfnamefont {T.~M.}\ \bibnamefont {{Tauris}}}, \bibinfo {author} {\bibfnamefont {E.~P.~J.}\ \bibnamefont {{van den Heuvel}}}, \ and\ \bibinfo {author} {\bibfnamefont {R.~F.}\ \bibnamefont {{Webbink}}},\ }\href {\doibase 10.1007/s00159-013-0059-2} {\bibfield  {journal} {\bibinfo  {journal} {\aapr}\ }\textbf {\bibinfo {volume} {21}},\ \bibinfo {eid} {59} (\bibinfo {year} {2013})},\ \Eprint {http://arxiv.org/abs/1209.4302} {arXiv:1209.4302 [astro-ph.HE]} \BibitemShut {NoStop}%
\bibitem [{\citenamefont {{Weidemann}}(2000)}]{2000weidemann}%
  \BibitemOpen
  \bibfield  {author} {\bibinfo {author} {\bibfnamefont {V.}~\bibnamefont {{Weidemann}}},\ }\href@noop {} {\bibfield  {journal} {\bibinfo  {journal} {\aap}\ }\textbf {\bibinfo {volume} {363}},\ \bibinfo {pages} {647} (\bibinfo {year} {2000})}\BibitemShut {NoStop}%
\bibitem [{\citenamefont {{El-Badry}}\ \emph {et~al.}(2018)\citenamefont {{El-Badry}}, \citenamefont {{Rix}},\ and\ \citenamefont {{Weisz}}}]{2018elbadry}%
  \BibitemOpen
  \bibfield  {author} {\bibinfo {author} {\bibfnamefont {K.}~\bibnamefont {{El-Badry}}}, \bibinfo {author} {\bibfnamefont {H.-W.}\ \bibnamefont {{Rix}}}, \ and\ \bibinfo {author} {\bibfnamefont {D.~R.}\ \bibnamefont {{Weisz}}},\ }\href {\doibase 10.3847/2041-8213/aaca9c} {\bibfield  {journal} {\bibinfo  {journal} {\apjl}\ }\textbf {\bibinfo {volume} {860}},\ \bibinfo {eid} {L17} (\bibinfo {year} {2018})},\ \Eprint {http://arxiv.org/abs/1805.05849} {arXiv:1805.05849 [astro-ph.SR]} \BibitemShut {NoStop}%
\bibitem [{\citenamefont {{Shariat}}\ and\ \citenamefont {{El-Badry}}(2026)}]{2026shariat}%
  \BibitemOpen
  \bibfield  {author} {\bibinfo {author} {\bibfnamefont {C.}~\bibnamefont {{Shariat}}}\ and\ \bibinfo {author} {\bibfnamefont {K.}~\bibnamefont {{El-Badry}}},\ }\href {\doibase 10.1088/1538-3873/ae453b} {\bibfield  {journal} {\bibinfo  {journal} {\pasp}\ }\textbf {\bibinfo {volume} {138}},\ \bibinfo {eid} {034202} (\bibinfo {year} {2026})},\ \Eprint {http://arxiv.org/abs/2601.00439} {arXiv:2601.00439 [astro-ph.SR]} \BibitemShut {NoStop}%
\bibitem [{\citenamefont {{Schwope}}\ \emph {et~al.}(2002)\citenamefont {{Schwope}}, \citenamefont {{Brunner}}, \citenamefont {{Hambaryan}},\ and\ \citenamefont {{Schwarz}}}]{2002schwope}%
  \BibitemOpen
  \bibfield  {author} {\bibinfo {author} {\bibfnamefont {A.~D.}\ \bibnamefont {{Schwope}}}, \bibinfo {author} {\bibfnamefont {H.}~\bibnamefont {{Brunner}}}, \bibinfo {author} {\bibfnamefont {V.}~\bibnamefont {{Hambaryan}}}, \ and\ \bibinfo {author} {\bibfnamefont {R.}~\bibnamefont {{Schwarz}}},\ }in\ \href@noop {} {\emph {\bibinfo {booktitle} {The Physics of Cataclysmic Variables and Related Objects}}},\ \bibinfo {series} {Astronomical Society of the Pacific Conference Series}, Vol.\ \bibinfo {volume} {261},\ \bibinfo {editor} {edited by\ \bibinfo {editor} {\bibfnamefont {B.~T.}\ \bibnamefont {{G{\"a}nsicke}}}, \bibinfo {editor} {\bibfnamefont {K.}~\bibnamefont {{Beuermann}}}, \ and\ \bibinfo {editor} {\bibfnamefont {K.}~\bibnamefont {{Reinsch}}}}\ (\bibinfo {year} {2002})\ p.\ \bibinfo {pages} {102}\BibitemShut {NoStop}%
\bibitem [{\citenamefont {{van Roestel}}\ \emph {et~al.}(2025)\citenamefont {{van Roestel}}, \citenamefont {{Rodriguez}}, \citenamefont {{Szkody}}, \citenamefont {{Brown}}, \citenamefont {{Caiazzo}}, \citenamefont {{Drake}}, \citenamefont {{El-Badry}}, \citenamefont {{Prince}}, \citenamefont {{Rich}}, \citenamefont {{Neill}}, \citenamefont {{Vanderbosch}}, \citenamefont {{Bellm}}, \citenamefont {{Dekany}}, \citenamefont {{Feinstein}}, \citenamefont {{Graham}}, \citenamefont {{Groom}}, \citenamefont {{Helou}}, \citenamefont {{Kulkarni}}, \citenamefont {{du Laz}}, \citenamefont {{Mahabal}}, \citenamefont {{Sharma}}, \citenamefont {{Sollerman}},\ and\ \citenamefont {{Wold}}}]{2025vanroestel}%
  \BibitemOpen
  \bibfield  {author} {\bibinfo {author} {\bibfnamefont {J.}~\bibnamefont {{van Roestel}}}, \bibinfo {author} {\bibfnamefont {A.~C.}\ \bibnamefont {{Rodriguez}}}, \bibinfo {author} {\bibfnamefont {P.}~\bibnamefont {{Szkody}}}, \bibinfo {author} {\bibfnamefont {A.~J.}\ \bibnamefont {{Brown}}}, \bibinfo {author} {\bibfnamefont {I.}~\bibnamefont {{Caiazzo}}}, \bibinfo {author} {\bibfnamefont {A.}~\bibnamefont {{Drake}}}, \bibinfo {author} {\bibfnamefont {K.}~\bibnamefont {{El-Badry}}}, \bibinfo {author} {\bibfnamefont {T.}~\bibnamefont {{Prince}}}, \bibinfo {author} {\bibfnamefont {R.~M.~R.}\ \bibnamefont {{Rich}}}, \bibinfo {author} {\bibfnamefont {J.~D.}\ \bibnamefont {{Neill}}}, \bibinfo {author} {\bibfnamefont {Z.}~\bibnamefont {{Vanderbosch}}}, \bibinfo {author} {\bibfnamefont {E.~C.}\ \bibnamefont {{Bellm}}}, \bibinfo {author} {\bibfnamefont {R.}~\bibnamefont {{Dekany}}}, \bibinfo {author} {\bibfnamefont {F.}~\bibnamefont {{Feinstein}}}, \bibinfo {author} {\bibfnamefont {M.}~\bibnamefont {{Graham}}},
  \bibinfo {author} {\bibfnamefont {S.~L.}\ \bibnamefont {{Groom}}}, \bibinfo {author} {\bibfnamefont {G.}~\bibnamefont {{Helou}}}, \bibinfo {author} {\bibfnamefont {S.~R.}\ \bibnamefont {{Kulkarni}}}, \bibinfo {author} {\bibfnamefont {T.}~\bibnamefont {{du Laz}}}, \bibinfo {author} {\bibfnamefont {A.}~\bibnamefont {{Mahabal}}}, \bibinfo {author} {\bibfnamefont {Y.}~\bibnamefont {{Sharma}}}, \bibinfo {author} {\bibfnamefont {J.}~\bibnamefont {{Sollerman}}}, \ and\ \bibinfo {author} {\bibfnamefont {A.}~\bibnamefont {{Wold}}},\ }\href {\doibase 10.1051/0004-6361/202451945} {\bibfield  {journal} {\bibinfo  {journal} {\aap}\ }\textbf {\bibinfo {volume} {696}},\ \bibinfo {eid} {A242} (\bibinfo {year} {2025})},\ \Eprint {http://arxiv.org/abs/2412.15153} {arXiv:2412.15153 [astro-ph.SR]} \BibitemShut {NoStop}%
\bibitem [{\citenamefont {{Paxton}}\ \emph {et~al.}(2011)\citenamefont {{Paxton}}, \citenamefont {{Bildsten}}, \citenamefont {{Dotter}}, \citenamefont {{Herwig}}, \citenamefont {{Lesaffre}},\ and\ \citenamefont {{Timmes}}}]{2011mesa}%
  \BibitemOpen
  \bibfield  {author} {\bibinfo {author} {\bibfnamefont {B.}~\bibnamefont {{Paxton}}}, \bibinfo {author} {\bibfnamefont {L.}~\bibnamefont {{Bildsten}}}, \bibinfo {author} {\bibfnamefont {A.}~\bibnamefont {{Dotter}}}, \bibinfo {author} {\bibfnamefont {F.}~\bibnamefont {{Herwig}}}, \bibinfo {author} {\bibfnamefont {P.}~\bibnamefont {{Lesaffre}}}, \ and\ \bibinfo {author} {\bibfnamefont {F.}~\bibnamefont {{Timmes}}},\ }\href {\doibase 10.1088/0067-0049/192/1/3} {\bibfield  {journal} {\bibinfo  {journal} {\apjs}\ }\textbf {\bibinfo {volume} {192}},\ \bibinfo {eid} {3} (\bibinfo {year} {2011})},\ \Eprint {http://arxiv.org/abs/1009.1622} {arXiv:1009.1622 [astro-ph.SR]} \BibitemShut {NoStop}%
\bibitem [{\citenamefont {{Paxton}}\ \emph {et~al.}(2013)\citenamefont {{Paxton}}, \citenamefont {{Cantiello}}, \citenamefont {{Arras}}, \citenamefont {{Bildsten}}, \citenamefont {{Brown}}, \citenamefont {{Dotter}}, \citenamefont {{Mankovich}}, \citenamefont {{Montgomery}}, \citenamefont {{Stello}}, \citenamefont {{Timmes}},\ and\ \citenamefont {{Townsend}}}]{2013mesa}%
  \BibitemOpen
  \bibfield  {author} {\bibinfo {author} {\bibfnamefont {B.}~\bibnamefont {{Paxton}}}, \bibinfo {author} {\bibfnamefont {M.}~\bibnamefont {{Cantiello}}}, \bibinfo {author} {\bibfnamefont {P.}~\bibnamefont {{Arras}}}, \bibinfo {author} {\bibfnamefont {L.}~\bibnamefont {{Bildsten}}}, \bibinfo {author} {\bibfnamefont {E.~F.}\ \bibnamefont {{Brown}}}, \bibinfo {author} {\bibfnamefont {A.}~\bibnamefont {{Dotter}}}, \bibinfo {author} {\bibfnamefont {C.}~\bibnamefont {{Mankovich}}}, \bibinfo {author} {\bibfnamefont {M.~H.}\ \bibnamefont {{Montgomery}}}, \bibinfo {author} {\bibfnamefont {D.}~\bibnamefont {{Stello}}}, \bibinfo {author} {\bibfnamefont {F.~X.}\ \bibnamefont {{Timmes}}}, \ and\ \bibinfo {author} {\bibfnamefont {R.}~\bibnamefont {{Townsend}}},\ }\href {\doibase 10.1088/0067-0049/208/1/4} {\bibfield  {journal} {\bibinfo  {journal} {\apjs}\ }\textbf {\bibinfo {volume} {208}},\ \bibinfo {eid} {4} (\bibinfo {year} {2013})},\ \Eprint {http://arxiv.org/abs/1301.0319} {arXiv:1301.0319 [astro-ph.SR]} \BibitemShut
  {NoStop}%
\bibitem [{\citenamefont {{Paxton}}\ \emph {et~al.}(2015)\citenamefont {{Paxton}}, \citenamefont {{Marchant}}, \citenamefont {{Schwab}}, \citenamefont {{Bauer}}, \citenamefont {{Bildsten}}, \citenamefont {{Cantiello}}, \citenamefont {{Dessart}}, \citenamefont {{Farmer}}, \citenamefont {{Hu}}, \citenamefont {{Langer}}, \citenamefont {{Townsend}}, \citenamefont {{Townsley}},\ and\ \citenamefont {{Timmes}}}]{2015mesa}%
  \BibitemOpen
  \bibfield  {author} {\bibinfo {author} {\bibfnamefont {B.}~\bibnamefont {{Paxton}}}, \bibinfo {author} {\bibfnamefont {P.}~\bibnamefont {{Marchant}}}, \bibinfo {author} {\bibfnamefont {J.}~\bibnamefont {{Schwab}}}, \bibinfo {author} {\bibfnamefont {E.~B.}\ \bibnamefont {{Bauer}}}, \bibinfo {author} {\bibfnamefont {L.}~\bibnamefont {{Bildsten}}}, \bibinfo {author} {\bibfnamefont {M.}~\bibnamefont {{Cantiello}}}, \bibinfo {author} {\bibfnamefont {L.}~\bibnamefont {{Dessart}}}, \bibinfo {author} {\bibfnamefont {R.}~\bibnamefont {{Farmer}}}, \bibinfo {author} {\bibfnamefont {H.}~\bibnamefont {{Hu}}}, \bibinfo {author} {\bibfnamefont {N.}~\bibnamefont {{Langer}}}, \bibinfo {author} {\bibfnamefont {R.~H.~D.}\ \bibnamefont {{Townsend}}}, \bibinfo {author} {\bibfnamefont {D.~M.}\ \bibnamefont {{Townsley}}}, \ and\ \bibinfo {author} {\bibfnamefont {F.~X.}\ \bibnamefont {{Timmes}}},\ }\href {\doibase 10.1088/0067-0049/220/1/15} {\bibfield  {journal} {\bibinfo  {journal} {\apjs}\ }\textbf {\bibinfo {volume} {220}},\
  \bibinfo {eid} {15} (\bibinfo {year} {2015})},\ \Eprint {http://arxiv.org/abs/1506.03146} {arXiv:1506.03146 [astro-ph.SR]} \BibitemShut {NoStop}%
\bibitem [{\citenamefont {{Paxton}}\ \emph {et~al.}(2018)\citenamefont {{Paxton}}, \citenamefont {{Schwab}}, \citenamefont {{Bauer}}, \citenamefont {{Bildsten}}, \citenamefont {{Blinnikov}}, \citenamefont {{Duffell}}, \citenamefont {{Farmer}}, \citenamefont {{Goldberg}}, \citenamefont {{Marchant}}, \citenamefont {{Sorokina}}, \citenamefont {{Thoul}}, \citenamefont {{Townsend}},\ and\ \citenamefont {{Timmes}}}]{2018mesa}%
  \BibitemOpen
  \bibfield  {author} {\bibinfo {author} {\bibfnamefont {B.}~\bibnamefont {{Paxton}}}, \bibinfo {author} {\bibfnamefont {J.}~\bibnamefont {{Schwab}}}, \bibinfo {author} {\bibfnamefont {E.~B.}\ \bibnamefont {{Bauer}}}, \bibinfo {author} {\bibfnamefont {L.}~\bibnamefont {{Bildsten}}}, \bibinfo {author} {\bibfnamefont {S.}~\bibnamefont {{Blinnikov}}}, \bibinfo {author} {\bibfnamefont {P.}~\bibnamefont {{Duffell}}}, \bibinfo {author} {\bibfnamefont {R.}~\bibnamefont {{Farmer}}}, \bibinfo {author} {\bibfnamefont {J.~A.}\ \bibnamefont {{Goldberg}}}, \bibinfo {author} {\bibfnamefont {P.}~\bibnamefont {{Marchant}}}, \bibinfo {author} {\bibfnamefont {E.}~\bibnamefont {{Sorokina}}}, \bibinfo {author} {\bibfnamefont {A.}~\bibnamefont {{Thoul}}}, \bibinfo {author} {\bibfnamefont {R.~H.~D.}\ \bibnamefont {{Townsend}}}, \ and\ \bibinfo {author} {\bibfnamefont {F.~X.}\ \bibnamefont {{Timmes}}},\ }\href {\doibase 10.3847/1538-4365/aaa5a8} {\bibfield  {journal} {\bibinfo  {journal} {\apjs}\ }\textbf {\bibinfo {volume}
  {234}},\ \bibinfo {eid} {34} (\bibinfo {year} {2018})},\ \Eprint {http://arxiv.org/abs/1710.08424} {arXiv:1710.08424 [astro-ph.SR]} \BibitemShut {NoStop}%
\bibitem [{\citenamefont {{El-Badry}}\ \emph {et~al.}(2021{\natexlab{a}})\citenamefont {{El-Badry}}, \citenamefont {{Quataert}}, \citenamefont {{Rix}}, \citenamefont {{Weisz}}, \citenamefont {{Kupfer}}, \citenamefont {{Shen}}, \citenamefont {{Xiang}}, \citenamefont {{Yang}},\ and\ \citenamefont {{Liu}}}]{2021el-badry}%
  \BibitemOpen
  \bibfield  {author} {\bibinfo {author} {\bibfnamefont {K.}~\bibnamefont {{El-Badry}}}, \bibinfo {author} {\bibfnamefont {E.}~\bibnamefont {{Quataert}}}, \bibinfo {author} {\bibfnamefont {H.-W.}\ \bibnamefont {{Rix}}}, \bibinfo {author} {\bibfnamefont {D.~R.}\ \bibnamefont {{Weisz}}}, \bibinfo {author} {\bibfnamefont {T.}~\bibnamefont {{Kupfer}}}, \bibinfo {author} {\bibfnamefont {K.~J.}\ \bibnamefont {{Shen}}}, \bibinfo {author} {\bibfnamefont {M.}~\bibnamefont {{Xiang}}}, \bibinfo {author} {\bibfnamefont {Y.}~\bibnamefont {{Yang}}}, \ and\ \bibinfo {author} {\bibfnamefont {X.}~\bibnamefont {{Liu}}},\ }\href {\doibase 10.1093/mnras/stab1318} {\bibfield  {journal} {\bibinfo  {journal} {\mnras}\ }\textbf {\bibinfo {volume} {505}},\ \bibinfo {pages} {2051} (\bibinfo {year} {2021}{\natexlab{a}})},\ \Eprint {http://arxiv.org/abs/2104.07033} {arXiv:2104.07033 [astro-ph.SR]} \BibitemShut {NoStop}%
\bibitem [{\citenamefont {{Rappaport}}\ \emph {et~al.}(1983)\citenamefont {{Rappaport}}, \citenamefont {{Verbunt}},\ and\ \citenamefont {{Joss}}}]{1983rappaport}%
  \BibitemOpen
  \bibfield  {author} {\bibinfo {author} {\bibfnamefont {S.}~\bibnamefont {{Rappaport}}}, \bibinfo {author} {\bibfnamefont {F.}~\bibnamefont {{Verbunt}}}, \ and\ \bibinfo {author} {\bibfnamefont {P.~C.}\ \bibnamefont {{Joss}}},\ }\href {\doibase 10.1086/161569} {\bibfield  {journal} {\bibinfo  {journal} {\apj}\ }\textbf {\bibinfo {volume} {275}},\ \bibinfo {pages} {713} (\bibinfo {year} {1983})}\BibitemShut {NoStop}%
\bibitem [{\citenamefont {{Knigge}}\ \emph {et~al.}(2011)\citenamefont {{Knigge}}, \citenamefont {{Baraffe}},\ and\ \citenamefont {{Patterson}}}]{2011knigge}%
  \BibitemOpen
  \bibfield  {author} {\bibinfo {author} {\bibfnamefont {C.}~\bibnamefont {{Knigge}}}, \bibinfo {author} {\bibfnamefont {I.}~\bibnamefont {{Baraffe}}}, \ and\ \bibinfo {author} {\bibfnamefont {J.}~\bibnamefont {{Patterson}}},\ }\href {\doibase 10.1088/0067-0049/194/2/28} {\bibfield  {journal} {\bibinfo  {journal} {\apjs}\ }\textbf {\bibinfo {volume} {194}},\ \bibinfo {eid} {28} (\bibinfo {year} {2011})},\ \Eprint {http://arxiv.org/abs/1102.2440} {arXiv:1102.2440 [astro-ph.SR]} \BibitemShut {NoStop}%
\bibitem [{\citenamefont {{El-Badry}}\ \emph {et~al.}(2021{\natexlab{b}})\citenamefont {{El-Badry}}, \citenamefont {{Rix}}, \citenamefont {{Quataert}}, \citenamefont {{Kupfer}},\ and\ \citenamefont {{Shen}}}]{2021elbadry_survey}%
  \BibitemOpen
  \bibfield  {author} {\bibinfo {author} {\bibfnamefont {K.}~\bibnamefont {{El-Badry}}}, \bibinfo {author} {\bibfnamefont {H.-W.}\ \bibnamefont {{Rix}}}, \bibinfo {author} {\bibfnamefont {E.}~\bibnamefont {{Quataert}}}, \bibinfo {author} {\bibfnamefont {T.}~\bibnamefont {{Kupfer}}}, \ and\ \bibinfo {author} {\bibfnamefont {K.~J.}\ \bibnamefont {{Shen}}},\ }\href {\doibase 10.1093/mnras/stab2583} {\bibfield  {journal} {\bibinfo  {journal} {\mnras}\ }\textbf {\bibinfo {volume} {508}},\ \bibinfo {pages} {4106} (\bibinfo {year} {2021}{\natexlab{b}})},\ \Eprint {http://arxiv.org/abs/2108.04255} {arXiv:2108.04255 [astro-ph.SR]} \BibitemShut {NoStop}%
\bibitem [{\citenamefont {{Rebassa-Mansergas}}\ \emph {et~al.}(2021)\citenamefont {{Rebassa-Mansergas}}, \citenamefont {{Solano}}, \citenamefont {{Jim{\'e}nez-Esteban}}, \citenamefont {{Torres}}, \citenamefont {{Rodrigo}}, \citenamefont {{Ferrer-Burjachs}}, \citenamefont {{Calcaferro}}, \citenamefont {{Althaus}},\ and\ \citenamefont {{C{\'o}rsico}}}]{2021rebassa}%
  \BibitemOpen
  \bibfield  {author} {\bibinfo {author} {\bibfnamefont {A.}~\bibnamefont {{Rebassa-Mansergas}}}, \bibinfo {author} {\bibfnamefont {E.}~\bibnamefont {{Solano}}}, \bibinfo {author} {\bibfnamefont {F.~M.}\ \bibnamefont {{Jim{\'e}nez-Esteban}}}, \bibinfo {author} {\bibfnamefont {S.}~\bibnamefont {{Torres}}}, \bibinfo {author} {\bibfnamefont {C.}~\bibnamefont {{Rodrigo}}}, \bibinfo {author} {\bibfnamefont {A.}~\bibnamefont {{Ferrer-Burjachs}}}, \bibinfo {author} {\bibfnamefont {L.~M.}\ \bibnamefont {{Calcaferro}}}, \bibinfo {author} {\bibfnamefont {L.~G.}\ \bibnamefont {{Althaus}}}, \ and\ \bibinfo {author} {\bibfnamefont {A.~H.}\ \bibnamefont {{C{\'o}rsico}}},\ }\href {\doibase 10.1093/mnras/stab2039} {\bibfield  {journal} {\bibinfo  {journal} {\mnras}\ }\textbf {\bibinfo {volume} {506}},\ \bibinfo {pages} {5201} (\bibinfo {year} {2021})},\ \Eprint {http://arxiv.org/abs/2107.06303} {arXiv:2107.06303 [astro-ph.SR]} \BibitemShut {NoStop}%
\bibitem [{\citenamefont {{Ivezic}}\ \emph {et~al.}(2019)\citenamefont {{Ivezic}}, \citenamefont {{Kahn}}, \citenamefont {{Tyson}}, \citenamefont {{Abel}}, \citenamefont {{Acosta}}, \citenamefont {{Allsman}}, \citenamefont {{Alonso}}, \citenamefont {{AlSayyad}}, \citenamefont {{Anderson}}, \citenamefont {{Andrew}}, \citenamefont {{Angel}}, \citenamefont {{Angeli}}, \citenamefont {{Ansari}}, \citenamefont {{Antilogus}}, \citenamefont {{Araujo}}, \citenamefont {{Armstrong}}, \citenamefont {{Arndt}}, \citenamefont {{Astier}}, \citenamefont {{Aubourg}}, \citenamefont {{Auza}}, \citenamefont {{Axelrod}}, \citenamefont {{Bard}}, \citenamefont {{Barr}}, \citenamefont {{Barrau}}, \citenamefont {{Bartlett}}, \citenamefont {{Bauer}}, \citenamefont {{Bauman}}, \citenamefont {{Baumont}}, \citenamefont {{Bechtol}}, \citenamefont {{Bechtol}}, \citenamefont {{Becker}}, \citenamefont {{Becla}}, \citenamefont {{Beldica}}, \citenamefont {{Bellavia}}, \citenamefont {{Bianco}}, \citenamefont {{Biswas}}, \citenamefont {{Blanc}},
  \citenamefont {{Blazek}}, \citenamefont {{Blandford}}, \citenamefont {{Bloom}}, \citenamefont {{Bogart}}, \citenamefont {{Bond}}, \citenamefont {{Booth}}, \citenamefont {{Borgland}}, \citenamefont {{Borne}}, \citenamefont {{Bosch}}, \citenamefont {{Boutigny}}, \citenamefont {{Brackett}}, \citenamefont {{Bradshaw}}, \citenamefont {{Brandt}}, \citenamefont {{Brown}}, \citenamefont {{Bullock}}, \citenamefont {{Burchat}}, \citenamefont {{Burke}}, \citenamefont {{Cagnoli}}, \citenamefont {{Calabrese}}, \citenamefont {{Callahan}}, \citenamefont {{Callen}}, \citenamefont {{Carlin}}, \citenamefont {{Carlson}}, \citenamefont {{Chandrasekharan}}, \citenamefont {{Charles-Emerson}}, \citenamefont {{Chesley}}, \citenamefont {{Cheu}}, \citenamefont {{Chiang}}, \citenamefont {{Chiang}}, \citenamefont {{Chirino}}, \citenamefont {{Chow}}, \citenamefont {{Ciardi}}, \citenamefont {{Claver}}, \citenamefont {{Cohen-Tanugi}}, \citenamefont {{Cockrum}}, \citenamefont {{Coles}}, \citenamefont {{Connolly}}, \citenamefont {{Cook}},
  \citenamefont {{Cooray}}, \citenamefont {{Covey}}, \citenamefont {{Cribbs}}, \citenamefont {{Cui}}, \citenamefont {{Cutri}}, \citenamefont {{Daly}}, \citenamefont {{Daniel}}, \citenamefont {{Daruich}}, \citenamefont {{Daubard}}, \citenamefont {{Daues}}, \citenamefont {{Dawson}}, \citenamefont {{Delgado}}, \citenamefont {{Dellapenna}}, \citenamefont {{de Peyster}}, \citenamefont {{de Val-Borro}}, \citenamefont {{Digel}}, \citenamefont {{Doherty}}, \citenamefont {{Dubois}}, \citenamefont {{Dubois-Felsmann}}, \citenamefont {{Durech}}, \citenamefont {{Economou}}, \citenamefont {{Eifler}}, \citenamefont {{Eracleous}}, \citenamefont {{Emmons}}, \citenamefont {{Fausti Neto}}, \citenamefont {{Ferguson}}, \citenamefont {{Figueroa}}, \citenamefont {{Fisher-Levine}}, \citenamefont {{Focke}}, \citenamefont {{Foss}}, \citenamefont {{Frank}}, \citenamefont {{Freemon}}, \citenamefont {{Gangler}}, \citenamefont {{Gawiser}}, \citenamefont {{Geary}}, \citenamefont {{Gee}}, \citenamefont {{Geha}}, \citenamefont {{Gessner}},
  \citenamefont {{Gibson}}, \citenamefont {{Gilmore}}, \citenamefont {{Glanzman}}, \citenamefont {{Glick}}, \citenamefont {{Goldina}}, \citenamefont {{Goldstein}}, \citenamefont {{Goodenow}}, \citenamefont {{Graham}}, \citenamefont {{Gressler}}, \citenamefont {{Gris}}, \citenamefont {{Guy}}, \citenamefont {{Guyonnet}}, \citenamefont {{Haller}}, \citenamefont {{Harris}}, \citenamefont {{Hascall}}, \citenamefont {{Haupt}}, \citenamefont {{Hernandez}}, \citenamefont {{Herrmann}}, \citenamefont {{Hileman}}, \citenamefont {{Hoblitt}}, \citenamefont {{Hodgson}}, \citenamefont {{Hogan}}, \citenamefont {{Howard}}, \citenamefont {{Huang}}, \citenamefont {{Huffer}}, \citenamefont {{Ingraham}}, \citenamefont {{Innes}}, \citenamefont {{Jacoby}}, \citenamefont {{Jain}}, \citenamefont {{Jammes}}, \citenamefont {{Jee}}, \citenamefont {{Jenness}}, \citenamefont {{Jernigan}}, \citenamefont {{Jevremovi{\'c}}}, \citenamefont {{Johns}}, \citenamefont {{Johnson}}, \citenamefont {{Johnson}}, \citenamefont {{Jones}}, \citenamefont
  {{Juramy-Gilles}}, \citenamefont {{Juri{\'c}}}, \citenamefont {{Kalirai}}, \citenamefont {{Kallivayalil}}, \citenamefont {{Kalmbach}}, \citenamefont {{Kantor}}, \citenamefont {{Karst}}, \citenamefont {{Kasliwal}}, \citenamefont {{Kelly}}, \citenamefont {{Kessler}}, \citenamefont {{Kinnison}}, \citenamefont {{Kirkby}}, \citenamefont {{Knox}}, \citenamefont {{Kotov}}, \citenamefont {{Krabbendam}}, \citenamefont {{Krughoff}}, \citenamefont {{Kub{\'a}nek}}, \citenamefont {{Kuczewski}}, \citenamefont {{Kulkarni}}, \citenamefont {{Ku}}, \citenamefont {{Kurita}}, \citenamefont {{Lage}}, \citenamefont {{Lambert}}, \citenamefont {{Lange}}, \citenamefont {{Langton}}, \citenamefont {{Le Guillou}}, \citenamefont {{Levine}}, \citenamefont {{Liang}}, \citenamefont {{Lim}}, \citenamefont {{Lintott}}, \citenamefont {{Long}}, \citenamefont {{Lopez}}, \citenamefont {{Lotz}}, \citenamefont {{Lupton}}, \citenamefont {{Lust}}, \citenamefont {{MacArthur}}, \citenamefont {{Mahabal}}, \citenamefont {{Mandelbaum}}, \citenamefont
  {{Markiewicz}}, \citenamefont {{Marsh}}, \citenamefont {{Marshall}}, \citenamefont {{Marshall}}, \citenamefont {{May}}, \citenamefont {{McKercher}}, \citenamefont {{McQueen}}, \citenamefont {{Meyers}}, \citenamefont {{Migliore}}, \citenamefont {{Miller}}, \citenamefont {{Mills}}, \citenamefont {{Miraval}}, \citenamefont {{Moeyens}}, \citenamefont {{Moolekamp}}, \citenamefont {{Monet}}, \citenamefont {{Moniez}}, \citenamefont {{Monkewitz}}, \citenamefont {{Montgomery}}, \citenamefont {{Morrison}}, \citenamefont {{Mueller}}, \citenamefont {{Muller}}, \citenamefont {{Mu{\~n}oz Arancibia}}, \citenamefont {{Neill}}, \citenamefont {{Newbry}}, \citenamefont {{Nief}}, \citenamefont {{Nomerotski}}, \citenamefont {{Nordby}}, \citenamefont {{O'Connor}}, \citenamefont {{Oliver}}, \citenamefont {{Olivier}}, \citenamefont {{Olsen}}, \citenamefont {{O'Mullane}}, \citenamefont {{Ortiz}}, \citenamefont {{Osier}}, \citenamefont {{Owen}}, \citenamefont {{Pain}}, \citenamefont {{Palecek}}, \citenamefont {{Parejko}},
  \citenamefont {{Parsons}}, \citenamefont {{Pease}}, \citenamefont {{Peterson}}, \citenamefont {{Peterson}}, \citenamefont {{Petravick}}, \citenamefont {{Libby Petrick}}, \citenamefont {{Petry}}, \citenamefont {{Pierfederici}}, \citenamefont {{Pietrowicz}}, \citenamefont {{Pike}}, \citenamefont {{Pinto}}, \citenamefont {{Plante}}, \citenamefont {{Plate}}, \citenamefont {{Plutchak}}, \citenamefont {{Price}}, \citenamefont {{Prouza}}, \citenamefont {{Radeka}}, \citenamefont {{Rajagopal}}, \citenamefont {{Rasmussen}}, \citenamefont {{Regnault}}, \citenamefont {{Reil}}, \citenamefont {{Reiss}}, \citenamefont {{Reuter}}, \citenamefont {{Ridgway}}, \citenamefont {{Riot}}, \citenamefont {{Ritz}}, \citenamefont {{Robinson}}, \citenamefont {{Roby}}, \citenamefont {{Roodman}}, \citenamefont {{Rosing}}, \citenamefont {{Roucelle}}, \citenamefont {{Rumore}}, \citenamefont {{Russo}}, \citenamefont {{Saha}}, \citenamefont {{Sassolas}}, \citenamefont {{Schalk}}, \citenamefont {{Schellart}}, \citenamefont {{Schindler}},
  \citenamefont {{Schmidt}}, \citenamefont {{Schneider}}, \citenamefont {{Schneider}}, \citenamefont {{Schoening}}, \citenamefont {{Schumacher}}, \citenamefont {{Schwamb}}, \citenamefont {{Sebag}}, \citenamefont {{Selvy}}, \citenamefont {{Sembroski}}, \citenamefont {{Seppala}}, \citenamefont {{Serio}}, \citenamefont {{Serrano}}, \citenamefont {{Shaw}}, \citenamefont {{Shipsey}}, \citenamefont {{Sick}}, \citenamefont {{Silvestri}}, \citenamefont {{Slater}}, \citenamefont {{Smith}}, \citenamefont {{Smith}}, \citenamefont {{Sobhani}}, \citenamefont {{Soldahl}}, \citenamefont {{Storrie-Lombardi}}, \citenamefont {{Stover}}, \citenamefont {{Strauss}}, \citenamefont {{Street}}, \citenamefont {{Stubbs}}, \citenamefont {{Sullivan}}, \citenamefont {{Sweeney}}, \citenamefont {{Swinbank}}, \citenamefont {{Szalay}}, \citenamefont {{Takacs}}, \citenamefont {{Tether}}, \citenamefont {{Thaler}}, \citenamefont {{Thayer}}, \citenamefont {{Thomas}}, \citenamefont {{Thornton}}, \citenamefont {{Thukral}}, \citenamefont {{Tice}},
  \citenamefont {{Trilling}}, \citenamefont {{Turri}}, \citenamefont {{Van Berg}}, \citenamefont {{Vanden Berk}}, \citenamefont {{Vetter}}, \citenamefont {{Virieux}}, \citenamefont {{Vucina}}, \citenamefont {{Wahl}}, \citenamefont {{Walkowicz}}, \citenamefont {{Walsh}}, \citenamefont {{Walter}}, \citenamefont {{Wang}}, \citenamefont {{Wang}}, \citenamefont {{Warner}}, \citenamefont {{Wiecha}}, \citenamefont {{Willman}}, \citenamefont {{Winters}}, \citenamefont {{Wittman}}, \citenamefont {{Wolff}}, \citenamefont {{Wood-Vasey}}, \citenamefont {{Wu}}, \citenamefont {{Xin}}, \citenamefont {{Yoachim}},\ and\ \citenamefont {{Zhan}}}]{2019rubin}%
  \BibitemOpen
  \bibfield  {author} {\bibinfo {author} {\bibfnamefont {Z.}~\bibnamefont {{Ivezic}}}, \bibinfo {author} {\bibfnamefont {S.~M.}\ \bibnamefont {{Kahn}}}, \bibinfo {author} {\bibfnamefont {J.~A.}\ \bibnamefont {{Tyson}}}, \bibinfo {author} {\bibfnamefont {B.}~\bibnamefont {{Abel}}}, \bibinfo {author} {\bibfnamefont {E.}~\bibnamefont {{Acosta}}}, \bibinfo {author} {\bibfnamefont {R.}~\bibnamefont {{Allsman}}}, \bibinfo {author} {\bibfnamefont {D.}~\bibnamefont {{Alonso}}}, \bibinfo {author} {\bibfnamefont {Y.}~\bibnamefont {{AlSayyad}}}, \bibinfo {author} {\bibfnamefont {S.~F.}\ \bibnamefont {{Anderson}}}, \bibinfo {author} {\bibfnamefont {J.}~\bibnamefont {{Andrew}}}, \bibinfo {author} {\bibfnamefont {J.~R.~P.}\ \bibnamefont {{Angel}}}, \bibinfo {author} {\bibfnamefont {G.~Z.}\ \bibnamefont {{Angeli}}}, \bibinfo {author} {\bibfnamefont {R.}~\bibnamefont {{Ansari}}}, \bibinfo {author} {\bibfnamefont {P.}~\bibnamefont {{Antilogus}}}, \bibinfo {author} {\bibfnamefont {C.}~\bibnamefont {{Araujo}}}, \bibinfo
  {author} {\bibfnamefont {R.}~\bibnamefont {{Armstrong}}}, \bibinfo {author} {\bibfnamefont {K.~T.}\ \bibnamefont {{Arndt}}}, \bibinfo {author} {\bibfnamefont {P.}~\bibnamefont {{Astier}}}, \bibinfo {author} {\bibfnamefont {{\'E}.}~\bibnamefont {{Aubourg}}}, \bibinfo {author} {\bibfnamefont {N.}~\bibnamefont {{Auza}}}, \bibinfo {author} {\bibfnamefont {T.~S.}\ \bibnamefont {{Axelrod}}}, \bibinfo {author} {\bibfnamefont {D.~J.}\ \bibnamefont {{Bard}}}, \bibinfo {author} {\bibfnamefont {J.~D.}\ \bibnamefont {{Barr}}}, \bibinfo {author} {\bibfnamefont {A.}~\bibnamefont {{Barrau}}}, \bibinfo {author} {\bibfnamefont {J.~G.}\ \bibnamefont {{Bartlett}}}, \bibinfo {author} {\bibfnamefont {A.~E.}\ \bibnamefont {{Bauer}}}, \bibinfo {author} {\bibfnamefont {B.~J.}\ \bibnamefont {{Bauman}}}, \bibinfo {author} {\bibfnamefont {S.}~\bibnamefont {{Baumont}}}, \bibinfo {author} {\bibfnamefont {E.}~\bibnamefont {{Bechtol}}}, \bibinfo {author} {\bibfnamefont {K.}~\bibnamefont {{Bechtol}}}, \bibinfo {author} {\bibfnamefont
  {A.~C.}\ \bibnamefont {{Becker}}}, \bibinfo {author} {\bibfnamefont {J.}~\bibnamefont {{Becla}}}, \bibinfo {author} {\bibfnamefont {C.}~\bibnamefont {{Beldica}}}, \bibinfo {author} {\bibfnamefont {S.}~\bibnamefont {{Bellavia}}}, \bibinfo {author} {\bibfnamefont {F.~B.}\ \bibnamefont {{Bianco}}}, \bibinfo {author} {\bibfnamefont {R.}~\bibnamefont {{Biswas}}}, \bibinfo {author} {\bibfnamefont {G.}~\bibnamefont {{Blanc}}}, \bibinfo {author} {\bibfnamefont {J.}~\bibnamefont {{Blazek}}}, \bibinfo {author} {\bibfnamefont {R.~D.}\ \bibnamefont {{Blandford}}}, \bibinfo {author} {\bibfnamefont {J.~S.}\ \bibnamefont {{Bloom}}}, \bibinfo {author} {\bibfnamefont {J.}~\bibnamefont {{Bogart}}}, \bibinfo {author} {\bibfnamefont {T.~W.}\ \bibnamefont {{Bond}}}, \bibinfo {author} {\bibfnamefont {M.~T.}\ \bibnamefont {{Booth}}}, \bibinfo {author} {\bibfnamefont {A.~W.}\ \bibnamefont {{Borgland}}}, \bibinfo {author} {\bibfnamefont {K.}~\bibnamefont {{Borne}}}, \bibinfo {author} {\bibfnamefont {J.~F.}\ \bibnamefont {{Bosch}}},
  \bibinfo {author} {\bibfnamefont {D.}~\bibnamefont {{Boutigny}}}, \bibinfo {author} {\bibfnamefont {C.~A.}\ \bibnamefont {{Brackett}}}, \bibinfo {author} {\bibfnamefont {A.}~\bibnamefont {{Bradshaw}}}, \bibinfo {author} {\bibfnamefont {W.~N.}\ \bibnamefont {{Brandt}}}, \bibinfo {author} {\bibfnamefont {M.~E.}\ \bibnamefont {{Brown}}}, \bibinfo {author} {\bibfnamefont {J.~S.}\ \bibnamefont {{Bullock}}}, \bibinfo {author} {\bibfnamefont {P.}~\bibnamefont {{Burchat}}}, \bibinfo {author} {\bibfnamefont {D.~L.}\ \bibnamefont {{Burke}}}, \bibinfo {author} {\bibfnamefont {G.}~\bibnamefont {{Cagnoli}}}, \bibinfo {author} {\bibfnamefont {D.}~\bibnamefont {{Calabrese}}}, \bibinfo {author} {\bibfnamefont {S.}~\bibnamefont {{Callahan}}}, \bibinfo {author} {\bibfnamefont {A.~L.}\ \bibnamefont {{Callen}}}, \bibinfo {author} {\bibfnamefont {J.~L.}\ \bibnamefont {{Carlin}}}, \bibinfo {author} {\bibfnamefont {E.~L.}\ \bibnamefont {{Carlson}}}, \bibinfo {author} {\bibfnamefont {S.}~\bibnamefont {{Chandrasekharan}}}, \bibinfo
  {author} {\bibfnamefont {G.}~\bibnamefont {{Charles-Emerson}}}, \bibinfo {author} {\bibfnamefont {S.}~\bibnamefont {{Chesley}}}, \bibinfo {author} {\bibfnamefont {E.~C.}\ \bibnamefont {{Cheu}}}, \bibinfo {author} {\bibfnamefont {H.-F.}\ \bibnamefont {{Chiang}}}, \bibinfo {author} {\bibfnamefont {J.}~\bibnamefont {{Chiang}}}, \bibinfo {author} {\bibfnamefont {C.}~\bibnamefont {{Chirino}}}, \bibinfo {author} {\bibfnamefont {D.}~\bibnamefont {{Chow}}}, \bibinfo {author} {\bibfnamefont {D.~R.}\ \bibnamefont {{Ciardi}}}, \bibinfo {author} {\bibfnamefont {C.~F.}\ \bibnamefont {{Claver}}}, \bibinfo {author} {\bibfnamefont {J.}~\bibnamefont {{Cohen-Tanugi}}}, \bibinfo {author} {\bibfnamefont {J.~J.}\ \bibnamefont {{Cockrum}}}, \bibinfo {author} {\bibfnamefont {R.}~\bibnamefont {{Coles}}}, \bibinfo {author} {\bibfnamefont {A.~J.}\ \bibnamefont {{Connolly}}}, \bibinfo {author} {\bibfnamefont {K.~H.}\ \bibnamefont {{Cook}}}, \bibinfo {author} {\bibfnamefont {A.}~\bibnamefont {{Cooray}}}, \bibinfo {author}
  {\bibfnamefont {K.~R.}\ \bibnamefont {{Covey}}}, \bibinfo {author} {\bibfnamefont {C.}~\bibnamefont {{Cribbs}}}, \bibinfo {author} {\bibfnamefont {W.}~\bibnamefont {{Cui}}}, \bibinfo {author} {\bibfnamefont {R.}~\bibnamefont {{Cutri}}}, \bibinfo {author} {\bibfnamefont {P.~N.}\ \bibnamefont {{Daly}}}, \bibinfo {author} {\bibfnamefont {S.~F.}\ \bibnamefont {{Daniel}}}, \bibinfo {author} {\bibfnamefont {F.}~\bibnamefont {{Daruich}}}, \bibinfo {author} {\bibfnamefont {G.}~\bibnamefont {{Daubard}}}, \bibinfo {author} {\bibfnamefont {G.}~\bibnamefont {{Daues}}}, \bibinfo {author} {\bibfnamefont {W.}~\bibnamefont {{Dawson}}}, \bibinfo {author} {\bibfnamefont {F.}~\bibnamefont {{Delgado}}}, \bibinfo {author} {\bibfnamefont {A.}~\bibnamefont {{Dellapenna}}}, \bibinfo {author} {\bibfnamefont {R.}~\bibnamefont {{de Peyster}}}, \bibinfo {author} {\bibfnamefont {M.}~\bibnamefont {{de Val-Borro}}}, \bibinfo {author} {\bibfnamefont {S.~W.}\ \bibnamefont {{Digel}}}, \bibinfo {author} {\bibfnamefont {P.}~\bibnamefont
  {{Doherty}}}, \bibinfo {author} {\bibfnamefont {R.}~\bibnamefont {{Dubois}}}, \bibinfo {author} {\bibfnamefont {G.~P.}\ \bibnamefont {{Dubois-Felsmann}}}, \bibinfo {author} {\bibfnamefont {J.}~\bibnamefont {{Durech}}}, \bibinfo {author} {\bibfnamefont {F.}~\bibnamefont {{Economou}}}, \bibinfo {author} {\bibfnamefont {T.}~\bibnamefont {{Eifler}}}, \bibinfo {author} {\bibfnamefont {M.}~\bibnamefont {{Eracleous}}}, \bibinfo {author} {\bibfnamefont {B.~L.}\ \bibnamefont {{Emmons}}}, \bibinfo {author} {\bibfnamefont {A.}~\bibnamefont {{Fausti Neto}}}, \bibinfo {author} {\bibfnamefont {H.}~\bibnamefont {{Ferguson}}}, \bibinfo {author} {\bibfnamefont {E.}~\bibnamefont {{Figueroa}}}, \bibinfo {author} {\bibfnamefont {M.}~\bibnamefont {{Fisher-Levine}}}, \bibinfo {author} {\bibfnamefont {W.}~\bibnamefont {{Focke}}}, \bibinfo {author} {\bibfnamefont {M.~D.}\ \bibnamefont {{Foss}}}, \bibinfo {author} {\bibfnamefont {J.}~\bibnamefont {{Frank}}}, \bibinfo {author} {\bibfnamefont {M.~D.}\ \bibnamefont {{Freemon}}},
  \bibinfo {author} {\bibfnamefont {E.}~\bibnamefont {{Gangler}}}, \bibinfo {author} {\bibfnamefont {E.}~\bibnamefont {{Gawiser}}}, \bibinfo {author} {\bibfnamefont {J.~C.}\ \bibnamefont {{Geary}}}, \bibinfo {author} {\bibfnamefont {P.}~\bibnamefont {{Gee}}}, \bibinfo {author} {\bibfnamefont {M.}~\bibnamefont {{Geha}}}, \bibinfo {author} {\bibfnamefont {C.~J.~B.}\ \bibnamefont {{Gessner}}}, \bibinfo {author} {\bibfnamefont {R.~R.}\ \bibnamefont {{Gibson}}}, \bibinfo {author} {\bibfnamefont {D.~K.}\ \bibnamefont {{Gilmore}}}, \bibinfo {author} {\bibfnamefont {T.}~\bibnamefont {{Glanzman}}}, \bibinfo {author} {\bibfnamefont {W.}~\bibnamefont {{Glick}}}, \bibinfo {author} {\bibfnamefont {T.}~\bibnamefont {{Goldina}}}, \bibinfo {author} {\bibfnamefont {D.~A.}\ \bibnamefont {{Goldstein}}}, \bibinfo {author} {\bibfnamefont {I.}~\bibnamefont {{Goodenow}}}, \bibinfo {author} {\bibfnamefont {M.~L.}\ \bibnamefont {{Graham}}}, \bibinfo {author} {\bibfnamefont {W.~J.}\ \bibnamefont {{Gressler}}}, \bibinfo {author}
  {\bibfnamefont {P.}~\bibnamefont {{Gris}}}, \bibinfo {author} {\bibfnamefont {L.~P.}\ \bibnamefont {{Guy}}}, \bibinfo {author} {\bibfnamefont {A.}~\bibnamefont {{Guyonnet}}}, \bibinfo {author} {\bibfnamefont {G.}~\bibnamefont {{Haller}}}, \bibinfo {author} {\bibfnamefont {R.}~\bibnamefont {{Harris}}}, \bibinfo {author} {\bibfnamefont {P.~A.}\ \bibnamefont {{Hascall}}}, \bibinfo {author} {\bibfnamefont {J.}~\bibnamefont {{Haupt}}}, \bibinfo {author} {\bibfnamefont {F.}~\bibnamefont {{Hernandez}}}, \bibinfo {author} {\bibfnamefont {S.}~\bibnamefont {{Herrmann}}}, \bibinfo {author} {\bibfnamefont {E.}~\bibnamefont {{Hileman}}}, \bibinfo {author} {\bibfnamefont {J.}~\bibnamefont {{Hoblitt}}}, \bibinfo {author} {\bibfnamefont {J.~A.}\ \bibnamefont {{Hodgson}}}, \bibinfo {author} {\bibfnamefont {C.}~\bibnamefont {{Hogan}}}, \bibinfo {author} {\bibfnamefont {J.~D.}\ \bibnamefont {{Howard}}}, \bibinfo {author} {\bibfnamefont {D.}~\bibnamefont {{Huang}}}, \bibinfo {author} {\bibfnamefont {M.~E.}\ \bibnamefont
  {{Huffer}}}, \bibinfo {author} {\bibfnamefont {P.}~\bibnamefont {{Ingraham}}}, \bibinfo {author} {\bibfnamefont {W.~R.}\ \bibnamefont {{Innes}}}, \bibinfo {author} {\bibfnamefont {S.~H.}\ \bibnamefont {{Jacoby}}}, \bibinfo {author} {\bibfnamefont {B.}~\bibnamefont {{Jain}}}, \bibinfo {author} {\bibfnamefont {F.}~\bibnamefont {{Jammes}}}, \bibinfo {author} {\bibfnamefont {M.~J.}\ \bibnamefont {{Jee}}}, \bibinfo {author} {\bibfnamefont {T.}~\bibnamefont {{Jenness}}}, \bibinfo {author} {\bibfnamefont {G.}~\bibnamefont {{Jernigan}}}, \bibinfo {author} {\bibfnamefont {D.}~\bibnamefont {{Jevremovi{\'c}}}}, \bibinfo {author} {\bibfnamefont {K.}~\bibnamefont {{Johns}}}, \bibinfo {author} {\bibfnamefont {A.~S.}\ \bibnamefont {{Johnson}}}, \bibinfo {author} {\bibfnamefont {M.~W.~G.}\ \bibnamefont {{Johnson}}}, \bibinfo {author} {\bibfnamefont {R.~L.}\ \bibnamefont {{Jones}}}, \bibinfo {author} {\bibfnamefont {C.}~\bibnamefont {{Juramy-Gilles}}}, \bibinfo {author} {\bibfnamefont {M.}~\bibnamefont {{Juri{\'c}}}},
  \bibinfo {author} {\bibfnamefont {J.~S.}\ \bibnamefont {{Kalirai}}}, \bibinfo {author} {\bibfnamefont {N.~J.}\ \bibnamefont {{Kallivayalil}}}, \bibinfo {author} {\bibfnamefont {B.}~\bibnamefont {{Kalmbach}}}, \bibinfo {author} {\bibfnamefont {J.~P.}\ \bibnamefont {{Kantor}}}, \bibinfo {author} {\bibfnamefont {P.}~\bibnamefont {{Karst}}}, \bibinfo {author} {\bibfnamefont {M.~M.}\ \bibnamefont {{Kasliwal}}}, \bibinfo {author} {\bibfnamefont {H.}~\bibnamefont {{Kelly}}}, \bibinfo {author} {\bibfnamefont {R.}~\bibnamefont {{Kessler}}}, \bibinfo {author} {\bibfnamefont {V.}~\bibnamefont {{Kinnison}}}, \bibinfo {author} {\bibfnamefont {D.}~\bibnamefont {{Kirkby}}}, \bibinfo {author} {\bibfnamefont {L.}~\bibnamefont {{Knox}}}, \bibinfo {author} {\bibfnamefont {I.~V.}\ \bibnamefont {{Kotov}}}, \bibinfo {author} {\bibfnamefont {V.~L.}\ \bibnamefont {{Krabbendam}}}, \bibinfo {author} {\bibfnamefont {K.~S.}\ \bibnamefont {{Krughoff}}}, \bibinfo {author} {\bibfnamefont {P.}~\bibnamefont {{Kub{\'a}nek}}}, \bibinfo
  {author} {\bibfnamefont {J.}~\bibnamefont {{Kuczewski}}}, \bibinfo {author} {\bibfnamefont {S.}~\bibnamefont {{Kulkarni}}}, \bibinfo {author} {\bibfnamefont {J.}~\bibnamefont {{Ku}}}, \bibinfo {author} {\bibfnamefont {N.~R.}\ \bibnamefont {{Kurita}}}, \bibinfo {author} {\bibfnamefont {C.~S.}\ \bibnamefont {{Lage}}}, \bibinfo {author} {\bibfnamefont {R.}~\bibnamefont {{Lambert}}}, \bibinfo {author} {\bibfnamefont {T.}~\bibnamefont {{Lange}}}, \bibinfo {author} {\bibfnamefont {J.~B.}\ \bibnamefont {{Langton}}}, \bibinfo {author} {\bibfnamefont {L.}~\bibnamefont {{Le Guillou}}}, \bibinfo {author} {\bibfnamefont {D.}~\bibnamefont {{Levine}}}, \bibinfo {author} {\bibfnamefont {M.}~\bibnamefont {{Liang}}}, \bibinfo {author} {\bibfnamefont {K.-T.}\ \bibnamefont {{Lim}}}, \bibinfo {author} {\bibfnamefont {C.~J.}\ \bibnamefont {{Lintott}}}, \bibinfo {author} {\bibfnamefont {K.~E.}\ \bibnamefont {{Long}}}, \bibinfo {author} {\bibfnamefont {M.}~\bibnamefont {{Lopez}}}, \bibinfo {author} {\bibfnamefont {P.~J.}\
  \bibnamefont {{Lotz}}}, \bibinfo {author} {\bibfnamefont {R.~H.}\ \bibnamefont {{Lupton}}}, \bibinfo {author} {\bibfnamefont {N.~B.}\ \bibnamefont {{Lust}}}, \bibinfo {author} {\bibfnamefont {L.~A.}\ \bibnamefont {{MacArthur}}}, \bibinfo {author} {\bibfnamefont {A.}~\bibnamefont {{Mahabal}}}, \bibinfo {author} {\bibfnamefont {R.}~\bibnamefont {{Mandelbaum}}}, \bibinfo {author} {\bibfnamefont {T.~W.}\ \bibnamefont {{Markiewicz}}}, \bibinfo {author} {\bibfnamefont {D.~S.}\ \bibnamefont {{Marsh}}}, \bibinfo {author} {\bibfnamefont {P.~J.}\ \bibnamefont {{Marshall}}}, \bibinfo {author} {\bibfnamefont {S.}~\bibnamefont {{Marshall}}}, \bibinfo {author} {\bibfnamefont {M.}~\bibnamefont {{May}}}, \bibinfo {author} {\bibfnamefont {R.}~\bibnamefont {{McKercher}}}, \bibinfo {author} {\bibfnamefont {M.}~\bibnamefont {{McQueen}}}, \bibinfo {author} {\bibfnamefont {J.}~\bibnamefont {{Meyers}}}, \bibinfo {author} {\bibfnamefont {M.}~\bibnamefont {{Migliore}}}, \bibinfo {author} {\bibfnamefont {M.}~\bibnamefont
  {{Miller}}}, \bibinfo {author} {\bibfnamefont {D.~J.}\ \bibnamefont {{Mills}}}, \bibinfo {author} {\bibfnamefont {C.}~\bibnamefont {{Miraval}}}, \bibinfo {author} {\bibfnamefont {J.}~\bibnamefont {{Moeyens}}}, \bibinfo {author} {\bibfnamefont {F.~E.}\ \bibnamefont {{Moolekamp}}}, \bibinfo {author} {\bibfnamefont {D.~G.}\ \bibnamefont {{Monet}}}, \bibinfo {author} {\bibfnamefont {M.}~\bibnamefont {{Moniez}}}, \bibinfo {author} {\bibfnamefont {S.}~\bibnamefont {{Monkewitz}}}, \bibinfo {author} {\bibfnamefont {C.}~\bibnamefont {{Montgomery}}}, \bibinfo {author} {\bibfnamefont {C.~B.}\ \bibnamefont {{Morrison}}}, \bibinfo {author} {\bibfnamefont {F.}~\bibnamefont {{Mueller}}}, \bibinfo {author} {\bibfnamefont {G.~P.}\ \bibnamefont {{Muller}}}, \bibinfo {author} {\bibfnamefont {F.}~\bibnamefont {{Mu{\~n}oz Arancibia}}}, \bibinfo {author} {\bibfnamefont {D.~R.}\ \bibnamefont {{Neill}}}, \bibinfo {author} {\bibfnamefont {S.~P.}\ \bibnamefont {{Newbry}}}, \bibinfo {author} {\bibfnamefont {J.-Y.}\ \bibnamefont
  {{Nief}}}, \bibinfo {author} {\bibfnamefont {A.}~\bibnamefont {{Nomerotski}}}, \bibinfo {author} {\bibfnamefont {M.}~\bibnamefont {{Nordby}}}, \bibinfo {author} {\bibfnamefont {P.}~\bibnamefont {{O'Connor}}}, \bibinfo {author} {\bibfnamefont {J.}~\bibnamefont {{Oliver}}}, \bibinfo {author} {\bibfnamefont {S.~S.}\ \bibnamefont {{Olivier}}}, \bibinfo {author} {\bibfnamefont {K.}~\bibnamefont {{Olsen}}}, \bibinfo {author} {\bibfnamefont {W.}~\bibnamefont {{O'Mullane}}}, \bibinfo {author} {\bibfnamefont {S.}~\bibnamefont {{Ortiz}}}, \bibinfo {author} {\bibfnamefont {S.}~\bibnamefont {{Osier}}}, \bibinfo {author} {\bibfnamefont {R.~E.}\ \bibnamefont {{Owen}}}, \bibinfo {author} {\bibfnamefont {R.}~\bibnamefont {{Pain}}}, \bibinfo {author} {\bibfnamefont {P.~E.}\ \bibnamefont {{Palecek}}}, \bibinfo {author} {\bibfnamefont {J.~K.}\ \bibnamefont {{Parejko}}}, \bibinfo {author} {\bibfnamefont {J.~B.}\ \bibnamefont {{Parsons}}}, \bibinfo {author} {\bibfnamefont {N.~M.}\ \bibnamefont {{Pease}}}, \bibinfo {author}
  {\bibfnamefont {J.~M.}\ \bibnamefont {{Peterson}}}, \bibinfo {author} {\bibfnamefont {J.~R.}\ \bibnamefont {{Peterson}}}, \bibinfo {author} {\bibfnamefont {D.~L.}\ \bibnamefont {{Petravick}}}, \bibinfo {author} {\bibfnamefont {M.~E.}\ \bibnamefont {{Libby Petrick}}}, \bibinfo {author} {\bibfnamefont {C.~E.}\ \bibnamefont {{Petry}}}, \bibinfo {author} {\bibfnamefont {F.}~\bibnamefont {{Pierfederici}}}, \bibinfo {author} {\bibfnamefont {S.}~\bibnamefont {{Pietrowicz}}}, \bibinfo {author} {\bibfnamefont {R.}~\bibnamefont {{Pike}}}, \bibinfo {author} {\bibfnamefont {P.~A.}\ \bibnamefont {{Pinto}}}, \bibinfo {author} {\bibfnamefont {R.}~\bibnamefont {{Plante}}}, \bibinfo {author} {\bibfnamefont {S.}~\bibnamefont {{Plate}}}, \bibinfo {author} {\bibfnamefont {J.~P.}\ \bibnamefont {{Plutchak}}}, \bibinfo {author} {\bibfnamefont {P.~A.}\ \bibnamefont {{Price}}}, \bibinfo {author} {\bibfnamefont {M.}~\bibnamefont {{Prouza}}}, \bibinfo {author} {\bibfnamefont {V.}~\bibnamefont {{Radeka}}}, \bibinfo {author}
  {\bibfnamefont {J.}~\bibnamefont {{Rajagopal}}}, \bibinfo {author} {\bibfnamefont {A.~P.}\ \bibnamefont {{Rasmussen}}}, \bibinfo {author} {\bibfnamefont {N.}~\bibnamefont {{Regnault}}}, \bibinfo {author} {\bibfnamefont {K.~A.}\ \bibnamefont {{Reil}}}, \bibinfo {author} {\bibfnamefont {D.~J.}\ \bibnamefont {{Reiss}}}, \bibinfo {author} {\bibfnamefont {M.~A.}\ \bibnamefont {{Reuter}}}, \bibinfo {author} {\bibfnamefont {S.~T.}\ \bibnamefont {{Ridgway}}}, \bibinfo {author} {\bibfnamefont {V.~J.}\ \bibnamefont {{Riot}}}, \bibinfo {author} {\bibfnamefont {S.}~\bibnamefont {{Ritz}}}, \bibinfo {author} {\bibfnamefont {S.}~\bibnamefont {{Robinson}}}, \bibinfo {author} {\bibfnamefont {W.}~\bibnamefont {{Roby}}}, \bibinfo {author} {\bibfnamefont {A.}~\bibnamefont {{Roodman}}}, \bibinfo {author} {\bibfnamefont {W.}~\bibnamefont {{Rosing}}}, \bibinfo {author} {\bibfnamefont {C.}~\bibnamefont {{Roucelle}}}, \bibinfo {author} {\bibfnamefont {M.~R.}\ \bibnamefont {{Rumore}}}, \bibinfo {author} {\bibfnamefont
  {S.}~\bibnamefont {{Russo}}}, \bibinfo {author} {\bibfnamefont {A.}~\bibnamefont {{Saha}}}, \bibinfo {author} {\bibfnamefont {B.}~\bibnamefont {{Sassolas}}}, \bibinfo {author} {\bibfnamefont {T.~L.}\ \bibnamefont {{Schalk}}}, \bibinfo {author} {\bibfnamefont {P.}~\bibnamefont {{Schellart}}}, \bibinfo {author} {\bibfnamefont {R.~H.}\ \bibnamefont {{Schindler}}}, \bibinfo {author} {\bibfnamefont {S.}~\bibnamefont {{Schmidt}}}, \bibinfo {author} {\bibfnamefont {D.~P.}\ \bibnamefont {{Schneider}}}, \bibinfo {author} {\bibfnamefont {M.~D.}\ \bibnamefont {{Schneider}}}, \bibinfo {author} {\bibfnamefont {W.}~\bibnamefont {{Schoening}}}, \bibinfo {author} {\bibfnamefont {G.}~\bibnamefont {{Schumacher}}}, \bibinfo {author} {\bibfnamefont {M.~E.}\ \bibnamefont {{Schwamb}}}, \bibinfo {author} {\bibfnamefont {J.}~\bibnamefont {{Sebag}}}, \bibinfo {author} {\bibfnamefont {B.}~\bibnamefont {{Selvy}}}, \bibinfo {author} {\bibfnamefont {G.~H.}\ \bibnamefont {{Sembroski}}}, \bibinfo {author} {\bibfnamefont {L.~G.}\
  \bibnamefont {{Seppala}}}, \bibinfo {author} {\bibfnamefont {A.}~\bibnamefont {{Serio}}}, \bibinfo {author} {\bibfnamefont {E.}~\bibnamefont {{Serrano}}}, \bibinfo {author} {\bibfnamefont {R.~A.}\ \bibnamefont {{Shaw}}}, \bibinfo {author} {\bibfnamefont {I.}~\bibnamefont {{Shipsey}}}, \bibinfo {author} {\bibfnamefont {J.}~\bibnamefont {{Sick}}}, \bibinfo {author} {\bibfnamefont {N.}~\bibnamefont {{Silvestri}}}, \bibinfo {author} {\bibfnamefont {C.~T.}\ \bibnamefont {{Slater}}}, \bibinfo {author} {\bibfnamefont {J.~A.}\ \bibnamefont {{Smith}}}, \bibinfo {author} {\bibfnamefont {R.~C.}\ \bibnamefont {{Smith}}}, \bibinfo {author} {\bibfnamefont {S.}~\bibnamefont {{Sobhani}}}, \bibinfo {author} {\bibfnamefont {C.}~\bibnamefont {{Soldahl}}}, \bibinfo {author} {\bibfnamefont {L.}~\bibnamefont {{Storrie-Lombardi}}}, \bibinfo {author} {\bibfnamefont {E.}~\bibnamefont {{Stover}}}, \bibinfo {author} {\bibfnamefont {M.~A.}\ \bibnamefont {{Strauss}}}, \bibinfo {author} {\bibfnamefont {R.~A.}\ \bibnamefont {{Street}}},
  \bibinfo {author} {\bibfnamefont {C.~W.}\ \bibnamefont {{Stubbs}}}, \bibinfo {author} {\bibfnamefont {I.~S.}\ \bibnamefont {{Sullivan}}}, \bibinfo {author} {\bibfnamefont {D.}~\bibnamefont {{Sweeney}}}, \bibinfo {author} {\bibfnamefont {J.~D.}\ \bibnamefont {{Swinbank}}}, \bibinfo {author} {\bibfnamefont {A.}~\bibnamefont {{Szalay}}}, \bibinfo {author} {\bibfnamefont {P.}~\bibnamefont {{Takacs}}}, \bibinfo {author} {\bibfnamefont {S.~A.}\ \bibnamefont {{Tether}}}, \bibinfo {author} {\bibfnamefont {J.~J.}\ \bibnamefont {{Thaler}}}, \bibinfo {author} {\bibfnamefont {J.~G.}\ \bibnamefont {{Thayer}}}, \bibinfo {author} {\bibfnamefont {S.}~\bibnamefont {{Thomas}}}, \bibinfo {author} {\bibfnamefont {A.~J.}\ \bibnamefont {{Thornton}}}, \bibinfo {author} {\bibfnamefont {V.}~\bibnamefont {{Thukral}}}, \bibinfo {author} {\bibfnamefont {J.}~\bibnamefont {{Tice}}}, \bibinfo {author} {\bibfnamefont {D.~E.}\ \bibnamefont {{Trilling}}}, \bibinfo {author} {\bibfnamefont {M.}~\bibnamefont {{Turri}}}, \bibinfo {author}
  {\bibfnamefont {R.}~\bibnamefont {{Van Berg}}}, \bibinfo {author} {\bibfnamefont {D.}~\bibnamefont {{Vanden Berk}}}, \bibinfo {author} {\bibfnamefont {K.}~\bibnamefont {{Vetter}}}, \bibinfo {author} {\bibfnamefont {F.}~\bibnamefont {{Virieux}}}, \bibinfo {author} {\bibfnamefont {T.}~\bibnamefont {{Vucina}}}, \bibinfo {author} {\bibfnamefont {W.}~\bibnamefont {{Wahl}}}, \bibinfo {author} {\bibfnamefont {L.}~\bibnamefont {{Walkowicz}}}, \bibinfo {author} {\bibfnamefont {B.}~\bibnamefont {{Walsh}}}, \bibinfo {author} {\bibfnamefont {C.~W.}\ \bibnamefont {{Walter}}}, \bibinfo {author} {\bibfnamefont {D.~L.}\ \bibnamefont {{Wang}}}, \bibinfo {author} {\bibfnamefont {S.-Y.}\ \bibnamefont {{Wang}}}, \bibinfo {author} {\bibfnamefont {M.}~\bibnamefont {{Warner}}}, \bibinfo {author} {\bibfnamefont {O.}~\bibnamefont {{Wiecha}}}, \bibinfo {author} {\bibfnamefont {B.}~\bibnamefont {{Willman}}}, \bibinfo {author} {\bibfnamefont {S.~E.}\ \bibnamefont {{Winters}}}, \bibinfo {author} {\bibfnamefont {D.}~\bibnamefont
  {{Wittman}}}, \bibinfo {author} {\bibfnamefont {S.~C.}\ \bibnamefont {{Wolff}}}, \bibinfo {author} {\bibfnamefont {W.~M.}\ \bibnamefont {{Wood-Vasey}}}, \bibinfo {author} {\bibfnamefont {X.}~\bibnamefont {{Wu}}}, \bibinfo {author} {\bibfnamefont {B.}~\bibnamefont {{Xin}}}, \bibinfo {author} {\bibfnamefont {P.}~\bibnamefont {{Yoachim}}}, \ and\ \bibinfo {author} {\bibfnamefont {H.}~\bibnamefont {{Zhan}}},\ }\href {\doibase 10.3847/1538-4357/ab042c} {\bibfield  {journal} {\bibinfo  {journal} {\apj}\ }\textbf {\bibinfo {volume} {873}},\ \bibinfo {eid} {111} (\bibinfo {year} {2019})},\ \Eprint {http://arxiv.org/abs/0805.2366} {arXiv:0805.2366 [astro-ph]} \BibitemShut {NoStop}%
\bibitem [{\citenamefont {{Sherman}}\ \emph {et~al.}(2025)\citenamefont {{Sherman}}, \citenamefont {{Kosogorov}}, \citenamefont {{Law}}, \citenamefont {{Ravi}}, \citenamefont {{Faber}}, \citenamefont {{Ocker}}, \citenamefont {{Connor}}, \citenamefont {{Qu}}, \citenamefont {{Shin}}, \citenamefont {{Sharma}}, \citenamefont {{Sanghavi}}, \citenamefont {{Hallinan}},\ and\ \citenamefont {{Hodges}}}]{2025sherman}%
  \BibitemOpen
  \bibfield  {author} {\bibinfo {author} {\bibfnamefont {M.~B.}\ \bibnamefont {{Sherman}}}, \bibinfo {author} {\bibfnamefont {N.}~\bibnamefont {{Kosogorov}}}, \bibinfo {author} {\bibfnamefont {C.}~\bibnamefont {{Law}}}, \bibinfo {author} {\bibfnamefont {V.}~\bibnamefont {{Ravi}}}, \bibinfo {author} {\bibfnamefont {J.~T.}\ \bibnamefont {{Faber}}}, \bibinfo {author} {\bibfnamefont {S.~K.}\ \bibnamefont {{Ocker}}}, \bibinfo {author} {\bibfnamefont {L.}~\bibnamefont {{Connor}}}, \bibinfo {author} {\bibfnamefont {Y.}~\bibnamefont {{Qu}}}, \bibinfo {author} {\bibfnamefont {K.}~\bibnamefont {{Shin}}}, \bibinfo {author} {\bibfnamefont {K.}~\bibnamefont {{Sharma}}}, \bibinfo {author} {\bibfnamefont {P.}~\bibnamefont {{Sanghavi}}}, \bibinfo {author} {\bibfnamefont {G.}~\bibnamefont {{Hallinan}}}, \ and\ \bibinfo {author} {\bibfnamefont {M.}~\bibnamefont {{Hodges}}},\ }\href {\doibase 10.48550/arXiv.2510.18136} {\bibfield  {journal} {\bibinfo  {journal} {arXiv e-prints}\ ,\ \bibinfo {eid} {arXiv:2510.18136}} (\bibinfo
  {year} {2025})},\ \Eprint {http://arxiv.org/abs/2510.18136} {arXiv:2510.18136 [astro-ph.HE]} \BibitemShut {NoStop}%
\bibitem [{\citenamefont {{Lee}}\ \emph {et~al.}(2026)\citenamefont {{Lee}}, \citenamefont {{Wang}}, \citenamefont {{Caleb}}, \citenamefont {{Murphy}}, \citenamefont {{An}}, \citenamefont {{Das}}, \citenamefont {{Dobie}}, \citenamefont {{Driessen}}, \citenamefont {{Kaplan}}, \citenamefont {{Lenc}}, \citenamefont {{Pritchard}}, \citenamefont {{Wadiasingh}},\ and\ \citenamefont {{Xu}}}]{2026lee_askap}%
  \BibitemOpen
  \bibfield  {author} {\bibinfo {author} {\bibfnamefont {Y.~W.~J.}\ \bibnamefont {{Lee}}}, \bibinfo {author} {\bibfnamefont {Y.}~\bibnamefont {{Wang}}}, \bibinfo {author} {\bibfnamefont {M.}~\bibnamefont {{Caleb}}}, \bibinfo {author} {\bibfnamefont {T.}~\bibnamefont {{Murphy}}}, \bibinfo {author} {\bibfnamefont {T.}~\bibnamefont {{An}}}, \bibinfo {author} {\bibfnamefont {B.}~\bibnamefont {{Das}}}, \bibinfo {author} {\bibfnamefont {D.}~\bibnamefont {{Dobie}}}, \bibinfo {author} {\bibfnamefont {L.~N.}\ \bibnamefont {{Driessen}}}, \bibinfo {author} {\bibfnamefont {D.~L.}\ \bibnamefont {{Kaplan}}}, \bibinfo {author} {\bibfnamefont {E.}~\bibnamefont {{Lenc}}}, \bibinfo {author} {\bibfnamefont {J.}~\bibnamefont {{Pritchard}}}, \bibinfo {author} {\bibfnamefont {Z.}~\bibnamefont {{Wadiasingh}}}, \ and\ \bibinfo {author} {\bibfnamefont {Z.}~\bibnamefont {{Xu}}},\ }\href {\doibase 10.1093/mnras/staf2008} {\bibfield  {journal} {\bibinfo  {journal} {\mnras}\ }\textbf {\bibinfo {volume} {545}},\ \bibinfo {eid} {staf2008}
  (\bibinfo {year} {2026})},\ \Eprint {http://arxiv.org/abs/2511.09770} {arXiv:2511.09770 [astro-ph.HE]} \BibitemShut {NoStop}%
\bibitem [{\citenamefont {{Connor}}\ and\ \citenamefont {{Ravi}}(2023)}]{2023connor}%
  \BibitemOpen
  \bibfield  {author} {\bibinfo {author} {\bibfnamefont {L.}~\bibnamefont {{Connor}}}\ and\ \bibinfo {author} {\bibfnamefont {V.}~\bibnamefont {{Ravi}}},\ }\href {\doibase 10.1093/mnras/stad667} {\bibfield  {journal} {\bibinfo  {journal} {\mnras}\ }\textbf {\bibinfo {volume} {521}},\ \bibinfo {pages} {4024} (\bibinfo {year} {2023})},\ \Eprint {http://arxiv.org/abs/2206.14310} {arXiv:2206.14310 [astro-ph.CO]} \BibitemShut {NoStop}%
\bibitem [{\citenamefont {{Connor}}\ \emph {et~al.}(2026)\citenamefont {{Connor}}, \citenamefont {{Ravi}}, \citenamefont {{Sanghavi}}, \citenamefont {{Balakrishan}}, \citenamefont {{Chung}}, \citenamefont {{Daghlian}}, \citenamefont {{Dunn}}, \citenamefont {{Griffin}}, \citenamefont {{Harnach}}, \citenamefont {{Hodges}}, \citenamefont {{Jameson}}, \citenamefont {{Gutierrez}}, \citenamefont {{Leung}}, \citenamefont {{Lin}}, \citenamefont {{Mehla}}, \citenamefont {{Modilim}}, \citenamefont {{Patel}}, \citenamefont {{Smith}},\ and\ \citenamefont {{Zeng}}}]{2026casm}%
  \BibitemOpen
  \bibfield  {author} {\bibinfo {author} {\bibfnamefont {L.}~\bibnamefont {{Connor}}}, \bibinfo {author} {\bibfnamefont {V.}~\bibnamefont {{Ravi}}}, \bibinfo {author} {\bibfnamefont {P.}~\bibnamefont {{Sanghavi}}}, \bibinfo {author} {\bibfnamefont {V.}~\bibnamefont {{Balakrishan}}}, \bibinfo {author} {\bibfnamefont {L.}~\bibnamefont {{Chung}}}, \bibinfo {author} {\bibfnamefont {S.}~\bibnamefont {{Daghlian}}}, \bibinfo {author} {\bibfnamefont {L.}~\bibnamefont {{Dunn}}}, \bibinfo {author} {\bibfnamefont {A.}~\bibnamefont {{Griffin}}}, \bibinfo {author} {\bibfnamefont {C.}~\bibnamefont {{Harnach}}}, \bibinfo {author} {\bibfnamefont {M.}~\bibnamefont {{Hodges}}}, \bibinfo {author} {\bibfnamefont {A.}~\bibnamefont {{Jameson}}}, \bibinfo {author} {\bibfnamefont {M.}~\bibnamefont {{Gutierrez}}}, \bibinfo {author} {\bibfnamefont {C.}~\bibnamefont {{Leung}}}, \bibinfo {author} {\bibfnamefont {M.}~\bibnamefont {{Lin}}}, \bibinfo {author} {\bibfnamefont {A.}~\bibnamefont {{Mehla}}}, \bibinfo {author} {\bibfnamefont
  {O.}~\bibnamefont {{Modilim}}}, \bibinfo {author} {\bibfnamefont {N.}~\bibnamefont {{Patel}}}, \bibinfo {author} {\bibfnamefont {K.}~\bibnamefont {{Smith}}}, \ and\ \bibinfo {author} {\bibfnamefont {L.}~\bibnamefont {{Zeng}}},\ }\href {\doibase 10.48550/arXiv.2604.13903} {\bibfield  {journal} {\bibinfo  {journal} {arXiv e-prints}\ ,\ \bibinfo {eid} {arXiv:2604.13903}} (\bibinfo {year} {2026})},\ \Eprint {http://arxiv.org/abs/2604.13903} {arXiv:2604.13903 [astro-ph.IM]} \BibitemShut {NoStop}%
\bibitem [{\citenamefont {{Lin}}\ \emph {et~al.}(2022)\citenamefont {{Lin}}, \citenamefont {{Lin}}, \citenamefont {{Li}}, \citenamefont {{Tseng}}, \citenamefont {{Jiang}}, \citenamefont {{Wang}}, \citenamefont {{Cheng}}, \citenamefont {{Pen}}, \citenamefont {{Chen}}, \citenamefont {{Chen}}, \citenamefont {{Chen}}, \citenamefont {{Goto}}, \citenamefont {{Hashimoto}}, \citenamefont {{Hwang}}, \citenamefont {{King}}, \citenamefont {{Kubo}}, \citenamefont {{Kuo}}, \citenamefont {{Mills}}, \citenamefont {{Nam}}, \citenamefont {{Oshiro}}, \citenamefont {{Shen}}, \citenamefont {{Tseng}}, \citenamefont {{Wang}}, \citenamefont {{Wu}}, \citenamefont {{Bower}}, \citenamefont {{Chang}}, \citenamefont {{Chen}}, \citenamefont {{Chen}}, \citenamefont {{Chiang}}, \citenamefont {{Fedynitch}}, \citenamefont {{Gusinskaia}}, \citenamefont {{Ho}}, \citenamefont {{Hsiao}}, \citenamefont {{Hu}}, \citenamefont {{Huang}}, \citenamefont {{J{\'a}uregui Garc{\'\i}a}}, \citenamefont {{Kim}}, \citenamefont {{Kuo}}, \citenamefont {{Ling}},
  \citenamefont {{On}}, \citenamefont {{Peterson}}, \citenamefont {{R. Raquel}}, \citenamefont {{Su}}, \citenamefont {{Uno}}, \citenamefont {{Wu}}, \citenamefont {{Yamasaki}},\ and\ \citenamefont {{Zhu}}}]{2022burstt}%
  \BibitemOpen
  \bibfield  {author} {\bibinfo {author} {\bibfnamefont {H.-H.}\ \bibnamefont {{Lin}}}, \bibinfo {author} {\bibfnamefont {K.-y.}\ \bibnamefont {{Lin}}}, \bibinfo {author} {\bibfnamefont {C.-T.}\ \bibnamefont {{Li}}}, \bibinfo {author} {\bibfnamefont {Y.-H.}\ \bibnamefont {{Tseng}}}, \bibinfo {author} {\bibfnamefont {H.}~\bibnamefont {{Jiang}}}, \bibinfo {author} {\bibfnamefont {J.-H.}\ \bibnamefont {{Wang}}}, \bibinfo {author} {\bibfnamefont {J.-C.}\ \bibnamefont {{Cheng}}}, \bibinfo {author} {\bibfnamefont {U.-L.}\ \bibnamefont {{Pen}}}, \bibinfo {author} {\bibfnamefont {M.-T.}\ \bibnamefont {{Chen}}}, \bibinfo {author} {\bibfnamefont {P.}~\bibnamefont {{Chen}}}, \bibinfo {author} {\bibfnamefont {Y.}~\bibnamefont {{Chen}}}, \bibinfo {author} {\bibfnamefont {T.}~\bibnamefont {{Goto}}}, \bibinfo {author} {\bibfnamefont {T.}~\bibnamefont {{Hashimoto}}}, \bibinfo {author} {\bibfnamefont {Y.-J.}\ \bibnamefont {{Hwang}}}, \bibinfo {author} {\bibfnamefont {S.-K.}\ \bibnamefont {{King}}}, \bibinfo {author}
  {\bibfnamefont {D.}~\bibnamefont {{Kubo}}}, \bibinfo {author} {\bibfnamefont {C.-Y.}\ \bibnamefont {{Kuo}}}, \bibinfo {author} {\bibfnamefont {A.}~\bibnamefont {{Mills}}}, \bibinfo {author} {\bibfnamefont {J.}~\bibnamefont {{Nam}}}, \bibinfo {author} {\bibfnamefont {P.}~\bibnamefont {{Oshiro}}}, \bibinfo {author} {\bibfnamefont {C.-S.}\ \bibnamefont {{Shen}}}, \bibinfo {author} {\bibfnamefont {H.-C.}\ \bibnamefont {{Tseng}}}, \bibinfo {author} {\bibfnamefont {S.-H.}\ \bibnamefont {{Wang}}}, \bibinfo {author} {\bibfnamefont {V.~F.-S.}\ \bibnamefont {{Wu}}}, \bibinfo {author} {\bibfnamefont {G.}~\bibnamefont {{Bower}}}, \bibinfo {author} {\bibfnamefont {S.-H.}\ \bibnamefont {{Chang}}}, \bibinfo {author} {\bibfnamefont {P.-A.}\ \bibnamefont {{Chen}}}, \bibinfo {author} {\bibfnamefont {Y.-C.}\ \bibnamefont {{Chen}}}, \bibinfo {author} {\bibfnamefont {Y.-K.}\ \bibnamefont {{Chiang}}}, \bibinfo {author} {\bibfnamefont {A.}~\bibnamefont {{Fedynitch}}}, \bibinfo {author} {\bibfnamefont {N.}~\bibnamefont
  {{Gusinskaia}}}, \bibinfo {author} {\bibfnamefont {S.~C.-C.}\ \bibnamefont {{Ho}}}, \bibinfo {author} {\bibfnamefont {T.~Y.-Y.}\ \bibnamefont {{Hsiao}}}, \bibinfo {author} {\bibfnamefont {C.-P.}\ \bibnamefont {{Hu}}}, \bibinfo {author} {\bibfnamefont {Y.~D.}\ \bibnamefont {{Huang}}}, \bibinfo {author} {\bibfnamefont {J.~M.}\ \bibnamefont {{J{\'a}uregui Garc{\'\i}a}}}, \bibinfo {author} {\bibfnamefont {S.~J.}\ \bibnamefont {{Kim}}}, \bibinfo {author} {\bibfnamefont {C.-Y.}\ \bibnamefont {{Kuo}}}, \bibinfo {author} {\bibfnamefont {D.~F.-J.}\ \bibnamefont {{Ling}}}, \bibinfo {author} {\bibfnamefont {A.~Y.~L.}\ \bibnamefont {{On}}}, \bibinfo {author} {\bibfnamefont {J.~B.}\ \bibnamefont {{Peterson}}}, \bibinfo {author} {\bibfnamefont {B.~J.}\ \bibnamefont {{R. Raquel}}}, \bibinfo {author} {\bibfnamefont {S.-C.}\ \bibnamefont {{Su}}}, \bibinfo {author} {\bibfnamefont {Y.}~\bibnamefont {{Uno}}}, \bibinfo {author} {\bibfnamefont {C.~K.-W.}\ \bibnamefont {{Wu}}}, \bibinfo {author} {\bibfnamefont {S.}~\bibnamefont
  {{Yamasaki}}}, \ and\ \bibinfo {author} {\bibfnamefont {H.-M.}\ \bibnamefont {{Zhu}}},\ }\href {\doibase 10.1088/1538-3873/ac8f71} {\bibfield  {journal} {\bibinfo  {journal} {\pasp}\ }\textbf {\bibinfo {volume} {134}},\ \bibinfo {eid} {094106} (\bibinfo {year} {2022})},\ \Eprint {http://arxiv.org/abs/2206.08983} {arXiv:2206.08983 [astro-ph.IM]} \BibitemShut {NoStop}%
\bibitem [{\citenamefont {{Vanderlinde}}\ \emph {et~al.}(2019)\citenamefont {{Vanderlinde}}, \citenamefont {{Liu}}, \citenamefont {{Gaensler}}, \citenamefont {{Bond}}, \citenamefont {{Hinshaw}}, \citenamefont {{Ng}}, \citenamefont {{Chiang}}, \citenamefont {{Stairs}}, \citenamefont {{Brown}}, \citenamefont {{Sievers}}, \citenamefont {{Mena}}, \citenamefont {{Smith}}, \citenamefont {{Bandura}}, \citenamefont {{Masui}}, \citenamefont {{Spekkens}}, \citenamefont {{Belostotski}}, \citenamefont {{Dobbs}}, \citenamefont {{Turok}}, \citenamefont {{Boyle}}, \citenamefont {{Rupen}}, \citenamefont {{Landecker}}, \citenamefont {{Pen}},\ and\ \citenamefont {{Kaspi}}}]{2019chord}%
  \BibitemOpen
  \bibfield  {author} {\bibinfo {author} {\bibfnamefont {K.}~\bibnamefont {{Vanderlinde}}}, \bibinfo {author} {\bibfnamefont {A.}~\bibnamefont {{Liu}}}, \bibinfo {author} {\bibfnamefont {B.}~\bibnamefont {{Gaensler}}}, \bibinfo {author} {\bibfnamefont {D.}~\bibnamefont {{Bond}}}, \bibinfo {author} {\bibfnamefont {G.}~\bibnamefont {{Hinshaw}}}, \bibinfo {author} {\bibfnamefont {C.}~\bibnamefont {{Ng}}}, \bibinfo {author} {\bibfnamefont {C.}~\bibnamefont {{Chiang}}}, \bibinfo {author} {\bibfnamefont {I.}~\bibnamefont {{Stairs}}}, \bibinfo {author} {\bibfnamefont {J.-A.}\ \bibnamefont {{Brown}}}, \bibinfo {author} {\bibfnamefont {J.}~\bibnamefont {{Sievers}}}, \bibinfo {author} {\bibfnamefont {J.}~\bibnamefont {{Mena}}}, \bibinfo {author} {\bibfnamefont {K.}~\bibnamefont {{Smith}}}, \bibinfo {author} {\bibfnamefont {K.}~\bibnamefont {{Bandura}}}, \bibinfo {author} {\bibfnamefont {K.}~\bibnamefont {{Masui}}}, \bibinfo {author} {\bibfnamefont {K.}~\bibnamefont {{Spekkens}}}, \bibinfo {author} {\bibfnamefont
  {L.}~\bibnamefont {{Belostotski}}}, \bibinfo {author} {\bibfnamefont {M.}~\bibnamefont {{Dobbs}}}, \bibinfo {author} {\bibfnamefont {N.}~\bibnamefont {{Turok}}}, \bibinfo {author} {\bibfnamefont {P.}~\bibnamefont {{Boyle}}}, \bibinfo {author} {\bibfnamefont {M.}~\bibnamefont {{Rupen}}}, \bibinfo {author} {\bibfnamefont {T.}~\bibnamefont {{Landecker}}}, \bibinfo {author} {\bibfnamefont {U.-L.}\ \bibnamefont {{Pen}}}, \ and\ \bibinfo {author} {\bibfnamefont {V.}~\bibnamefont {{Kaspi}}},\ }in\ \href {\doibase 10.5281/zenodo.3765414} {\emph {\bibinfo {booktitle} {Canadian Long Range Plan for Astronomy and Astrophysics White Papers}}},\ Vol.\ \bibinfo {volume} {2020}\ (\bibinfo {year} {2019})\ p.~\bibinfo {pages} {28},\ \Eprint {http://arxiv.org/abs/1911.01777} {arXiv:1911.01777 [astro-ph.IM]} \BibitemShut {NoStop}%
\bibitem [{\citenamefont {{Hallinan}}\ \emph {et~al.}(2019)\citenamefont {{Hallinan}}, \citenamefont {{Ravi}}, \citenamefont {{Weinreb}}, \citenamefont {{Kocz}}, \citenamefont {{Huang}}, \citenamefont {{Woody}}, \citenamefont {{Lamb}}, \citenamefont {{D'Addario}}, \citenamefont {{Catha}}, \citenamefont {{Law}}, \citenamefont {{Kulkarni}}, \citenamefont {{Phinney}}, \citenamefont {{Eastwood}}, \citenamefont {{Bouman}}, \citenamefont {{McLaughlin}}, \citenamefont {{Ransom}}, \citenamefont {{Siemens}}, \citenamefont {{Cordes}}, \citenamefont {{Lynch}}, \citenamefont {{Kaplan}}, \citenamefont {{Brazier}}, \citenamefont {{Bhatnagar}}, \citenamefont {{Myers}}, \citenamefont {{Walter}},\ and\ \citenamefont {{Gaensler}}}]{2019hallinan}%
  \BibitemOpen
  \bibfield  {author} {\bibinfo {author} {\bibfnamefont {G.}~\bibnamefont {{Hallinan}}}, \bibinfo {author} {\bibfnamefont {V.}~\bibnamefont {{Ravi}}}, \bibinfo {author} {\bibfnamefont {S.}~\bibnamefont {{Weinreb}}}, \bibinfo {author} {\bibfnamefont {J.}~\bibnamefont {{Kocz}}}, \bibinfo {author} {\bibfnamefont {Y.}~\bibnamefont {{Huang}}}, \bibinfo {author} {\bibfnamefont {D.~P.}\ \bibnamefont {{Woody}}}, \bibinfo {author} {\bibfnamefont {J.}~\bibnamefont {{Lamb}}}, \bibinfo {author} {\bibfnamefont {L.}~\bibnamefont {{D'Addario}}}, \bibinfo {author} {\bibfnamefont {M.}~\bibnamefont {{Catha}}}, \bibinfo {author} {\bibfnamefont {C.}~\bibnamefont {{Law}}}, \bibinfo {author} {\bibfnamefont {S.~R.}\ \bibnamefont {{Kulkarni}}}, \bibinfo {author} {\bibfnamefont {E.~S.}\ \bibnamefont {{Phinney}}}, \bibinfo {author} {\bibfnamefont {M.~W.}\ \bibnamefont {{Eastwood}}}, \bibinfo {author} {\bibfnamefont {K.}~\bibnamefont {{Bouman}}}, \bibinfo {author} {\bibfnamefont {M.}~\bibnamefont {{McLaughlin}}}, \bibinfo {author}
  {\bibfnamefont {S.}~\bibnamefont {{Ransom}}}, \bibinfo {author} {\bibfnamefont {X.}~\bibnamefont {{Siemens}}}, \bibinfo {author} {\bibfnamefont {J.}~\bibnamefont {{Cordes}}}, \bibinfo {author} {\bibfnamefont {R.}~\bibnamefont {{Lynch}}}, \bibinfo {author} {\bibfnamefont {D.}~\bibnamefont {{Kaplan}}}, \bibinfo {author} {\bibfnamefont {A.}~\bibnamefont {{Brazier}}}, \bibinfo {author} {\bibfnamefont {S.}~\bibnamefont {{Bhatnagar}}}, \bibinfo {author} {\bibfnamefont {S.}~\bibnamefont {{Myers}}}, \bibinfo {author} {\bibfnamefont {F.}~\bibnamefont {{Walter}}}, \ and\ \bibinfo {author} {\bibfnamefont {B.}~\bibnamefont {{Gaensler}}},\ }in\ \href {\doibase 10.48550/arXiv.1907.07648} {\emph {\bibinfo {booktitle} {Bulletin of the American Astronomical Society}}},\ Vol.~\bibinfo {volume} {51}\ (\bibinfo {year} {2019})\ p.\ \bibinfo {pages} {255},\ \Eprint {http://arxiv.org/abs/1907.07648} {arXiv:1907.07648 [astro-ph.IM]} \BibitemShut {NoStop}%
\bibitem [{\citenamefont {{Tingay}}\ \emph {et~al.}(2013)\citenamefont {{Tingay}}, \citenamefont {{Goeke}}, \citenamefont {{Bowman}}, \citenamefont {{Emrich}}, \citenamefont {{Ord}}, \citenamefont {{Mitchell}}, \citenamefont {{Morales}}, \citenamefont {{Booler}}, \citenamefont {{Crosse}}, \citenamefont {{Wayth}}, \citenamefont {{Lonsdale}}, \citenamefont {{Tremblay}}, \citenamefont {{Pallot}}, \citenamefont {{Colegate}}, \citenamefont {{Wicenec}}, \citenamefont {{Kudryavtseva}}, \citenamefont {{Arcus}}, \citenamefont {{Barnes}}, \citenamefont {{Bernardi}}, \citenamefont {{Briggs}}, \citenamefont {{Burns}}, \citenamefont {{Bunton}}, \citenamefont {{Cappallo}}, \citenamefont {{Corey}}, \citenamefont {{Deshpande}}, \citenamefont {{Desouza}}, \citenamefont {{Gaensler}}, \citenamefont {{Greenhill}}, \citenamefont {{Hall}}, \citenamefont {{Hazelton}}, \citenamefont {{Herne}}, \citenamefont {{Hewitt}}, \citenamefont {{Johnston-Hollitt}}, \citenamefont {{Kaplan}}, \citenamefont {{Kasper}}, \citenamefont {{Kincaid}},
  \citenamefont {{Koenig}}, \citenamefont {{Kratzenberg}}, \citenamefont {{Lynch}}, \citenamefont {{Mckinley}}, \citenamefont {{Mcwhirter}}, \citenamefont {{Morgan}}, \citenamefont {{Oberoi}}, \citenamefont {{Pathikulangara}}, \citenamefont {{Prabu}}, \citenamefont {{Remillard}}, \citenamefont {{Rogers}}, \citenamefont {{Roshi}}, \citenamefont {{Salah}}, \citenamefont {{Sault}}, \citenamefont {{Udaya-Shankar}}, \citenamefont {{Schlagenhaufer}}, \citenamefont {{Srivani}}, \citenamefont {{Stevens}}, \citenamefont {{Subrahmanyan}}, \citenamefont {{Waterson}}, \citenamefont {{Webster}}, \citenamefont {{Whitney}}, \citenamefont {{Williams}}, \citenamefont {{Williams}},\ and\ \citenamefont {{Wyithe}}}]{2013ska}%
  \BibitemOpen
  \bibfield  {author} {\bibinfo {author} {\bibfnamefont {S.~J.}\ \bibnamefont {{Tingay}}}, \bibinfo {author} {\bibfnamefont {R.}~\bibnamefont {{Goeke}}}, \bibinfo {author} {\bibfnamefont {J.~D.}\ \bibnamefont {{Bowman}}}, \bibinfo {author} {\bibfnamefont {D.}~\bibnamefont {{Emrich}}}, \bibinfo {author} {\bibfnamefont {S.~M.}\ \bibnamefont {{Ord}}}, \bibinfo {author} {\bibfnamefont {D.~A.}\ \bibnamefont {{Mitchell}}}, \bibinfo {author} {\bibfnamefont {M.~F.}\ \bibnamefont {{Morales}}}, \bibinfo {author} {\bibfnamefont {T.}~\bibnamefont {{Booler}}}, \bibinfo {author} {\bibfnamefont {B.}~\bibnamefont {{Crosse}}}, \bibinfo {author} {\bibfnamefont {R.~B.}\ \bibnamefont {{Wayth}}}, \bibinfo {author} {\bibfnamefont {C.~J.}\ \bibnamefont {{Lonsdale}}}, \bibinfo {author} {\bibfnamefont {S.}~\bibnamefont {{Tremblay}}}, \bibinfo {author} {\bibfnamefont {D.}~\bibnamefont {{Pallot}}}, \bibinfo {author} {\bibfnamefont {T.}~\bibnamefont {{Colegate}}}, \bibinfo {author} {\bibfnamefont {A.}~\bibnamefont {{Wicenec}}}, \bibinfo
  {author} {\bibfnamefont {N.}~\bibnamefont {{Kudryavtseva}}}, \bibinfo {author} {\bibfnamefont {W.}~\bibnamefont {{Arcus}}}, \bibinfo {author} {\bibfnamefont {D.}~\bibnamefont {{Barnes}}}, \bibinfo {author} {\bibfnamefont {G.}~\bibnamefont {{Bernardi}}}, \bibinfo {author} {\bibfnamefont {F.}~\bibnamefont {{Briggs}}}, \bibinfo {author} {\bibfnamefont {S.}~\bibnamefont {{Burns}}}, \bibinfo {author} {\bibfnamefont {J.~D.}\ \bibnamefont {{Bunton}}}, \bibinfo {author} {\bibfnamefont {R.~J.}\ \bibnamefont {{Cappallo}}}, \bibinfo {author} {\bibfnamefont {B.~E.}\ \bibnamefont {{Corey}}}, \bibinfo {author} {\bibfnamefont {A.}~\bibnamefont {{Deshpande}}}, \bibinfo {author} {\bibfnamefont {L.}~\bibnamefont {{Desouza}}}, \bibinfo {author} {\bibfnamefont {B.~M.}\ \bibnamefont {{Gaensler}}}, \bibinfo {author} {\bibfnamefont {L.~J.}\ \bibnamefont {{Greenhill}}}, \bibinfo {author} {\bibfnamefont {P.~J.}\ \bibnamefont {{Hall}}}, \bibinfo {author} {\bibfnamefont {B.~J.}\ \bibnamefont {{Hazelton}}}, \bibinfo {author}
  {\bibfnamefont {D.}~\bibnamefont {{Herne}}}, \bibinfo {author} {\bibfnamefont {J.~N.}\ \bibnamefont {{Hewitt}}}, \bibinfo {author} {\bibfnamefont {M.}~\bibnamefont {{Johnston-Hollitt}}}, \bibinfo {author} {\bibfnamefont {D.~L.}\ \bibnamefont {{Kaplan}}}, \bibinfo {author} {\bibfnamefont {J.~C.}\ \bibnamefont {{Kasper}}}, \bibinfo {author} {\bibfnamefont {B.~B.}\ \bibnamefont {{Kincaid}}}, \bibinfo {author} {\bibfnamefont {R.}~\bibnamefont {{Koenig}}}, \bibinfo {author} {\bibfnamefont {E.}~\bibnamefont {{Kratzenberg}}}, \bibinfo {author} {\bibfnamefont {M.~J.}\ \bibnamefont {{Lynch}}}, \bibinfo {author} {\bibfnamefont {B.}~\bibnamefont {{Mckinley}}}, \bibinfo {author} {\bibfnamefont {S.~R.}\ \bibnamefont {{Mcwhirter}}}, \bibinfo {author} {\bibfnamefont {E.}~\bibnamefont {{Morgan}}}, \bibinfo {author} {\bibfnamefont {D.}~\bibnamefont {{Oberoi}}}, \bibinfo {author} {\bibfnamefont {J.}~\bibnamefont {{Pathikulangara}}}, \bibinfo {author} {\bibfnamefont {T.}~\bibnamefont {{Prabu}}}, \bibinfo {author}
  {\bibfnamefont {R.~A.}\ \bibnamefont {{Remillard}}}, \bibinfo {author} {\bibfnamefont {A.~E.~E.}\ \bibnamefont {{Rogers}}}, \bibinfo {author} {\bibfnamefont {A.}~\bibnamefont {{Roshi}}}, \bibinfo {author} {\bibfnamefont {J.~E.}\ \bibnamefont {{Salah}}}, \bibinfo {author} {\bibfnamefont {R.~J.}\ \bibnamefont {{Sault}}}, \bibinfo {author} {\bibfnamefont {N.}~\bibnamefont {{Udaya-Shankar}}}, \bibinfo {author} {\bibfnamefont {F.}~\bibnamefont {{Schlagenhaufer}}}, \bibinfo {author} {\bibfnamefont {K.~S.}\ \bibnamefont {{Srivani}}}, \bibinfo {author} {\bibfnamefont {J.}~\bibnamefont {{Stevens}}}, \bibinfo {author} {\bibfnamefont {R.}~\bibnamefont {{Subrahmanyan}}}, \bibinfo {author} {\bibfnamefont {M.}~\bibnamefont {{Waterson}}}, \bibinfo {author} {\bibfnamefont {R.~L.}\ \bibnamefont {{Webster}}}, \bibinfo {author} {\bibfnamefont {A.~R.}\ \bibnamefont {{Whitney}}}, \bibinfo {author} {\bibfnamefont {A.}~\bibnamefont {{Williams}}}, \bibinfo {author} {\bibfnamefont {C.~L.}\ \bibnamefont {{Williams}}}, \ and\
  \bibinfo {author} {\bibfnamefont {J.~S.~B.}\ \bibnamefont {{Wyithe}}},\ }\href {\doibase 10.1017/pasa.2012.007} {\bibfield  {journal} {\bibinfo  {journal} {\pasa}\ }\textbf {\bibinfo {volume} {30}},\ \bibinfo {eid} {e007} (\bibinfo {year} {2013})},\ \Eprint {http://arxiv.org/abs/1206.6945} {arXiv:1206.6945 [astro-ph.IM]} \BibitemShut {NoStop}%
\end{thebibliography}%

\begin{appendix}

\section{Extended Data}

\begin{figure}
    \centering
    \includegraphics[width=\textwidth]{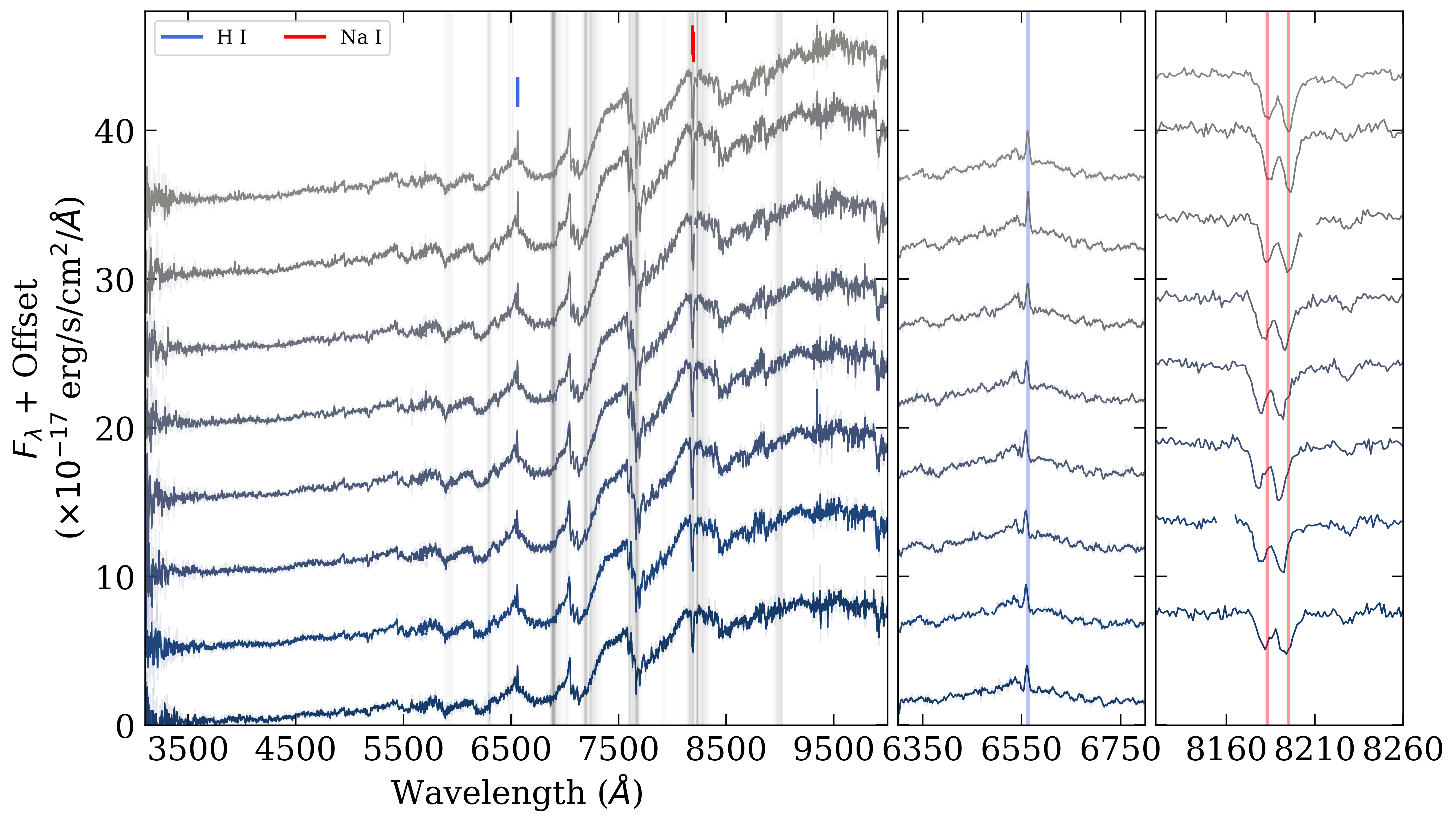}
    \caption{Individual spectra of ILT J1101 taken by Keck I/LRIS on 26 January 2025. H$\alpha$ emission is present in every spectrum, as is the Na I doublet. Gray bands indicate telluric features.}
    \label{fig:all_spec}
\end{figure}

\begin{figure}
    \centering
    \includegraphics[width=0.45\linewidth]{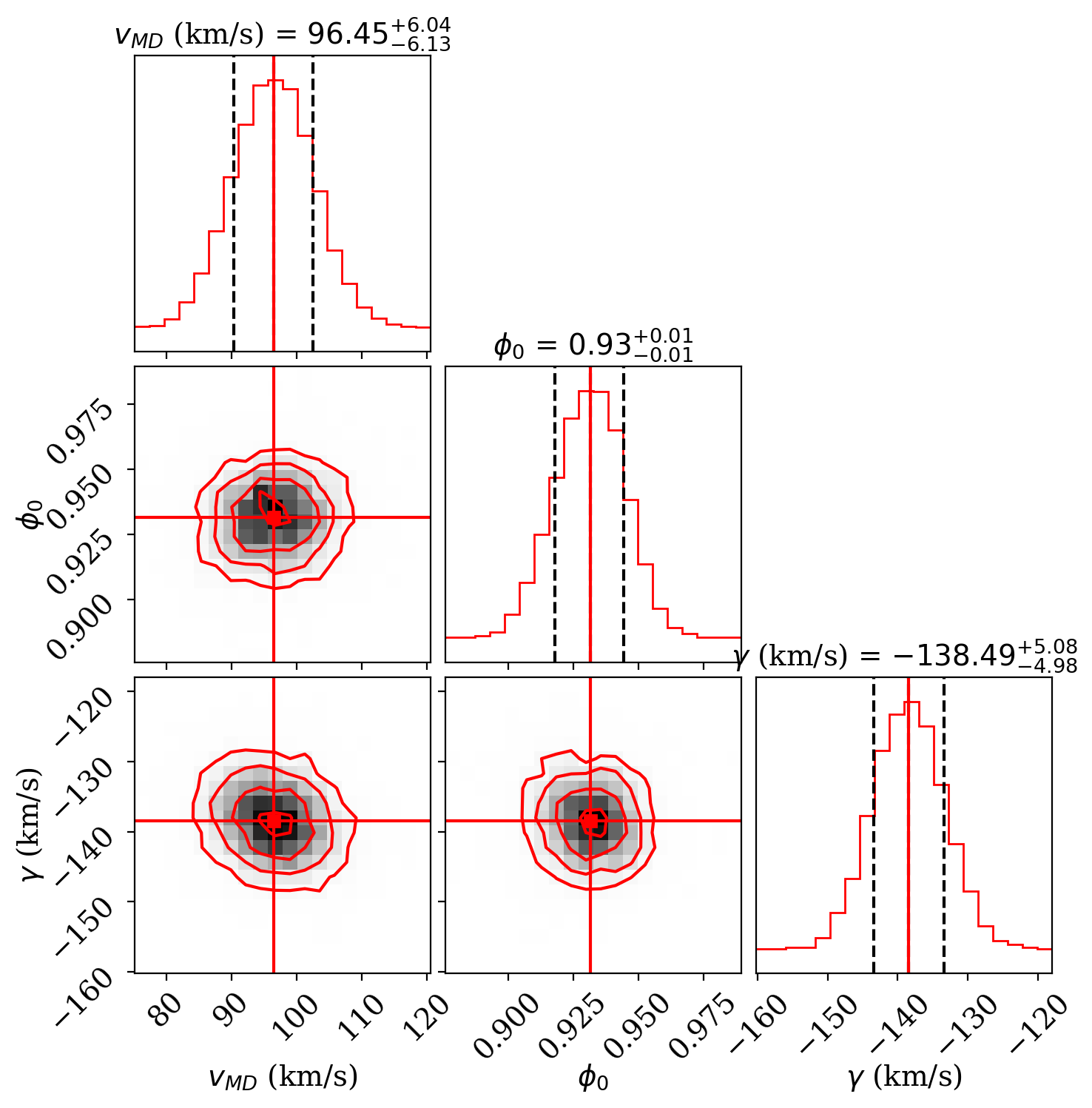}    \includegraphics[width=0.45\linewidth]{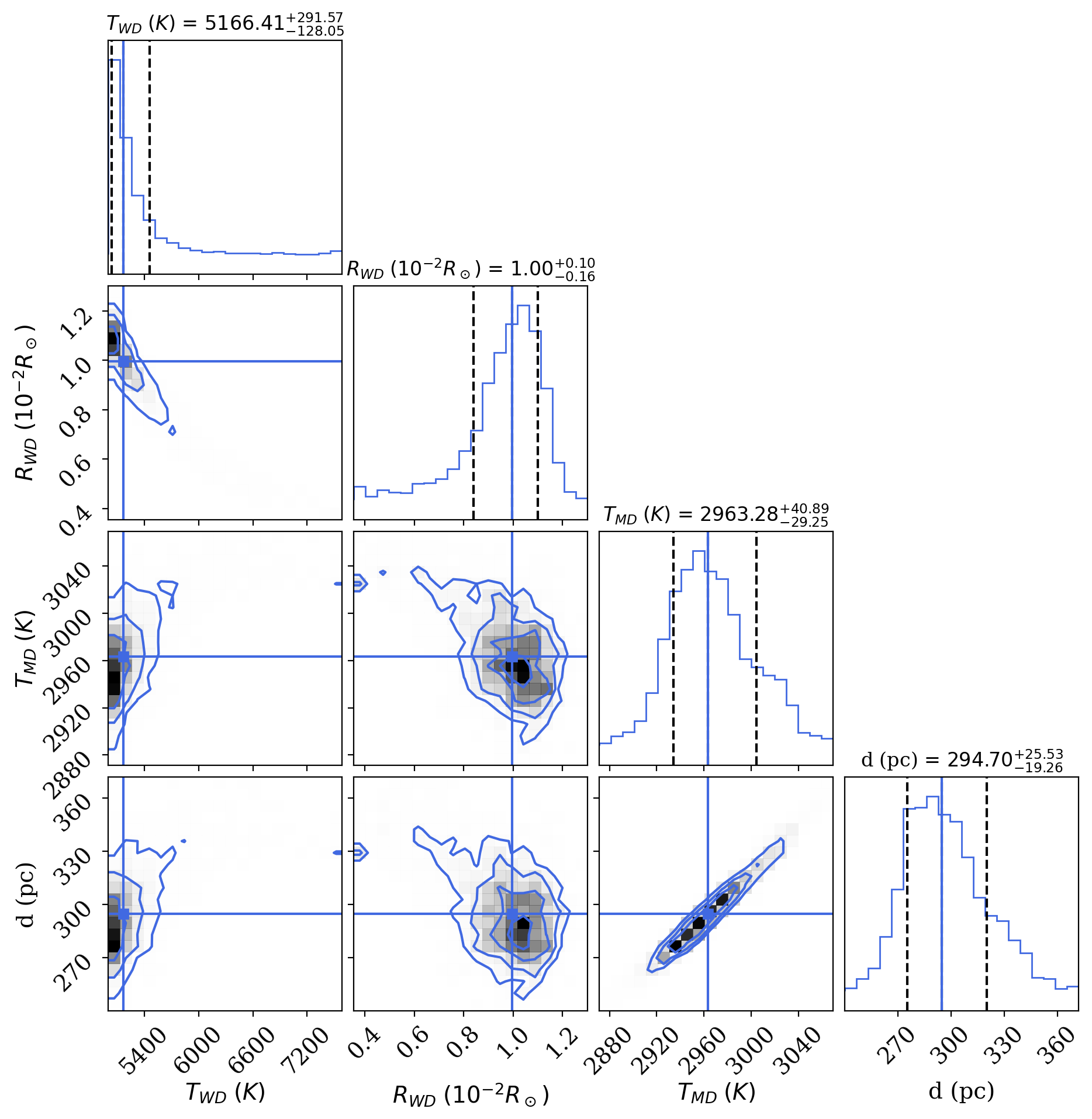}
    \caption{MCMC corner plots to constrain the orbital solution (\textit{left}) and binary parameters (\textit{right}) of ILT J1101. The average spectrum across all orbital phases was taken as the data, while the model adopted was a WD + M dwarf detached binary, as described in Section \ref{sec:parameters}.}
    \label{fig:corner}
\end{figure}

Figure \ref{fig:all_spec} shows each individual spectrum, offset by equal amounts, of ILT J1101 taken by Keck I/LRIS on 26 January 2025.
Figure \ref{fig:corner} shows the MCMC corner plots to constrain the orbital solution and binary parameters of ILT J1101.

\end{appendix}

\end{document}